\documentclass[aps, prd, twocolumn,preprintnumbers, amssymb,amsmath,aps,floatfix,nofootinbib,superscriptaddress,showpacs]{revtex4-1}

\usepackage{natbib}
\usepackage{epsfig}
\usepackage{bm}
\usepackage{amssymb}
\usepackage{amsmath}
\usepackage{color}
\usepackage{subfigure}
\usepackage[colorlinks,
            linkcolor=blue,
            anchorcolor=black,
            citecolor=blue
            ]{hyperref}

\usepackage{soul}

\setstcolor{blue}

\begin{document}

\title{Harmonics of Lepton-Jet Correlations in inclusive and diffractive scatterings}

\affiliation{School of Science and Engineering, The Chinese University of Hong Kong, Shenzhen, Shenzhen, Guangdong, 518172, P.R. China}

\author{Xuan-Bo Tong}\email{xuan.bo.tong@jyu.fi} 

\affiliation{Department of Physics, University of Jyväskylä, P.O. Box 35, 40014 University of Jyväskylä, Finland}
\affiliation{Helsinki Institute of Physics, P.O. Box 64, 00014 University of Helsinki, Finland}
\affiliation{School of Science and Engineering, The Chinese University of Hong Kong, Shenzhen, Shenzhen, Guangdong, 518172, P.R. China}

\author{Bo-Wen Xiao}\email{xiaobowen@cuhk.edu.cn} 
\affiliation{School of Science and Engineering, The Chinese University of Hong Kong, Shenzhen, Shenzhen, Guangdong, 518172, P.R. China}

\author{Yuan-Yuan Zhang}
\email{yuanyuan.zhang@iat.cn} 
\affiliation{Shandong Institute of Advanced Technology, Jinan 250100, China}
\affiliation{School of Science and Engineering, The Chinese University of Hong Kong, Shenzhen, Shenzhen, Guangdong, 518172, P.R. China}

\begin{abstract}
Based on our previous work, we study the harmonic coefficient of both inclusive and diffractive azimuthal angle dependent lepton-jet correlations in Hadron-Electron Ring Accelerator and the future electron-ion collider. Numerical calculations for inclusive and diffractive harmonics and the ratio of harmonics in $e+\text{Au}$~and $e+p$ indicate their strong discriminating power for non-saturation model and saturation model.  Additionally, we demonstrate that the t-dependent diffractive harmonics can serve as novel observables for nuclear density profile.
\end{abstract}  
\maketitle

\section{Introduction}
In a recent paper~\cite{Tong:2022zwp}, we have demonstrated how the harmonics of lepton-jet correlation can serve as a new probe for saturation phenomenon in deeply inelastic scattering~(DIS). In this paper, we present a more detailed elaboration on harmonics of lepton-jet correlation and extend the discussion to diffractive lepton-jet production. 

Gluon saturation~\cite{Gribov:1983ivg,Mueller:1985wy,Mueller:1989st,McLerran:1993ni,McLerran:1993ka,McLerran:1994vd} is a phenomenon observed in proton/nucleus of high energy collisions. Large-$x$ partons radiate small-$x$ gluons, resulting in an increase in the density of small-$x$ gluons. These small-$x$ gluons come into close proximity, interact and recombine. These two effects compete until the small-$x$ gluon density saturates. The typical transverse momentum associated with saturated gluons is refered to as the saturation scale $Q_s$. 

The Color Glass Condensate (CGC) effective theory 
is the theoretical framework used to describe saturated gluons. In the CGC effective theory, large-$x$ parton are treated as static and localized color sources, while small-$x$ partons are modeled as classical and dynamical fields. The relationship between the sources and fields is governed by the classical Yang-Mills equation. When considering the interaction of an energetic parton with the classical field, light-like Wilson lines emerge, which resum the multiple interactions between the high energy parton and the classical field. The two-point correlator of a quark wilson line and an anti-quark wilson line yields the dipole scattering matrix. The Wilson lines and dipole scattering matrix are the building blocks of small-$x$ physics. More detailed descriptions of the CGC framework can be found in the reviews~\cite{Morreale:2021pnn,Gelis:2010nm,Iancu:2003xm,Kovchegov:2012mbw,Albacete:2014fwa,Blaizot:2016qgz}.

\begin{figure}
\centering
\includegraphics[scale=0.5]{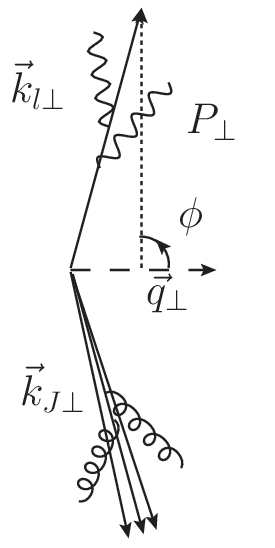}
 \caption{Lepton-jet in transverse plane perpendicular to beam direction. The final jet radiates soft gluons, while the final lepton emits soft photons.}  
 \label{fig:lepjet_gluonphoton}
\end{figure}

Two particle correlation, such as di-jet~\cite{Gelis:2002nn,Baier:2005dv,Dominguez:2010xd,Dominguez:2011wm,Mueller:2013wwa,Dumitru:2015gaa,Dumitru:2016jku,Boer:2016fqd,Dumitru:2018kuw,Zhao:2021kae,Boussarie:2021ybe,Taels:2022tza,Caucal:2022ulg,Caucal:2023nci,Caucal:2023fsf,Caucal:2021ent,Zhang:2021tcc,Metz:2011wb,Dominguez:2011br,Boussarie:2014lxa,Boussarie:2016ogo,Salazar:2019ncp,Boussarie:2019ero,Mantysaari:2019hkq,Boer:2021upt,Iancu:2021rup,Iancu:2022lcw,Hatta:2016dxp,Altinoluk:2015dpi,Mantysaari:2019csc,Fucilla:2022wcg,Rodriguez-Aguilar:2023ihz,Altinoluk:2022jkk,Altinoluk:2023qfr}, di-hadron~\cite{Hagiwara:2021xkf,Zheng:2014vka,Bergabo:2021woe,Bergabo:2022tcu,Iancu:2022gpw,Bergabo:2023wed,Fucilla:2022wcg} and jet plus color-neutral particle~\cite{Dominguez:2011wm,Kolbe:2020tlq} has been extensively utilized to employ various aspects of saturation in the future electron-ion collider (EIC)~\cite{Boer:2011fh,Accardi:2012qut,AbdulKhalek:2021gbh,AbdulKhalek:2022hcn,Abir:2023fpo}. These correlations allow for investigations into Weizs\"{a}cker-Williams gluon distributions~\cite{Dominguez:2010xd,Dominguez:2011wm,Mueller:2013wwa,Dumitru:2015gaa,Dumitru:2016jku,Boer:2016fqd,Dumitru:2018kuw,Zhao:2021kae,Boussarie:2021ybe,Caucal:2022ulg,Taels:2022tza,Caucal:2023nci,Caucal:2023fsf}, including the linear polarized one~\cite{Metz:2011wb, Dominguez:2011br,Dumitru:2015gaa,Dumitru:2016jku,Dumitru:2018kuw,Zhao:2021kae,Boussarie:2021ybe,Taels:2022tza,Caucal:2022ulg}. The measurement of unpolarized dipole gluon distribution~\cite{Dominguez:2010xd,Dominguez:2011wm} and its linear polarized counterpart~\cite{Metz:2011wb, Dominguez:2011br} is also possible. Furthermore, multi-gluon correlations within the nucleus target can be probed~\cite{Gelis:2002nn,Baier:2005dv,Mantysaari:2019hkq,Caucal:2021ent}, and the Wigner funtion can be investigated within the small-$x$ framework~\cite{Hatta:2016dxp,Mantysaari:2019csc, Hagiwara:2021xkf}.
The separation of Sudakov resummation and small-$x$ resummation has been elucidated~\cite{Mueller:2012uf,Mueller:2013wwa}.
Besides EIC, the two particle correlations have also been extensively discussed in LHC and RHIC~(see e.g.,~\cite{Dominguez:2010xd,Dominguez:2011wm,Mueller:2013wwa,Jalilian-Marian:2004vhw,Kharzeev:2004bw,Marquet:2007vb,Tuchin:2009nf,Dumitru:2010ak,Kutak:2012rf,vanHameren:2014ala,Kotko:2015ura,vanHameren:2016ftb,vanHameren:2019ysa,vanHameren:2020rqt,vanHameren:2023oiq,Marquet:2016cgx,Klein:2019qfb,Iancu:2020mos,Bolognino:2021mrc,Al-Mashad:2022zbq,Agostini:2022oge,Iancu:2023lel,Hagiwara:2017fye,Kotko:2017oxg,Bhattacharya:2018lgm,Boussarie:2018zwg,Albacete:2010pg,Stasto:2011ru,Lappi:2012nh,Iancu:2013dta,Albacete:2018ruq,Stasto:2018rci,Jalilian-Marian:2005qbq,Jalilian-Marian:2012wwi,Stasto:2012ru,Rezaeian:2012wa,Basso:2015pba,Rezaeian:2016szi,Basso:2016ulb,Boer:2017xpy,Benic:2017znu,Goncalves:2020tvh,Benic:2022ixp,Boer:2022njw,Gelis:2002fw,Kovner:2014qea,Kovner:2015rna,Marquet:2019ltn,Akcakaya:2012si,Marquet:2017xwy,Ju:2022wia,Ganguli:2023joy}).

Typically, two-particle correlation exhibit a back-to-back configuration in the transverse plane perpendicular to the beam direction. This is referred to as the correlation limit, where the imbalance momentum $|\vec{q}_{\perp}|=|\vec{k}_{1\perp}+\vec{k}_{2\perp}|$ is much softer than the relative momentum $|\vec{P}_{\perp}|=|(\vec{k}_{1\perp}-\vec{k}_{2\perp})/2|$. In this limit, the soft imbalanced momentum can reach the saturation region  $|\vec{q}_{\perp}|\lesssim Q_s$. Thus the two particle correlation in the correlation limit serves as a robust probe for saturation.

A recent addition to the repertoire of two-particle 
correlation is the lepton-jet correlation in DIS~\cite{Liu:2018trl,Liu:2020dct}. Fig.~\ref{fig:lepjet_gluonphoton} shows the lepton-jet production in the transverse plane, with final jet radiating gluons and final lepton emitting photons. Lepton-jet correlation offers a valuable avenue for studying the transverse momentum dependent~(TMD) quark distribution, as well as TMD quark Sivers function~\cite{Liu:2018trl,Liu:2020dct,Arratia:2020nxw,Kang:2021ffh,Arratia:2022oxd,Kang:2020fka,Yang:2023vyv,Yang:2023zod}, and Collins fragmentation function~\cite{Arratia:2020nxw,Kang:2021ffh,Arratia:2022oxd}. Notably, in the small-$x$ region, the TMD quark distribution within the small-$x$ framework contains critical information about gluon saturation. Consequently, the lepton-jet correlation emerges as an opportunity to probe this intriguing phenomenon~\cite{Tong:2022zwp}.

\par

Following the theoretical papers on the lepton-jet correlation, more and more experimental studies have emerged. The first measurement of lepton-jet correlation at Hadron-Electron Ring Accelerator (HERA)~\cite{Abramowicz:1998ii} with the H1 detector has been published~\cite{H1:2021wkz}. Additionally, event generation and detector-response simulations are conducted to investigate the lepton-jet production at the EIC kinematics~\cite{Arratia:2019vju,Arratia:2020nxw,Arratia:2022oxd}.  

The azimuthal angle anisotropy or harmonics of the lepton-jet correlation serves as the observable in the search for gluon saturation. Previous studies have indicated that the azimuthal angle anisotropy is caused by soft gluon radiation~\cite{Hatta:2021jcd, Hatta:2020bgy}, as the soft gluon radiation from the final jet tends to align with the final jet. In small-$x$ formalism, the initial TMD quark distribution can be seen from dipole-nucleus multiple scattering and small-$x$ gluon radiation in the schematic diagram of dipole picture. The transverse momentum of the initial quark, which is determined by gluon saturation, does not exhibit a preferred angle. Therefore, gluon saturation tend to suppress the anisotropy. 

We calculate the harmonics of the lepton-jet correlation within both saturation framework and non-saturation framework. The saturation framework is based on the small-$x$ factorization and resummation~\cite{Dominguez:2010xd, Dominguez:2011wm, Marquet:2009ca, Xiao:2010sa, Xiao:2010sp}, while non-saturation framework means TMD factorization~\cite{Hatta:2021jcd,Hatta:2020bgy,Catani:2014qha,Catani:2017tuc} with collinear PDFs. As the saturation scale is proportional to the nuclear size, $Q_s^2 \varpropto A^{1/3}$ with $A$ representing the nucleon number, the suppression of harmonics is more pronounced in large nucleus compared to proton. We observe this effect when comparing the harmonics in proton and gold nucleus.

Motivated by recent papers~\cite{Iancu:2021rup,Hatta:2022lzj}, we further study the harmonics of diffractive lepton-jet production in DIS. In the small-$x$ region, diffractive parton disitributions~\cite{Berera:1995fj} are related to color-dipole S-matrix~\cite{Hebecker:1997gp,Buchmuller:1998jv,Golec-Biernat:1999qor,Hautmann:1999ui,Hautmann:2000pw,Golec-Biernat:2001gyl,Iancu:2021rup,Hatta:2022lzj}, allowing us to probe gluon saturation through DPDFs. For comprehensive overviews and recent progresses of diffraction within the dipole picture, refer to reviews~\cite{Morreale:2021pnn,Mantysaari:2020axf,Frankfurt:2022jns} and Refs.~\cite{Munier:2003zb, Marquet:2007nf,Kowalski:2008sa, Kugeratski:2005ck, Cazaroto:2008iy, Kovchegov:1999ji,Kovchegov:2011aa, Kovner:2006ge, Lublinsky:2014bma, Levin:2001pr,Levin:2001yv,Levin:2002fj,Hatta:2006hs, Hatta:2006zz,Contreras:2018adl,Bendova:2020hkp,Le:2021afn,Beuf:2022kyp,Lappi:2023frf,Singh:2023yvj,Mantysaari:2016jaz,Mantysaari:2016ykx,Mantysaari:2020lhf,Mantysaari:2022kdm,Kumar:2022aly,Demirci:2022wuy,Mantysaari:2022bsp,Mantysaari:2023qsq,Boussarie:2014lxa,Boussarie:2016ogo,Salazar:2019ncp,Boussarie:2019ero,Boer:2021upt,Iancu:2021rup,Iancu:2022lcw,Rodriguez-Aguilar:2023ihz,Hatta:2016dxp,Altinoluk:2015dpi,Mantysaari:2019csc,Mantysaari:2019hkq,Hatta:2022lzj,Fucilla:2022wcg}. Similar to the discussion on semi-inclusive diffractive deep inelastic scattering (SIDDIS) at small-$x$~\cite{Hatta:2022lzj}, we expect that QCD factorization also holds for diffractive lepton-jet production. 
Although diffractive lepton-jet process is defined in lepton-nucleon center-of-mass frame, the rapidity gap is nearly the same as rapidity gap in the photon-nucleon center-of-mass frame $Y_{\text{IP}}\sim \ln(1/x_{\text{IP}})$. We calculate the harmonics for diffractive process, and observe the decrease of harmonics when transitioning from a proton target to a gold target. The t-dependent harmonics are found to be sensitive to different nuclear density profiles.

The rest of this paper is organized as follows. In Sec.~\ref{sec:LJC}, we discuss the inclusive lepton-jet correlation. In Sec.~\ref{subsec:azi_LJC}, we derive the azimuthal angle dependent lepton-jet correlation in the small-$x$ framework. Then, in Sec.~\ref{subsec:analytic}, we obtain the harmonics and its analytical expression by saddle point approximation. The QED correction to the harmonics is discussed in Sec.~\ref{subsec:QED_harmonics}. Comprehensive numerical calculations of the inclusive lepton-jet harmonics are presented in Sec.~\ref{subsec:numerical_harmonics}. In Sec.~\ref{sec:DLJC}, we explore the diffractive lepton-jet correlation. In Sec.~\ref{subsec:rapidity_diff}, we demonstrate that the rapidity of the diffractive lepton-jet production is the same as in semi-inclusive diffractive DIS process. The numerical calculation about for harmonics are presented in Sec.~\ref{subsec:numerical_harmonics}.

\section{lepton-jet correlation}
\label{sec:LJC}
In deeply inelastic scattering, an energetic lepton scatters off a proton or nucleus target 
\begin{align}
    \ell(k)+\textrm{A} (p)\rightarrow \ell^{\prime}\left(k_{\ell}\right)+\operatorname{Jet}\left(k_{J}\right)+X. 
\end{align}
In this process, we detect the scattered lepton and final jet, and measure the azimuthal angle between final lepton and jet. The momentum and rapidity of the outgoing lepton are denoted as $k_{\ell}$ and $y_{\ell}$, while the momentum and rapidity of the final jet are $k_{J}$ and $y_{J}$.

At leading order, the differential cross-section of lepton-jet correlation in the correlation limit can be expressed as
\begin{equation}
\frac{d^5\sigma ^{(0)}}{d y_l d^2P_{\perp} d^2 q_{\perp}}= \sigma_0 \int d^2 v_{\perp} \delta^{(2)}(q_{\perp}-v_{\perp}) x f_q(x,v_{\perp} )~
\label{equ:LJC_LO}
\end{equation}
where $\sigma_0 = ({\alpha_e^2}/{\hat{s} Q^2}) [{2(\hat{s}^2+\hat{u}^2)}/{Q^4}]$ with $\hat{s}$, $\hat{u}$ as Mandelstam variables of the partonic subprocess and $Q^2=-(k-k_\ell)^2$ as the virtuality of the photon. The $x$ is the longitudinal momentum fraction of the incoming quark with respect to the target proton or nucleus. At this order and considering the small initial quark transverse momentum $v_{\perp}$, the rapidities of two final particles are correlated due to the constraints $1=\frac{k_{\ell \perp}}{\sqrt{s_{e N} }}\left(e^{y_{\ell}}+e^{y_J}\right)$ and $x=\frac{k_{\ell \perp}}{\sqrt{s_{e N} }}\left(e^{-y_{\ell}}+e^{-y_J}\right) $,
where $\sqrt{s_{e N} }$ is the center-of-mass energy of the incoming lepton and nucleon. The $f_q(x,v_{\perp} )$ is the unintegrated quark distribution in the small-$x$ framework, its expression in coordinate space~\cite{McLerran:1998nk,Venugopalan:1999wu,Mueller:1999wm,Marquet:2009ca,Xiao:2017yya} after the Fourier transform reads 
\begin{align}
x f_q(& x, b_{\perp})
= \frac{N_{c}S_\perp}{8\pi^{4}}   \int  d \epsilon_f^2 d^2 r_{\perp} 
\frac{(\vec b_\perp +\vec r_\perp) \cdot \vec r_\perp }{ | \vec b_\perp +\vec r_\perp||\vec r_\perp| } 
\notag \\ 
&\times \epsilon_f^2  K_{1}(\epsilon_f | \vec b_\perp +\vec r_\perp|) K_{1}\left(\epsilon_f |\vec r_{\perp}|\right) 
\notag \\ 
&\times \Big [1+{\cal S}_x(b_\perp )-{\cal S}_x(b_\perp+r_\perp)-{\cal S}_x(r_\perp)\Big  ]~.
\label{eq:match}
\end{align}
In the above expression, $S_\perp$ represents the averaged transverse area of the target hadron, while ${\cal S}_x(r_{\perp})$ denotes the dipole scattering matrix with $r_{\perp}$ as the transeverse size of the dipole. The $\epsilon_f^2 = z(1-z)Q^2$ involves the momentum fraction $z$ for quark/anti-quark in the dipole. In CGC, the distribution of initial quark transverse momentum $\vec{v}_{\perp}$ is isotropic, which results in the leading order lepton-jet correlation Eq.(\ref{equ:LJC_LO}) being independent of the azimuthal angle.

\subsection{Azimuthal angle dependent lepton-jet correlation}
\label{subsec:azi_LJC}
\subsubsection{One soft gluon radiation}

Soft gluon radiations from the final jet introduce the azimuthal angle dependence. 
The azimuthal angle $\phi$ is defined as the angle between imbalanced momentum $\vec{q}_{\perp}$ and relative momentum $\vec{P}_{\perp}$. We start from one soft gluon radiation.

At one-loop order, one additional soft gluon radiation introduces the azimuthal angle dependence
\begin{equation}
\begin{aligned}
&\frac{d^5\sigma ^{(1)}}{d y_l d^2P_{\perp} d^2 q_{\perp}}= \sigma_0 \int  d^2v_{\perp}  xf_q(x,v_{\perp}) \\
&\quad\times \int d^2 k_{g\perp}S(k_{g\perp})\delta^{(2)}(q_\perp+k_{g\perp}-v_{\perp})~,
\label{eq:LJC_NLO}
\end{aligned}
\end{equation}
where $S(k_{g\perp})$ is the eikonal formula, which represents the probability of one gluon radiation from initial quark and final jet,
\begin{align}
S(k_{g\perp})=&g^2 C_F \int \frac{d y_g }{2(2\pi)^3}\frac{2k_J\cdot k_q }{k_J\cdot k_g \ k_q\cdot k_g}~.
\end{align}
Here, $k_q$ and $k_J$ refer to the momenta of the incoming quark and final jet, respectively. In the calculation of the eikonal formula, it is necessary to subtract the soft gluon inside the jet cone by imposing the constraint
 \begin{align}
 \Delta_{k_g k_J}=(y_g-y_J)+(\phi_g-\phi_J)^2>R^2.
 \end{align}
  The $y_{g},y_{J}$ and $\phi_{g},\phi_{J}$ represent the rapidities and azimuthal angles of the gluon and jet, respectively. The relative angle between one soft gluon and the jet $\phi_g-\phi_J$ is the azimuthal angle $\phi$ under the correlation limit.
 \begin{figure}[htbp]
\centering
\includegraphics[scale=0.7]{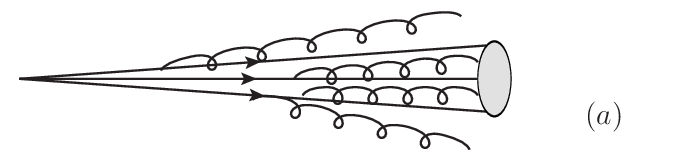}
\includegraphics[scale=0.65]{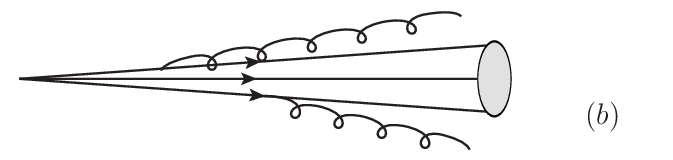}
 \caption{(a) Soft gluon radiation of the final jet before subtraction. (b) Soft gluon radiation of the final jet after the subtraction of contributions inside the jet cone.}  
 \label{fig:jetcone_subtract}
\end{figure}
 Fig. \ref{fig:jetcone_subtract} demonstrates the subtraction of soft gluons inside the jet cone. After the subtraction , the eikonal formula $S(k_{g\perp})$ is as follows:
 \begin{equation}
 \begin{aligned}
 &S(k_{g\perp})= \frac{g^2 C_F}{(2\pi)^3} \frac{1}{k_{g\perp}^2} \Big\lbrace \ln\frac{Q^4}{k_{g\perp}^2k_{J\perp}^2} + \frac{2\cos\phi}{\sin \phi} (\pi -\phi) -2y_{+}  \\ 
 & - \frac{2\cos\phi}{\sin\phi} [ \tan^{-1}(\frac{e^{y_+} - \cos\phi }{\sin\phi}) -  \tan^{-1}(\frac{e^{y_-} - \cos\phi }{\sin\phi}) ]\Big\rbrace~.
 \end{aligned}
 \label{equ:eikonal_subtract}
 \end{equation}
where $y_{\pm} = \pm \sqrt{R^2-\phi^2}$. By performing the harmonic analysis with respect to the azimuthal angle, the eikonal formula can be expressed as $S(k_{g\perp})=S_{\mathrm{iso}}(k_{g\perp})+S_{\mathrm{aniso}}(k_{g\perp})$. The isotropic and anisotropic components are given by 
 \begin{equation}
\begin{aligned}
S_{\mathrm{iso}}(k_{g\perp}) =& \frac{\alpha_s C_F}{2\pi^2 k_{g\perp}^2}\Big[\ln\frac{Q^2}{k_{g\perp}^2}+ \ln\frac{Q^2}{k_{J\perp}^2}+ c_0(R) \Big]~,\\
S_{\mathrm{aniso}}(k_{g\perp}) =& \frac{\alpha_s C_F}{2\pi^2 k_{g\perp}^2} 2 \sum_{n=1}^{\infty} c_n(R) \cos n\phi~.
\label{eq:onegluon}
\end{aligned}
\end{equation}
The harmonic coefficients of the final jet cone $R$ can be evaluated using the following formula
\begin{align}
c_n(R)= & \frac{2}{\pi} \int_0^R d \phi \left\lbrace \frac{\cos \phi}{\sin \phi}\left[(\pi-\phi)-\tan ^{-1}\left(\frac{e^{y_+}-\cos \phi}{\sin \phi}\right) \right.\right. \notag \\ 
&\left. \left. +\tan ^{-1}\left(\frac{e^{y_-}-\cos \phi}{\sin \phi}\right)\right]   - y_+\right\rbrace\cos n\phi   \notag  \\ 
& +\frac{2}{\pi} \int_R^\pi d \phi \frac{\cos \phi}{\sin \phi}(\pi-\phi) \cos n \phi~,
\end{align}  
The two integration regions come from constraint $\phi \leq R $ for some terms of Eq.(\ref{equ:eikonal_subtract}) containing $y_{\pm}$. These above equations are general expressions for large and small $R$. For a small jet cone with $R\ll 1$, the simplified expressions of $S_{\mathrm{iso}}(k_{g\perp})$ and $c_n(R)$ can be found in the paper by Hatta et al.~\cite{Hatta:2021jcd}. 

In this one-loop oder calculation, we assume the validity of small-$x$ factorization for the back-to-back lepton-jet production. A more rigorous demonstration of the small-$x$ factorization of the lepton-jet correlation would involve subtracting the rapidity divergences and ensuring the cancellation of the infrared divergences, as demonstrated in previous studies~\cite{Caucal:2021ent,Taels:2022tza,Caucal:2022ulg,Chirilli:2011km,Chirilli:2012jd}. The subtracted rapidity divergence from real and virtual diagrams is then renormalized into the small-$x$ parton distribution~\cite{Chirilli:2011km,Chirilli:2012jd,Xiao:2017yya}, following the procedures in Refs.~\cite{Mueller:2013wwa,Mueller:2012uf}. The infrared divergences cancel between real and virtual diagrams, leaving finite term that include Sudakov type logarithms. 
 
The delta function in Eq.~(\ref{eq:LJC_NLO}) facilitates the Fourier transform of the cross-section to the $b_{\perp}$-space
\begin{equation}
\begin{aligned}
\frac{d^5\sigma ^{(1)}}{d y_l d^2P_{\perp} d^2 q_{\perp}}= &\sigma_0 \int \frac{d^2b_{\perp}}{(2\pi)^2}  e^{i\vec{q}_{\perp}\cdot\vec{b}_{\perp}}  xf_q(x,b_{\perp})\\ 
&\times [S_{\mathrm{iso}}(b_{\perp})+S_{\mathrm{aniso}}(b_{\perp})]
\end{aligned}
\end{equation}
When fourier transforming to $b_{\perp}$-space, the isotropic part has the corresponding virtual diagram contribution which cancel the infrared divergences. Afterwards, we still encounter single and double logarithms in terms of $Q^2/\mu_{b}^2$ as follows
\begin{equation}
\begin{aligned}
S_{\mathrm{iso}}(b_{\perp}) =& - \int_{\mu_{b}}^{Q} \frac{d \mu}{\mu} \frac{\alpha_{s}(\mu) C_{F}}{\pi}
\Big[\ln \frac{Q^{2}}{\mu^{2}}
+\ln \frac{Q^{2}}{P_{\perp}^{2}}
 +c_{0}(R)\Big]~,
 \label{equ:iso_sudkov}
\end{aligned}
\end{equation}
where $\mu_b=b_0/b_\perp$ with $b_0\equiv 2e^{-\gamma_E}$ and $\gamma_E$ the Euler constant. 

The anisotropic part is convergent, which can be seen in $b_{\perp}$-space via the Fourier transform. By utilizing the Jacobi-Anger expansion formula
\begin{equation}
e^{i z \cos (\phi)}=J_0(z)+2 \sum_{n=1}^{\infty} i^n J_n(z) \cos (n \phi) 
\label{equ:Jacobi-Anger}
\end{equation}
and integration formula for Bessel function
\begin{equation}
\int_0^{\infty} \frac{d z }{z}J_n\left(z \left|b_{\perp}\right|\right)=\frac{1}{n}~,
\end{equation}
we obtain the expression for the anisotropic part in $b_{\perp}$-space
\begin{equation}
\begin{aligned}
S_{\mathrm{aniso}}(b_{\perp})=\frac{\alpha_s C_F}{\pi} \sum_n i^n c_n  \frac{2\cos n\phi_b }{n}~.
\end{aligned}
\end{equation}
Here, $\phi_b$ represents the angle between $b_\perp$ and $k_{J\perp}$.

\subsubsection{Multiple soft gluon resummation}

When considering contributions from soft gluon emissions to all orders, the isotropic part has been resummed into the exponential factor
\begin{align}
\frac{d^5\sigma }{d y_l d^2P_{\perp} d^2 q_{\perp}}
\approx&\sigma_0 \int \frac{d^2b_\perp}{(2\pi)^2} e^{i \vec{q}_{\perp}\cdot \vec{b}_{\perp}} xf_q(x,b_{\perp})  
\notag \\ 
&\times e^{S_{\mathrm{iso}}(b_{\perp})} \Big[1+ S_{\mathrm{aniso}}(b_{\perp})\Big]~.
\label{eq:app1}
\end{align}
The isotropic part corresponds to the Sudakov factor $S_{\mathrm{iso}}(b_{\perp})= -\text{Sud}(b_{\perp})$. The techniques of Sudakov resummation are developed in Refs.~\cite{Mueller:2013wwa,Mueller:2012uf} and \cite{Catani:2014qha,Catani:2017tuc,Hatta:2021jcd,Hatta:2020bgy}. Compared to the Sudakov factor in the collinear factorization framework~\cite{Hatta:2021jcd}, there is a difference of a single logarithmic term with the coefficient $-3/2$. In the TMD framework, this term arises from the collinear divergence. However, in small-$x$ framework being considered here, this term is absent.

By using Eq.~(\ref{equ:Jacobi-Anger}) and integrating over $\phi_b$, we get the azimuthal angle dependent lepton-jet correlation
\begin{equation}
\begin{aligned}
  & \frac{d^{5} \sigma(\ell P \rightarrow \ell^{\prime} J)}{d y_{\ell} d^{2} P_{ \perp} d^{2} q_{\perp}}=
\sigma_0\int  \frac{b_{\perp}db_{\perp}}{2\pi}xf_q(x,b_{\perp})  e^{-\text{Sud}(b_{\perp})} 
   \\ 
  &  
 \Big[J_0( q_\perp b_\perp) + 
\sum_{n=1}^\infty 2\cos( n\phi) \frac{ \alpha_s(\mu_{b}) C_Fc_n(R)}{n\pi} J_n(q_\perp b_\perp) \Big] ~.
\label{equ:LJC-Xsection}
\end{aligned}
\end{equation}
In the calculation, it is important to note that the angle between $b_{\perp}$ and $P_{\perp}$ is set to be $(\pi-\phi_b)$ and thus the phase factor $e^{i \vec{q}_{\perp}\cdot \vec{b}_{\perp}}$ can be written as $e^{i q_\perp b_\perp \cos[\phi - (\pi - \phi_b)] }$.

\subsection{Harmonics and its analytical expression}
\label{subsec:analytic}
To quantify the azimuthal anisotropy of the lepton-jet correlation, we define the harmonics or Fourier coefficient of the azimuthal angle dependent lepton-jet correlation as
\begin{equation}
\langle \cos n\phi \rangle 
  =\frac{\sigma_0  \int b_{\perp} d b_{\perp} J_{n}\left(q_{\perp}b_{\perp}\right) 
W(x,b_\perp)\frac{ \alpha_s(\mu_{b})C_{F}  c_{n}(R)}{n \pi} 
 }
{ \sigma_0 \int b_{\perp} d b_{\perp} J_{0}\left(q_{\perp}b_{\perp}\right)  W(x,b_\perp)  }~.
\label{eq:harmonics}
\end{equation}
$W$ function is defined as
\begin{equation}
W(x,b_\perp)  = xf_q(x,b_{\perp}) e^{-\text{Sud}(b_\perp)}~.
\end{equation}

In the small $q_{\perp}$ limit, we can expand the Bessel function by
\begin{equation}
J_{n}\left(q_{\perp}b_{\perp}\right) \sim {(q_\perp b_\perp/2)^{n}}/{\Gamma(n+1)}~.
\end{equation}
The $n$-th harmonic is proportional to $q_{\perp}^n$, $ \langle \cos n\phi \rangle \sim {\cal C}_n q_\perp^n $. We now elaborate on how to get an analytical expression for the power-law coefficient ${\cal C}_n$.

The Sudakov factor contains the large logarithm $\ln Q^2$ under the correlation limit $Q\geq P_\perp\gg q_{\perp}$, which serves as a large parameter for the saddle point approximation~\cite{Parisi:1979se,Collins:1981va,Collins:1984kg,Shi:2021hwx}. We evaluate the two integrals in the numerator and denominator using this formula
\begin{equation}
\begin{aligned}
&\int_{-\infty}^{+\infty} d z F(z) e^{- E(z)}\\ 
 \approx &\left[\frac{2 \pi}{E^{\prime \prime}\left(z^{\text{sp}}\right)}\right]^{1 / 2} F\left(z^{\text{sp}}\right) e^{- E\left(z^{\text{sp}}\right)},
\end{aligned}
\end{equation}
where $z = \ln (\Lambda_{\text{QCD}} b_{\perp})$. The saddle point can be determined by
\begin{equation}
 \frac {d E\left(z\right)}{d z} |_{z=z_{\text{sp} }}= 0 \quad \text{with} \quad E''(z)>0.
\end{equation}

The harmonics are
\begin{align}
\langle \cos n\phi \rangle 
  & \approx 
\left(\frac{q_{\perp} b_0 }{2\Lambda_{\text{QCD}}} \right)^n  
 \frac{\alpha_s(\mu^{\text{sp}}_n) C_F  c_{n} (R)}{\pi   n \Gamma(n+1)}\frac{
  f_q(x,b^{\text{sp}}_{\perp n})  }
{
 f_q(x, b^{\text{sp}}_{\perp 0})  
}~
\notag \\ &
\times  \left[\frac{ 2\beta_1+ C_F}{ (n+2)\beta_1+ C_F }  \right]
^{1+\frac{ C_F}{2\beta_{1}}\ln  \frac{e^{c_0(R)} Q^4}{ \Lambda_{\text{QCD}}^2 P_\perp^2} }
~,
\label{eq:smallqT}
\end{align}
where $\beta_1=(33-2 n_f)/12$, $\mu^{\text{sp}}_n= b_0/b^{\text{sp}}_{\perp n}$, $b_0\equiv2e^{-\gamma_E}$.The saddle points $b^{\text{sp}}_{\perp n}$ are 
\begin{align}
b^{\text{sp}}_{\perp n} =\frac{ b_0}{\Lambda_{\text{QCD}} } \Bigg[\frac{  e^{c_0(R)}Q^4}{\Lambda_{\text{QCD}}^2 P_\perp^2 }\Bigg]^{-\frac{C_F}{2(2+n)\beta_1+2C_F}}~,
\label{eq:saddleP}
\end{align}
The saddle point approximation is a widely used technique in high energy physics. In particular, the saddle point with $n=0$ in this context is similar to the saddle point discussed in Ref.~\cite{Collins:1984kg}. The typical values of the saddle points for the lepton-jet correlation are estimated to be around $1.5$ GeV$^{-1}$ for $b^{\text{sp}}_{\perp 0}$ and roughly $2.5$ GeV$^{-1}$ for the cases where $n$ equals $1$, $2$, or $3$ for EIC kinematics.

From the analytical expression of the harmonics given in Eq.~(\ref{eq:smallqT}), we know the information about the parton saturation is encoded in the ratio of unintegrated quark distribution $f_q(x,b_\perp^{\text{sp}})$. It is observed that
as $Q\geq P_\perp \rightarrow \infty$, the saddle points $b^{\text{sp}}_{\perp n}$ approach zero. Hence, we can employ small-$b_\perp$ approximation of $f_q(x,b_\perp^{\text{sp}})$ as follows:
\begin{equation}
f_q(x,b_\perp^{\text{sp}}) \propto Q_s^2\ln  \frac{1}{Q_s b_\perp^{\text{sp}}}
\end{equation}
as explained in \cite {Xiao:2017yya,Mueller:1999wm}.
Thus, the harmonics have the following asymptotic form
\begin{align}
 \langle \cos n\phi \rangle  \propto   \frac{
f_q(x,b^{\text{sp}}_{\perp n})  }
{  f_q(x, b^{\text{sp}}_{\perp 0}) }\approx\frac{\ln(Q_s b^{\text{sp}}_{\perp n})  }{\ln(Q_s b^{\text{sp}}_{\perp 0})}~.
\label{equ:asymptotic_harmonics}
\end{align}
The derivative of $\langle \cos n\phi \rangle$ with respect to $Q_s$ reads
\begin{equation}
\frac{\partial \langle \cos n\phi \rangle}{\partial Q_s} = \frac{\ln b^{\text{sp}}_{\perp 0}/b^{\text{sp}}_{\perp n} }{Q_s \ln^2(Q_s b^{\text{sp}}_{\perp 0})}~.
\label{equ:asymptotic_harmonics_deri}
\end{equation}
Since $b^{\text{sp}}_{\perp n}>b^{\text{sp}}_{\perp 0}$ according to Eq.~(\ref{eq:saddleP}), the derivative is negative. As the saturation momentum $Q_s$ increases, the harmonics decrease. We can observe this feature in the numerical calculations.

\subsection{QED radiation contribution to the harmonics}
\label{subsec:QED_harmonics}
Soft gluon radiations can occur in the QCD sector of lepton-jet scattering, while soft photon radiations can occur in the QED sector~\cite{Hatta:2021jcd,Liu:2020rvc,Liu:2021jfp}. 
Moreover, soft photon emissions from the final state lepton also contribute to the azimuthal anisotropy~\cite{Hatta:2021jcd,Tong:2022zwp}, as depicted in Fig.~\ref{fig:lepjet_gluonphoton}. 

 Soft photons tend to align with the final lepton, which is the away side of final jet direction. That may reduce the odd harmonics and increase the even harmonics, since $\cos n\phi$ with even $n$ exhibits a symmetric shape, while $\cos n\phi$ with odd $n$ show an asymmetric shape between $0$ and $\pi$ in the azimuthal angle. 

By calculating the similar eikonal formula 
\begin{equation}
S_{\gamma}(k_{g\perp})= e^2 \int \frac{d y_\gamma }{2(2\pi)^3}\frac{2k_J\cdot k_q }{k_J\cdot k_\gamma \ k_q\cdot k_\gamma},
\end{equation}
we obtain the isotropic and anisotropic part of eikonal formula for the one photon radiation
 \begin{equation}
\begin{aligned}
S^\gamma_{\mathrm{iso}}(b_{\perp}) =& -\int_{\mu_{b}}^{Q} \frac{d \mu}{\mu} \frac{\alpha_{e}}{\pi}
\Big[\ln \frac{Q^{2}}{\mu^{2}}
+\ln \frac{Q^{2}}{P_{\perp}^{2}} -\frac{3}{2}
 +c_{0}^\gamma\Big]~,\\
 S^\gamma_{\mathrm{aniso}}(b_{\perp})=&\frac{\alpha_e}{\pi} \sum_n i^n c_n^\gamma  \frac{2\cos n\phi_b }{n},
\label{eq:onegluon}
\end{aligned}
\end{equation}
with 
\begin{equation}
\begin{aligned}
c_{n}^\gamma=(-1)^n\Big [\ln \frac{P_\perp^2}{m_e^2}+\frac{2}{\pi}\int^\pi_0 d \phi (\pi- \phi)\frac{\cos \phi}{\sin \phi }(\cos n \phi-1)\Big]~,
\label{eq:cgamma}
\end{aligned}
\end{equation}
where $m_e$ is the electron mass and $\alpha_e$ is the QED coupling. When considering multiple soft photon radiations, the isotropic part can be resummed into QED Sudakov factor $\text{Sud}^{\gamma}(b_\perp) = -S^\gamma_{\mathrm{iso}}(b_{\perp}) $. Although there are large logarithms of ${P_\perp^2}/{m_e^2}$ present in the anisotropic part, we only retain the leading order contribution from anisotropic part, since the small QED coupling constant $\alpha_e\approx 1/137$ compensates the large logarithms. The harmonics with QED correction reads
\begin{equation}
\begin{aligned}
&\langle \cos n\phi \rangle_{\text{QED} } \\
  &\hspace{-0.2cm}=\frac{\sigma_0  \int b_{\perp} d b_{\perp} J_{n}\left(q_{\perp}b_{\perp}\right) 
W_{\text{QED}}(x,b_\perp)\frac{ [\alpha_s(\mu_{b})C_{F}  c_{n}(R)+ \alpha_e c_n^{\gamma}] }{n \pi} 
 }
{ \sigma_0 \int b_{\perp} d b_{\perp} J_{0}\left(q_{\perp}b_{\perp}\right)  W_{\text{QED}}(x,b_\perp)  }~,
\end{aligned}
\label{eq:harmonics_QED}
\end{equation}
with 
\begin{equation}
\begin{aligned}
W_{\text{QED}}(x,b_\perp) = xf_q(x,b_{\perp}) e^{-\text{Sud}(b_\perp)-\text{Sud}^{\gamma}(b_\perp)}~.
\end{aligned}
\end{equation}
The QED Sudakov factor is negligible compared to the QCD Sudakov factor, due to the smallness of the QED coupling constant $\alpha_e$. However, the QED correction to the coefficient $\alpha_s C_F c_n+\alpha_{e} c_{n}^\gamma$ is sizable. The numerical calculations in the next section will show these two features. 

\subsection{Numerical calculation of the harmonics}
\label{subsec:numerical_harmonics}

The first calculation involves computing the harmonics for both non-saturation and saturation models, considering both proton and nucleus target. 

The harmonics of non-saturation model can be calculated using the same formula as Eq.~(\ref{eq:harmonics}), but with a different $\widetilde{W}$ function
\begin{equation}
\widetilde{W} =  \sum_q e_q^2 xf_q(x,\mu_b) e^{-\widetilde{\text{Sud}}(b_\perp)}~.
\label{eq:W}
\end{equation}
 The $f_q(x,\mu_b)$ represents the collinear quark distribution, encompassing both valence and sea quarks. For the proton, we utilize the NLO PDF sets of CT18A \cite{Hou:2019efy}, while for the gold nucleus, we adopt the EPPS21 \cite{Eskola:2021nhw} PDF sets. Compared with the Sudakov factor of saturation model in Eq.(\ref{equ:iso_sudkov}), the Sudakov factor $\widetilde{\text{Sud}}(b_\perp)$
\begin{equation}
\widetilde{\text{Sud}}(b_\perp) = \int_{\mu_{b}}^{Q} \frac{d \mu}{\mu} \frac{\alpha_{s}(\mu) C_{F}}{\pi}
\Big[\ln \frac{Q^{2}}{\mu^{2}}
+\ln \frac{Q^{2}}{P_{\perp}^{2}} -\frac{3}{2}
 +c_{0}(R)\Big]
\end{equation}
has extra $-3/2$ term, which corresponds to the collinear divergence~\cite{Hatta:2021jcd}. In the numerical calculation, we introduce the non-perturbative Sudakov factor~\cite{Sun:2014dqm,Prokudin:2015ysa} 
\begin{equation}
\widetilde{\text{Sud}}(b_\perp) \rightarrow \widetilde{\text{Sud}}(b_*) + \widetilde{\text{Sud}}_{\text{NP}}^q(b_\perp)
\end{equation}
with $b_{*}$-prescription $b^*_\perp=b_\perp/\sqrt{1+b_\perp^2/b_\text{max}^2}$, and $b_\text{max}=1.5~$GeV$^{-1}$. Here we only include the non-perturbative Sudakov factor associated with the initial quark $\widetilde{\text{Sud}}_{\text{NP}}^q(b_\perp)$, ignoring that of the final jet $\widetilde{\text{Sud}}_{\text{NP}}^{\text{jet}}(b_\perp)$~\cite{Hatta:2021jcd}. This choice allows for a direct comparison with the saturation model.

For saturation model, we consider two parameterizations for the dipole scattering matrix ${\cal S}_x(r_\perp)$ in the unintegrated quark distribution $f_q(x,v_{\perp} )$ as given in Eq.~(\ref{eq:match}). The first one is the GBW model~\cite{Golec-Biernat:1998zce},
\begin{equation}
{\cal S}_x(r_{\perp})=e^{ - \frac{r_{\perp}^2Q_s^2(x)}{4} }~,
\label{equ:GBW_Smatrix}
\end{equation}
where saturation momentum squared for proton is $Q_{s,p}^2(x)=(x_0/x)^{0.28}~\text{GeV}^2$ with $x_0 = 3\times 10^{-4}$. The saturation momentum squared of gold nucleus is approximately $Q_{s,A}^2\approx 5Q_{s,p}^2$. The other parameterization for the dipole scattering matrix is the solution of the running-coupling Balitsky-Kovchegov (rcBK) equation~\cite{Balitsky:1995ub,Kovchegov:2006wf,Kovchegov:1999yj,Kovchegov:2006vj,Albacete:2010sy,Golec-Biernat:2001dqn,Albacete:2007yr,Balitsky:2006wa,Gardi:2006rp,Balitsky:2007feb,Berger:2010sh}, with the modified McLerran-Venugopalan (MV)~\cite{Albacete:2010sy,Fujii:2013gxa} model as the initial condition 
\begin{equation}
{\cal S}_{x_0}(r_{\perp})=e^{- \frac{ (r_{\perp}^{2} Q_{s 0}^{2})}{4} ^{\gamma}\ln \frac{1}{(\Lambda r_{\perp})+e}  }
\label{equ:rcBK_MV}
\end{equation}
with $\gamma=1.118$, $\Lambda = 0.241$ GeV, and $Q_{s0,p}^2=0.16$~GeV$^2$ at $x_0=0.01$. For $e+\text{Au}$ rcBK calculation, we solve the rcBK equation with $Q_{s,A}^2\approx 5Q_{s,p}^2$. More realistic initial condition for dipole-nucleus amplitude\cite{Lappi:2013zma} can be chosen.

Since the non-perturbative region is usually dominated by the small-x dipole distribution, we do not need to introduce the non-perturbative Sudakov factor for saturation model. Thus, we simply write
\begin{equation}
\text{Sud}(b_\perp) \rightarrow \text{Sud}(b_*)~,
\end{equation}
which limits the perturbative Sudakov factor in the small $b_\perp$ region. 

The kinematics bins for future EIC that we use to calculate the $q_\perp$-distribution of $\langle \cos n\phi \rangle$ are $\sqrt{s_{eN}}=89$ GeV, $y_\ell=2.41$,\ $0.008\leq x\leq 0.01,\ 4$ GeV$\leq P_\perp\leq4.4$ GeV, 5.6~GeV$\leq Q\leq$ 5.9~GeV. The choice of kinematic region aligns with the simulation study~\cite{Arratia:2019vju} of EIC. The lower cut for $x$ is determined by given $s_{eN}, P_\perp$, specifically $x_{\text{min}}\approx 4 P_\perp^2/s_{\textrm{eN}}$. The limited collision energy $s_{eN}$ makes it difficult to probe lower $x$ value ($x\leq 1\times 10^{-3}$) in the lepton-jet correlation.

\begin{figure*}[htpb]
\includegraphics[scale=0.8]{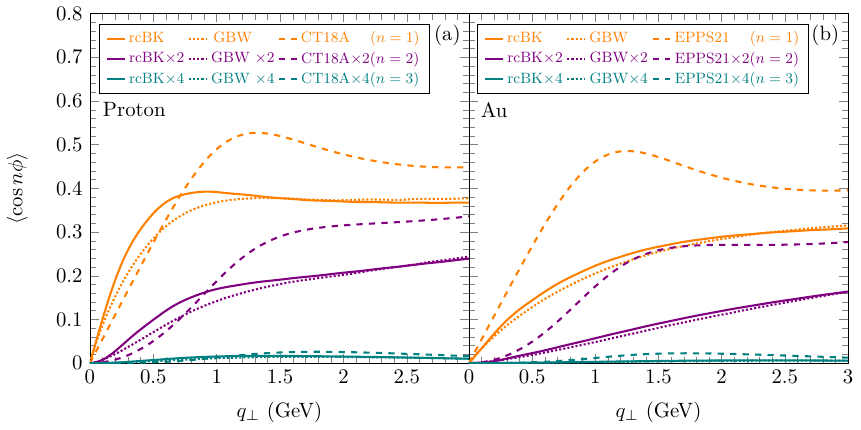}
 \caption{ (a) First three harmonics of inclusive lepton-jet production in $e+p$ collisions using inputs from the rcBK solution, GBW model, and CT18A PDFs.  (b) First three harmonics of inclusive lepton-jet production predicted for $e+\text{Au}$~collisions using inputs: the rcBK solution, GBW model, and EPPS21 PDFs. The calculation is for EIC kinematics: $\sqrt{s_{eN}}=89$ GeV, $0.008<x<0.01,\, y_\ell=2.41$ with jet cone size $R=0.4$. The QED corrections are included.}  
 \label{fig:Asymmetry_p+A}
\end{figure*}

Fig.~\ref{fig:Asymmetry_p+A} presents the $q_\perp$-distribution of $\langle \cos n\phi \rangle$ for different models, using both proton and nucleus targets with a jet cone size $R=0.4$. The results exhibit a common trend: all lines sharply rise from zero in the small-$q_\perp$ region and gradually approach a plateau in the large-$q_\perp$ region.  In the large-$q_\perp$ region, our results appear to be more flat than the lines shown in Fig. 4 of \cite{Hatta:2021jcd}. The discrepancy arises because our calculation is performed within the $4$ GeV$\leq P_\perp\leq 4.4$ GeV bin, while their calculation is specific to a single $P_\perp$ value. Besides the common trend, we also observe the hierarchy of harmonics with the harmonic number $n$. What's more, the harmonics of the saturation model show a sizable decrease from the proton to the gold nucleus target. First, Eq.~(\ref{equ:asymptotic_harmonics_deri}) shows $\partial \langle \cos n\phi \rangle/ \partial Q_s <0$ and it indicates that the harmonics decreases with an increase in $Q_s$. Furthermore, the saturation scale squared $Q_{s,A}^2\propto A^{1/3}Q_{s,p}^2$ is larger in the gold nucleus than in the proton. Therefore, we can explain the observed decrease in harmonics from proton to Au in Fig.~\ref{fig:Asymmetry_p+A}.  The results in Fig.~~\ref{fig:Asymmetry_p+A} have included the QED correction, different from similar plots in our previous paper\cite{Tong:2022zwp} .

\begin{figure}[htbp]
\centering
\includegraphics[scale=0.8]{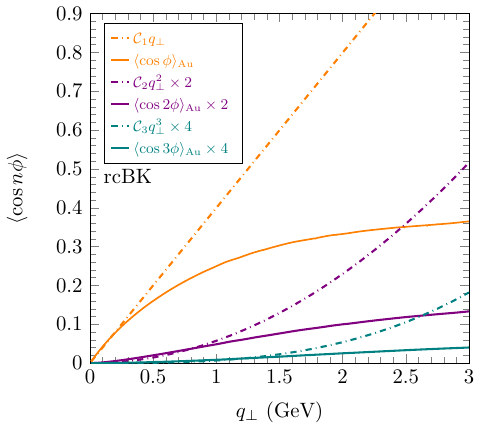}
 \caption{Comparison of the exact results of harmonics and their small-$q_\perp$ asymptotic behaviors. Solid lines stand for the exact results, while dash-dotted lines depict the small-$q_\perp$ asymptotic expansions given by Eq.~(\ref{eq:smallqT}). The EIC kinematics for calculation is $\sqrt{s_{eN}}=89$ GeV, $x=0.008$, $y_\ell=2.41$ with jet cone size $R=0.4$. The QED corrections are not included.}  
 \label{fig:analy_Au}
\end{figure}

In Fig.~\ref{fig:analy_Au}, we plot the analytical expression of harmonics for small-$q_\perp$ Eq.~(\ref{eq:smallqT}) from section~\ref{subsec:analytic} and compare it with harmonics obtained for the rcBK model with an Au taget. The harmonics are calculated for the specific value of $x=0.008$. The comparison validates the analytical expression of harmonics at small-$q_\perp$.

\begin{figure*}[htbp]
\centering
 \includegraphics[scale=0.8]{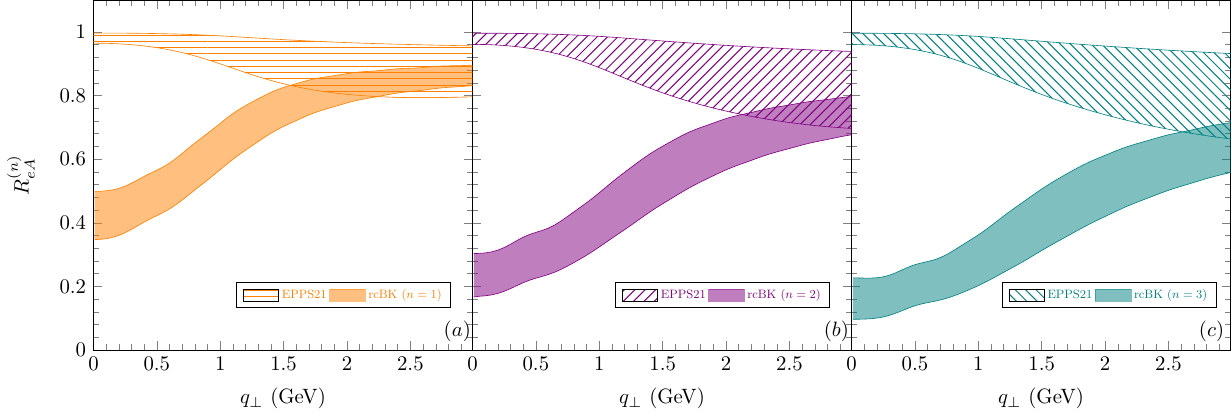}
   \caption{Nuclear modification factors of the inclusive lepton-jet harmonics for the cases where $n=1$, $2$, and $3$. The upper bands represent ${R}_{eA}^{(n)}$ based on the inputs of the EPPS21 gold nuclear PDFs with uncertainties. The lower bands yield ${R}_{eA}^{(n)}$ calculated with the rcBK solution, where the gold saturation scale $Q_{s,A}^2$ varies from $3Q_{s,p}^2$ (upper bound in each band) to $5Q_{s,p}^2$ (lower bound). The EIC kinematics for calculation is $\sqrt{s_{eN}}=89$ GeV, $0.008<x<0.01,\, y_\ell=2.41$ with jet cone size $R=0.4$.
   }
   \label{fig:ratio_QsBand}
\end{figure*}

To further quantify the suppression of the anisotropy in $e+ \text{Au}$ collisions compared to $e+p$ collisions, we define the nuclear modification factor as follows:
\begin{equation}
\begin{aligned}
   {R}_{eA}^{(n)}= \frac{\langle \cos n\phi  \rangle_{eA}}{\langle \cos n\phi  \rangle_{ep}}~.
   \label{eq:ratioReA}
\end{aligned}
\end{equation}
In Fig.~\ref{fig:ratio_QsBand}, we plot the nuclear modification factor for the non-saturation model and the saturation model. The non-saturation model untilizes the EPPS21 gold nucleus PDFs and the CT18A proton PDFs, with error band at $90\%$ confidence-level. We neglect the uncertainties from the baseline proton PDFs, as they are small. On the other hand, the saturation model employs the rcBK solution, where the gold saturation scale squared $Q_{s,A}^2$ varies from $3Q_{s,p}^2$ (upper bound in each band) to $5Q_{s,p}^2$ (lower bound). The ${R}_{eA}^{(n)}$ predicted from EPPS21 PDFs (non-saturation model) and rcBK solution (saturation model) show distinct behaviors at small-$q_\perp$ region and converge to unity at large-$q_\perp$ region. This difference justifies the nuclear modification factor as a tool to distinguish the saturation and non-saturation frameworks. 

In Fig.~\ref{fig:ratio_QsBand}, we also observe the hierarchy of the nuclear modification factor of the saturation models. This can be explained by the asymptotic expression of the harmonics, Eq.~(\ref{equ:asymptotic_harmonics}). By substituting the $Q_{s,A}^2 \approx  5 Q_{s,p}^2$ in Eq.~(\ref{equ:asymptotic_harmonics}), we find
\begin{equation}
{R}_{eA}^{(n)} \propto \frac{\ln(5Q_{s,p}^2 b^{\text{sp}}_{\perp n}) }{\ln(Q_{s,p}^2 b^{\text{sp}}_{\perp n})} \frac{\ln(Q_{s,p}^2 b^{\text{sp}}_{\perp 0}) }{\ln(5Q_{s,p}^2 b^{\text{sp}}_{\perp 0})}
\end{equation}
and knowing $b^{\text{sp}}_{\perp n}$ increasing with $n$, we understand why the nuclear modification factor ${R}_{eA}^{(n)}$ decreases with the increase of $n$ as shown in the numerical results.

\begin{figure}[htbp]
\centering
\includegraphics[scale=0.8]{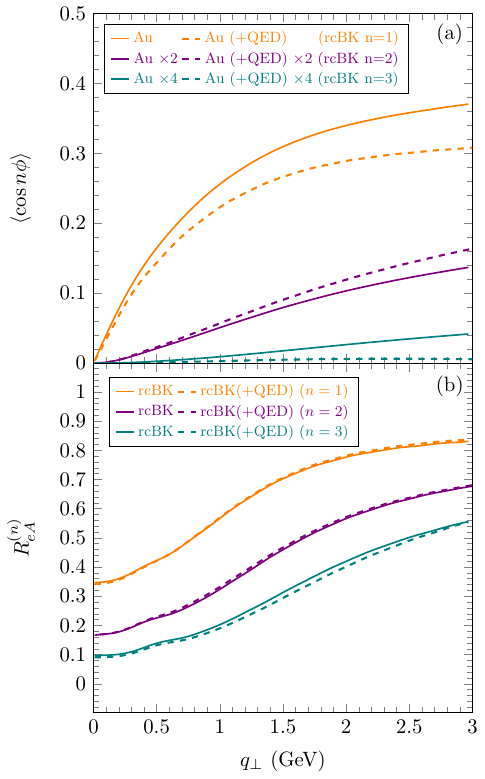}
 \caption{QED modification to the first three harmonics of the inclusive lepton-jet correlation and nuclear modification factors with the input from the rcBK solution for $e+ \text{Au}$ collisions. Solid and dashed lines represent the harmonics or nuclear modification factor without and with QED modifications, respectively. The calculation is performed in the following EIC kinematics: $\sqrt{s_{eN}}=89$ GeV, $0.008<x<0.01$,$y_l=2.41$ with jet cone size $R=0.4$. }  
 \label{fig:QED_asy_ratio}
\end{figure}

 Fig.~\ref{fig:QED_asy_ratio} shows the harmonics and nuclear modification factor with and without QED correction, using the rcBK solution as the input. The QED corrections for the harmonics are quite evident, reducing the odd harmonics and increasing the even harmonics. This evident correction to the harmonics can be explained by the sizable correction to the coefficient $\alpha_s C_F c_n+\alpha_{e} c_{n}^\gamma$ in Eq.~(\ref{eq:harmonics_QED}). However, the QED correction to the nuclear modification factor is found to be negligible, as the coefficient $\alpha_s C_F c_n+\alpha_{e} c_{n}^\gamma$ cancels between $\langle \cos n\phi  \rangle_{eA}$ and $\langle \cos n\phi  \rangle_{ep}$. In order to compare with experimental data, all subsequent calculations incorporate  QED correction.

\begin{figure*}[htpb]
\centering
\includegraphics[scale=0.8]{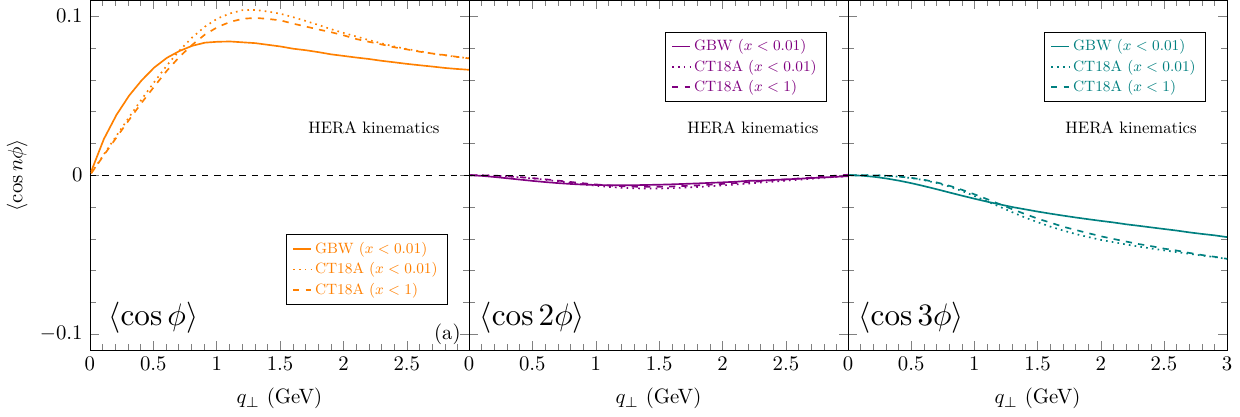}
 \caption{First three harmonics in $e+p$ collisions with inputs from the GBW model and CT18A PDFs in the HERA kinematics, with additional cut on the initial quark momentum fraction $x$. The HERA kinematics are $0.2<y<0.7$, $-1<\eta_{\text{lab}}<2.5$, $k_{J\perp}>10$ GeV, $Q^2>150$ GeV$^2$ with jet cone $R=1.0$.}  
 \label{fig:HERA_H1}
\end{figure*}

Our calculation can be compared with a recent experimental study~\cite{Torales-Acosta:202302}
at HERA, where electron and proton collide at energy of $27.6$ GeV and $920$ GeV, respectively. The kinematics bin are $0.2<y<0.7$, $-1<\eta_{\text{lab}}<2.5$, $k_{J\perp}>10$ GeV, $Q^2>150$ GeV$^2$. Here, $y=P\cdot q/P\cdot k$ represents the energy fraction taken by the photon from the lepton in lab frame, and $\eta_{\text{lab}}$ is the rapidity range that the detector can cover. The jet cone size is $R=1.0$. In our calculation, we compute the harmonics in this HERA kinematics and present the results in Fig.~\ref{fig:HERA_H1}. In the calculation, the kinematic restrictions constrain the rapidity $y_J$ (or $y_l$), $k_{J\perp}$, initial quark momentum fraction $x$ and their combination, since $Q^2, y, \eta_{\text{lab}}$ are expressed in terms of $y_J, k_{J\perp}, x$. We compute both saturation framework with the GBW model, and non-saturation framework with CT18A proton PDFs. For the saturation framework, we apply the extra cut $x<0.01$, while for the non-saturation framework, we have two different cuts $x<0.01$ and $x<1$. In our calculation, we also include the QED correction.

In Fig.~\ref{fig:HERA_H1}, we observe that the harmonics $\cos\phi$ are sizable, while the harmonics $\langle \cos 2\phi  \rangle$ and $\langle \cos 3\phi  \rangle$ are almost zero. This behavior can be attributed to the fact that the Fourier coefficient $c_n(R)$ decreases with $R$, as evident from Fig.~3 in Ref~\cite{Hatta:2021jcd}. The negative values of $\langle \cos 2\phi  \rangle$ and $\langle \cos 3\phi  \rangle$ also come from the Fourier coefficient $c_n(R)$ with $R=1.0$. The observed trend of these harmonics is consistent with the previous results obtained for the EIC kinematics, as shown in Fig.~\ref{fig:Asymmetry_p+A}.

\section{diffractive lepton-jet correlation}
\label{sec:DLJC}
\begin{figure}[htbp]
\centering
\includegraphics[scale=0.22]{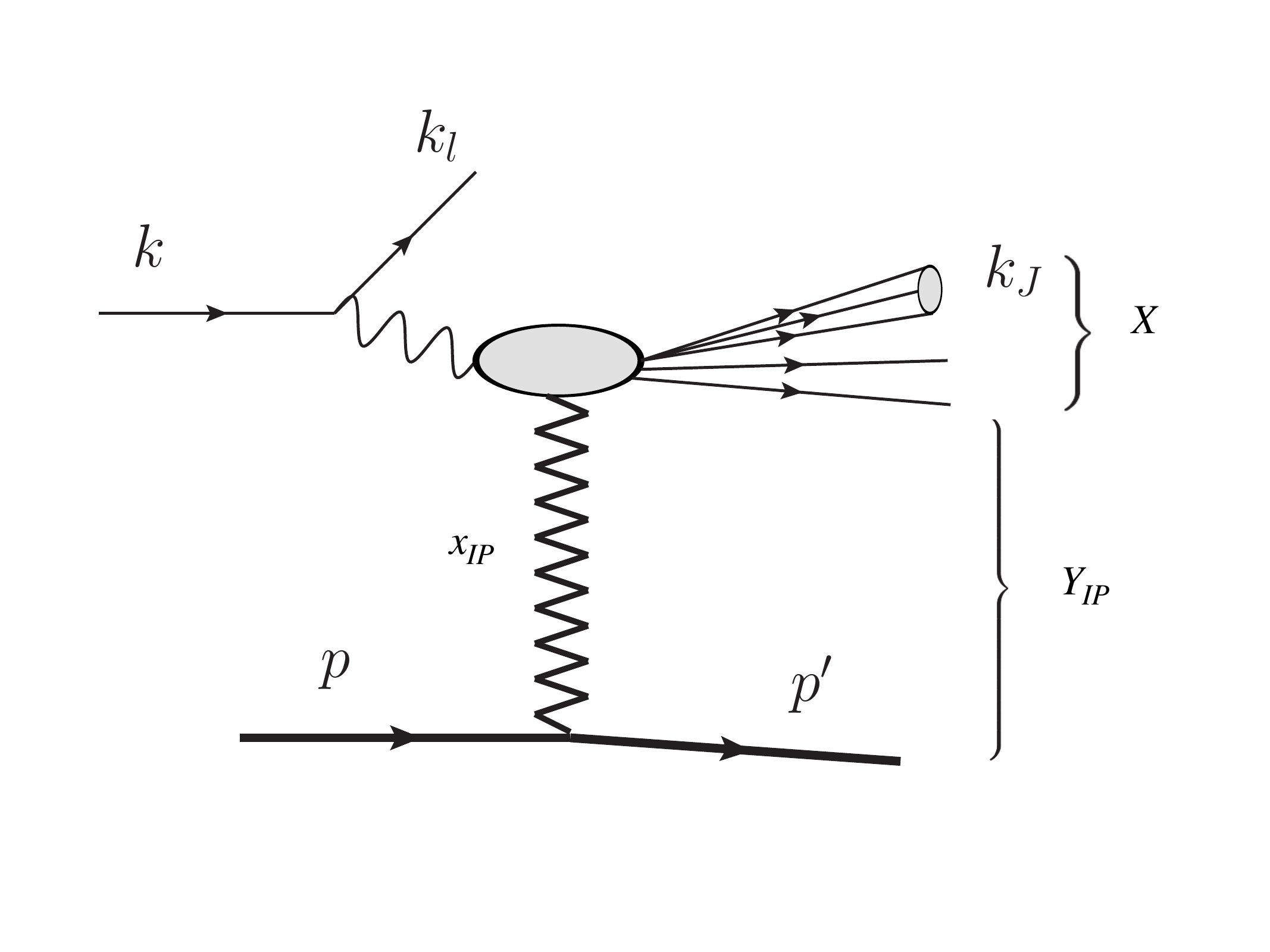}
 \caption{Diffractive lepton-jet production process in DIS. The final state lepton with one jet can be measured, while the incoming nucleon exchanges multiple gluon in a color singlet state with the virtual photon, with the exchanged longitudinal momentum fraction $x_{\text{IP}}$. A large rapidity gap exists between the nucleon remnant $p'$ and the hard interaction production $X$.}
 \label{fig:diffractive_lepjet}
\end{figure}

In high energy $ep$ and $eA$ collisions, the diffractive lepton-jet process occurs when we observe a large rapidity gap $Y_{\text{IP}}$ between the hard interaction part and the remnant proton/nucleus, in addition to measuring the scattered lepton and one jet, as shown in Fig.~\ref{fig:diffractive_lepjet}. 

The diffractive process can be understood as the proton/nucleus exchanging colorless multiple gluon with the hard interaction part. The momentum transfer in the diffraction is denoted as $t = (p'-p)^2= \Delta^2 \approx -\vec \Delta_\perp^2 $, while the momentum fraction carried by the pomeron from the incoming nucleon is $x_{\text{IP}} = n\cdot (p-p')/n\cdot p$, where $n=(0,1,0_\perp)$ . In Fig.~\ref{fig:diffractive_lepjet}, the hard interaction production is denoted as $X$, and its mass is defined as $M$. From the definition of the mass $(x_{\text{IP}} p + q)^2=M^2$, we obtain
\begin{equation}
x_{\text{IP}} = \frac{M^2+Q^2}{W^2+Q^2}~,
\end{equation}
where $ W^2 = (p+q)^2$ represents the center-of-mass energy squared of the photon-nucleon system.

The semi-inclusive diffractive DIS process~\cite{Hatta:2022lzj} has been shown to be factorized in terms of TMD diffractive parton distribution function (DPDF). We assume that the diffractive lepton-jet production can also be factorized in terms of quark TMD DPDF, where the longitudinal momentum fraction carried by quark from the pomeron is $\beta = x/ x_{\text{IP}}$. In the small-$x$ framework, the quark TMD DPDF~\cite{Berera:1995fj,Anselmino:2011ss} is related to dipole S-matrix and encode information about gluon saturation~\cite{Hatta:2022lzj}.

The expression of the quark TMD DPDF in $k_\perp$ space is taken from Ref.~\cite{Hatta:2022lzj}
\begin{equation}
\begin{aligned}
\frac{df_q^D(\beta, k_{\perp},t;x_{ \text{IP} })}{d Y_{\text{IP}} dt}  = &\frac{N_c \beta}{2\pi} \int d^2 k_{1\perp} d^2 k_{2\perp} \mathcal{F}_{x_{\text{IP}}}\left(k_{1\perp}, \Delta_\perp \right) \\  
& \times \mathcal{F}_{x_{\text{IP}}}\left(k_{2\perp}, \Delta_\perp \right) \mathcal{T}_{q}\left(k_{\perp}, k_{1\perp},k_{2\perp}\right)
\end{aligned}
\label{eq:qTMD_DPDF}
\end{equation}
with $\mathcal{T}_{q}$ defined as sum of four terms $\mathcal{T}_{q} \equiv T_q(k_\perp, k_{1\perp}, k_{2\perp}) -  T_q(k_\perp, 0, k_{2\perp}) - T_q(k_\perp, k_{1\perp}, 0) +T_q(k_\perp, 0, 0)$, where
\begin{equation}
\begin{aligned}
&T_q\left(k_{\perp}, k_{1 \perp}, k_{2 \perp}\right) \\
&\hspace{-0.2cm} =\frac{(k_{\perp} -k_{1 \perp}) \cdot (k_{\perp} - k_{2 \perp})  k_{\perp}^2}{\left[\beta k_{\perp}^2+(1-\beta) (k_{\perp} -k_{1 \perp})^2\right]\left[\beta k_{\perp}^2+(1-\beta) (k_{\perp} -k_{2 \perp})^2\right]},
\end{aligned}
\end{equation}
and $\mathcal{F}_{x_{\text{IP}}}\left(k_{1\perp}, \Delta_\perp \right)$ represents the Fourier transform of the dipole S-matrix in the fundamental representation
\begin{equation}
\mathcal{F}_{x_{\text{IP}}}\left(k_{1\perp}, \Delta_\perp \right) =  \int \frac{d^2 b_{\perp} d^2 r_{\perp}}{(2 \pi)^4} e^{i \vec{k}_{1\perp} \cdot \vec{r}_{\perp}+i \vec{\Delta}_{\perp} \cdot \vec{b}_{\perp}} {\cal S}_x(r_{\perp},b_{\perp})~, 
\end{equation}
where $r_\perp$ is the dipole separation, and $b_\perp$ is the impact parameter. 

The azimuthal angle dependent cross-section and harmonics of diffractive process have similar definitions as Eq.~(\ref{equ:LJC-Xsection}) and Eq.~(\ref{eq:harmonics}). By replacing the small-$x$ unintegrated quark distribution $f_q(x,b_\perp)$ with small-$x$ quark TMD DPDF, we get the angle dependent diffractive  lepton-jet cross section
\begin{equation}
\begin{aligned}
  & \frac{d^{5} \sigma(\ell P \rightarrow \ell^{\prime} J)}{d y_{\ell} d^{2} P_{ \perp} d^{2} q_{\perp} d Y_{\text{IP}} dt }=
\sigma_0\int  \frac{b_{\perp}db_{\perp}}{2\pi} W_{\text{diff}}
   \\ 
  &  
 \Big[J_0( q_\perp b_\perp) + 
\sum_{n=1}^\infty 2\cos( n\phi) \frac{ \alpha_s(\mu_{b}) C_Fc_n(R)}{n\pi} J_n(q_\perp b_\perp) \Big] ~.
\label{equ:LJC-Xsection}
\end{aligned}
\end{equation}
and diffractive harmonics
\begin{equation}
\begin{aligned}
&\langle \cos n\phi \rangle_{\text{diff}} \\ 
& =\frac{\sigma_0  \int b_{\perp} d b_{\perp} J_{n}\left(q_{\perp}b_{\perp}\right) 
W_{\text{diff}}\frac{ \alpha_s(\mu_{b})C_{F}  c_{n}(R)}{n \pi} 
 }
{ \sigma_0 \int b_{\perp} d b_{\perp} J_{0}\left(q_{\perp}b_{\perp}\right) W_{\text{diff}}  }~.
\end{aligned}
\label{eq:diff_harmonics}
\end{equation}
where $W_{\text{diff}}$ function is defined as
\begin{equation}
\begin{aligned}
&W_{\text{diff}}(x,\beta, b_\perp; x_{ \text{IP} } ) \\ 
& = e^{-\text{Sud}(b_\perp)}\int d^2k_{\perp} e^{i \vec{k}_{\perp} \cdot \vec{b}_{\perp}} x\frac{df_q^D(\beta, k_{\perp},t;x_{ \text{IP} })}{d Y_{\text{IP}} dt}.
\end{aligned}
\end{equation}

\subsection{The rapidity gap of diffractive lepton-jet production}
\label{subsec:rapidity_diff}

The rapidity gap for semi-inclusive diffractive DIS (SIDDIS) follows the traditional rapidity gap for the diffractive process $Y_{\text{IP}} \sim \ln{1}/{x_{\text{IP}}}$~\cite{Barone:2002cv}. SIDDIS is defined in the Breit frame, which is the photon-nucleon center-of-mass frame (frame $C$), while the diffractive lepton-jet process is measured in the lepton-nucleon center-of-mass frame (frame $A$). The rapidity gap of the diffractive lepton-jet is different from that of SIDDIS due to the Lorentz transformation between the two frames, which involves a Lorentz rotation.

The frame transformation from the lepton-nucleon center-of-mass frame (frame $A$) to the photon-nucleon center-of-mass frame (frame $C$) can be understood in three steps: (1) The Lorentz boost from frame A to the nucleon rest frame with the lepton moving in the $-z$ direction (frame $B$); (2) The rotation from frame $B$ to the nucleon rest frame with photon moving in the $-z'$ direction (frame  $B'$); (3) The Lorentz boost from frame $B'$ to frame $C$.


\begin{figure}[htbp]
\centering
\includegraphics[scale=0.35]{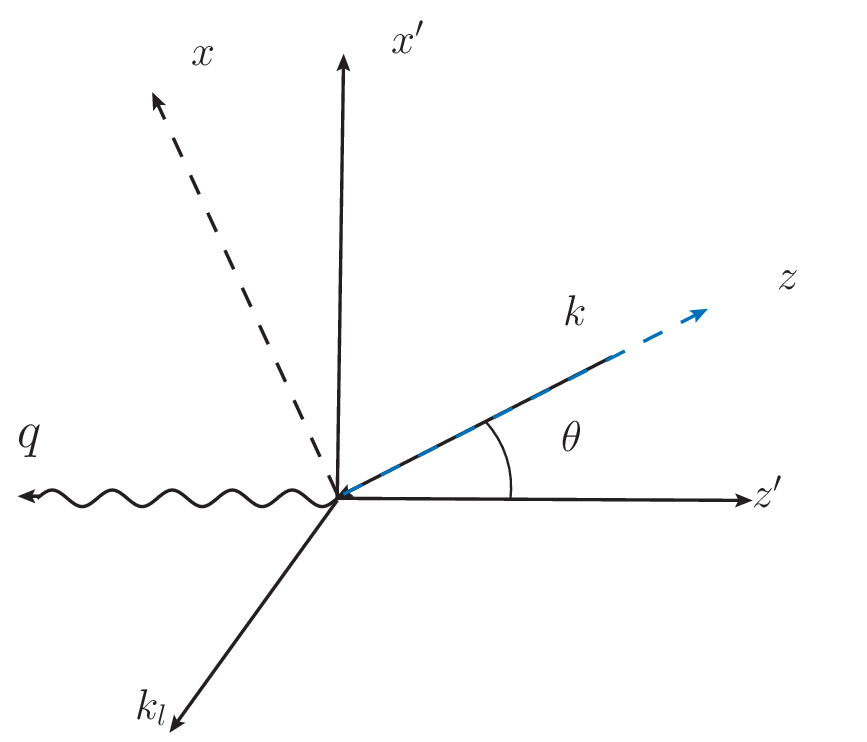}
 \caption{The rotation from the nucleon rest frame with the lepton in the $-z$ direction (frame $B$) to the nucleon rest frame with the photon in the $-z'$ direction (frame  $B'$), is depicted in the lepton plane defined by the incoming lepton with momentum $k$ and the outgoing lepton with momentum $k_l$.}  
 \label{fig:LorentzRotate}
\end{figure}
The demonstration of the Lorentz rotation~\cite{Diehl:2005pc} can be seen in Fig.~\ref{fig:LorentzRotate}, with rotation angle denoted as $\theta$. The rotation angle can be determined from the four-momentum of the virtual photon in these two frame,
\begin{equation}
\tan\theta = - \frac{q_B^1}{q_B^3}~,
\end{equation}
where $q_B=(q_B^0, q_B^1,0,q_B^3)$ is the photon four-momentum in frame $B$. By choosing the kinematics $\sqrt{s_{eN}}=89$ GeV, $y_\ell=y_J= 2.41$,$x=0.008$, $P_\perp=4$ GeV, $Q = 5.6$ ~GeV and $\beta = 0.94$, we find that $\theta=0.00187$ for the Lorentz rotation. The Lorentz rotation matrix is nearly an identity matrix, indicating that the rapidity is barely changed by the Lorentz rotation. Since the rapidity gap is invariant under Lorentz boost, the rapidity gap in the photon-nucleon center-of-mass frame (frame $A$) is almost the same value as the rapidity gap in the lepton-nucleon  center-of-mass frame (frame $C$). Therefore, we can use $Y_{\text{IP}} \sim \ln{1}/{x_{\text{IP}}}$ to represent the rapidity gap for diffractive lepton-jet production.

\subsection{Numerical calculation of diffractive harmonics}
\label{subsec:numerical_diffharmonics}
In the numerical calculation, we first neglect the impact parameter $b_\perp$ dependence of the dipole S-matrix and untilize two models for ${\cal S}_x(r_{\perp})$: the GBW model Eq.~(\ref{equ:GBW_Smatrix}) and the solution of the rcBK equation with the modified MV model as the initial condition, shown in Eq.~(\ref{equ:rcBK_MV}).  

We calculate the $q_\perp$-distribution of the harmonics $\langle \cos n\phi \rangle_{ \text{diff} }$. The kinematics bin of diffractive lepton-jet production at the future EIC is defined as follows: $\sqrt{s_{eN}}=89$ GeV, $y_\ell=2.41$, $0.008\leq x\leq 0.0094, \beta = 0.94, x_{\text{IP} } =x/\beta$, $ 4$ GeV$\leq P_\perp\leq4.32$ GeV, 5.6~GeV$\leq Q\leq$ 5.89~GeV. The value of $\beta$ can vary, but it should be chosen such that $x_{\text{IP} }$ falls within the range of $[0.008,0.01]$.

\begin{figure*}[htpb]
\centering
\includegraphics[scale=0.8]{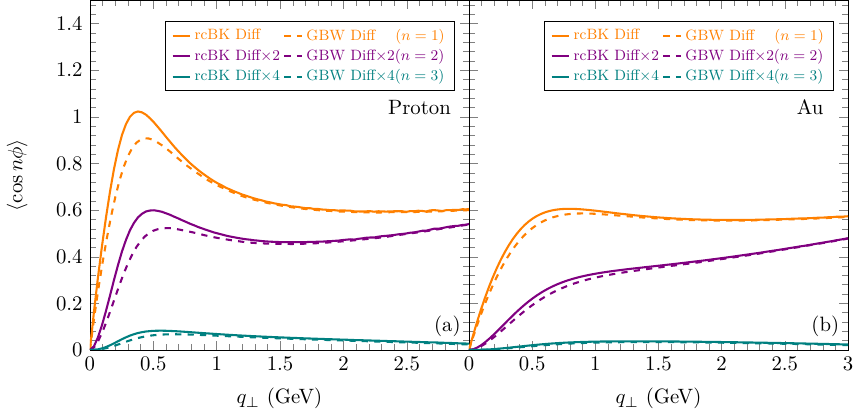}
 \caption{First three harmonics of diffractive lepton-jet production in (a) $e+p$ collisions and (b) $e+\text{Au}$ collisions with the inputs from the rcBK solution, GBW model. The EIC kinematics for calculation is $\sqrt{s_{eN}}=89$ GeV, $\beta=0.94$, $0.008<x<0.0094$, $y_l=2.41$ with jet cone size $R=0.4$.  }  
 \label{fig:Diff_harmonics}
\end{figure*}

In Fig.~\ref{fig:Diff_harmonics}, we plot the harmonics of diffractive lepton-jet production for saturation models, considering both proton and gold nucleus target, with jet cone size $R=0.4$. The decrease of harmonics from proton to gold nucleus target are also observed in Fig.~\ref{fig:Diff_harmonics}. Notably, the harmonics of the diffractive process are nearly two times the value of the harmonics of the inclusive lepton-jet process. This behavior can be explained by the asymptotic form of the harmonics, as given in Eq.~(\ref{equ:asymptotic_harmonics}). For the same choice of $Q, P_\perp$ for inclusive and diffractive lepton-jet process, the saddle point $b^{\text{sp}}_{\perp n}$ values are the same. For example, if $x=0.008$, $P_\perp=4$~GeV, $Q = 5.6$~GeV, the saddle points are $b^{\text{sp}}_{\perp 0} = 1.68$ GeV$^{-1}$, $b^{\text{sp}}_{\perp 1}=2.22$ GeV$^{-1}$, $b^{\text{sp}}_{\perp 2}=2.59$ GeV$^{-1}$, and $ b^{\text{sp}}_{\perp 3} = 2.87$ GeV$^{-1}$. 
\begin{figure}[htpb]
\centering
\includegraphics[scale=0.8]{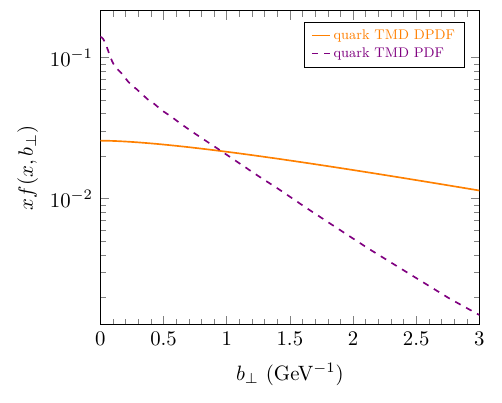}
 \caption{The comparison between quark diffractive TMD distribution $x\frac{df_q^D(\beta, b_{\perp};x_{ \text{IP} })}{d Y_{\text{IP}} dt}$ and quark TMD distribution $xf(x,b_{\perp})$ in coordinate space for $b_\perp \in [0,3]$ GeV$^{-1}$.}  
 \label{fig:DPDF_PDF}
\end{figure}
We plot the quark TMD DPDF and PDF for $b_\perp \in [0,   3]$ GeV$^{-1}$ in Fig.~\ref{fig:DPDF_PDF}. It is evident that in small $b_\perp$ region the flat DPDF gives
\begin{equation}
{\frac{df_q^D(\beta, b_{\perp n}^{\mathrm{sp}};x_{ \text{IP} })}{d Y_{\text{IP}} dt}}/{\frac{df_q^D(\beta, b_{\perp 0}^{\mathrm{sp}};x_{ \text{IP} })}{d Y_{\text{IP}} dt}} \approx 1~,
\end{equation}
while the steeply declining PDF results in 
\begin{equation}
\frac{f(x, b_{\perp n}^{\mathrm{sp}})}{f(x, b_{\perp 0}^{\mathrm{sp}})} \ll 1~.
\end{equation}
This difference causes $\langle \cos n\phi \rangle_{ \text{diff} }$  a couple times of $\langle \cos n\phi \rangle$.

\begin{figure}[htpb]
\centering
\includegraphics[scale=0.8]{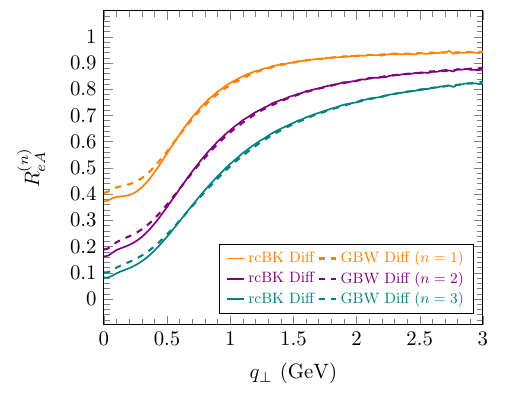}
 \caption{Nuclear modification factor of the first three harmonics for diffractive lepton-jet production, with the rcBK solution and GBW model as inputs. The EIC kinematics for calculation is $\sqrt{s_{eN}}=89$ GeV, $\beta=0.94$, $0.008<x<0.0094$, $y_l=2.41$ with jet cone size $R=0.4$. }  
 \label{fig:Diff_Ratio}
\end{figure}

We also plot the nuclear modification factor for diffractive harmonics in Fig.~\ref{fig:Diff_Ratio}, using both the GBW model and rcBK solution as inputs. Surprisingly, the nuclear modification factor of diffractive lepton-jet harmonics is nearly identical to that of inclusive lepton-jet harmonics in Fig.~\ref{fig:ratio_QsBand}. The larger harmonics and nearly identical nuclear modification factor, compared to inclusive lepton-jet production,  make them even better observables for studying saturation phenomenon.

We investigate the $t$ dependence of the harmonics, by restoring the impact factor $b_\perp$ dependence of the quark diffractive PDF. In essence, this calculation enables us to explore the sensitivity of diffractive harmonics to nuclear density profile. To demonstrate this feature, we choose two different density profiles for proton(nucleus): one being a uniform cylinder, the other a uniform sphere.

By considering the proton(nucleus) as a uniform cylinder with radius $r_p$($r_A$), we employ the GBW model for the dipole S-matrix. The Fourier transform of the dipole S-matrix reads
 \begin{equation}
\begin{aligned}
\mathcal{F}_{x_{\text{IP}}}\left(k_{1\perp}, \Delta_\perp \right)=& \int \frac{d^2 b_{\perp} d^2 r_{\perp}}{(2 \pi)^4} e^{i \vec{k}_{1\perp} \cdot \vec{r}_{\perp}+i \vec{\Delta}_{\perp} \cdot \vec{b}_{\perp}} e^{ - \frac{r_{\perp}^2Q_{s ,p}^2(x)}{4} } \\ 
=& \frac{r_p J_1(r_p\Delta_\perp)}{2\pi \Delta_\perp} \int \frac{d^2 r_{\perp}}{(2 \pi)^2} e^{i \vec{k}_{1\perp} \cdot \vec{r}_{\perp}} e^{ - \frac{r_{\perp}^2Q_{s ,p}^2(x)}{4} }
\end{aligned}
\label{eq:cylinder_PDF}
\end{equation}
In this case, the $t$($\Delta_\perp$) dependence factorizes. Therefore, it cancels out between numerator and denominator of the diffractive harmonics in Eq.(\ref{eq:diff_harmonics}). Consequently, the harmonics of "cylinder-like" proton do not exhibit $t$ dependence.

For a uniform sphere proton(nucleus), we employ the modified GBW model 
\begin{equation}
{\cal S}_x(r_{\perp}, b_\perp)=e^{ - \frac{r_{\perp}^2Q_{s ,p}^2(x,b_\perp)}{4} }~,
\end{equation}
where
\begin{equation}
Q_{s ,p}^2(x,b_\perp) = c_s \sqrt{1-\frac{b_\perp^2}{r_p^2}}~.
\end{equation}
The radius of the proton is $r_p = 4.2$ GeV$^{-1}$ (for the gold nucleus $r_A = 32.5$ GeV$^{-1}$). To compare with the above cylinder profile, we require that the impact parameter dependent saturation scale squared $Q_{s ,p}^2(x,b_\perp)$ satisfies the normalization condition
\begin{equation}
\int d^2b_\perp Q_{s ,p}^2(x,b_\perp) = \pi r_p^2 Q_{s,p}^2(x)~.
\end{equation}
 The right-hand side saturation scale squared of the traditional GBW model is given by $Q_{s,p}^2(x)=(x_0/x)^{0.28}$ GeV$^2$ with $x_0 =3\times 10^{-4}$. For the gold nucleus, we choose $Q_{s,A}^2(x) =  5Q_{s,p}^2(x)$. As the conjugate variable of $b_\perp$ in the Fourier transform, the $t$($\Delta_\perp$) dependence of diffractive harmonics opens a new dimension to distinguish different nuclear density profiles.

\begin{figure*}[htpb]
\centering
\includegraphics[scale=0.8]{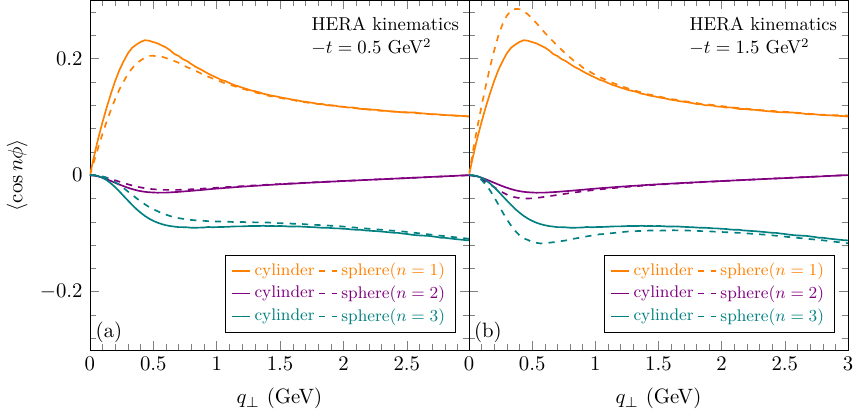}
 \caption{The comparison of diffractive harmonics of lepton-jet production in $e+p$ collisions in the HERA kinematics, for cylinder and sphere proton shape. The $t$ has two value (a) $-t=0.5$ GeV$^2$(b)  $-t=1.5$ GeV$^2$.  The HERA kinematics are $\sqrt{s}_{eN}=319$ GeV, $0.2<y<0.7$, $-1<\eta_{\text{lab}}<2.5$, $k_{J\perp}>10$ GeV, $Q^2>150$ GeV$^2$ with $\beta=0.94, x_{\text{IP}}<0.01, R=1.0$.  }  
 \label{fig:HERA_tDep}
\end{figure*}

\begin{figure*}[htpb]
\centering
\includegraphics[scale=0.8]{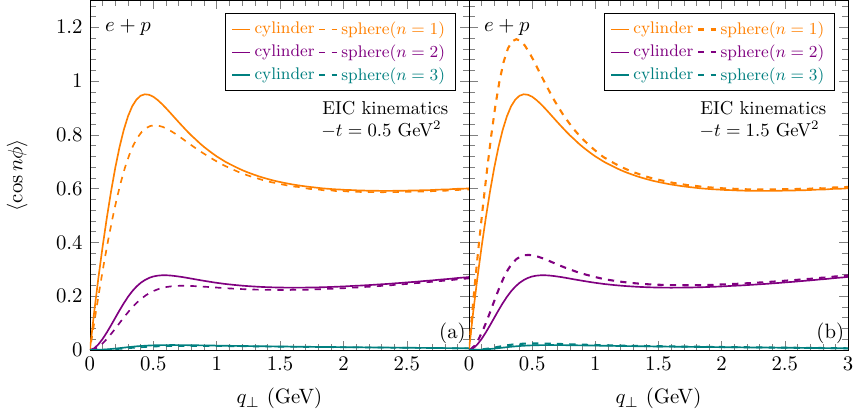}
 \caption{The comparison of diffractive harmonics of lepton-jet production in $e+p$ collisions in the EIC kinematics, for cylinder and sphere  proton shape. The $t$ has two value (a) $-t=0.5$ GeV$^2$(b)  $-t=1.5$ GeV$^2$. The $t$ dependent model assume a sphere shape of the proton. The EIC kinematics are $\sqrt{s}_{eN}=89$ GeV, $x=0.008$, $y_l=2.41$ with $\beta=0.94, x_{\text{IP}}<0.01, R=0.4$. }  
 \label{fig:EIC_tDep}
\end{figure*}

\begin{figure*}[htpb]
\centering
\includegraphics[scale=0.8]{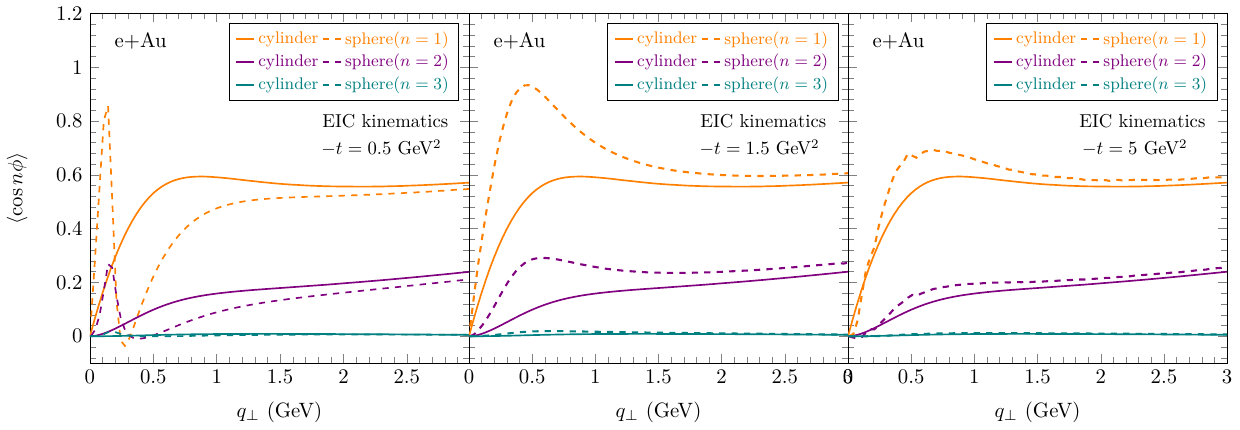}
 \caption{The comparison of diffractive harmonics of lepton-jet production in $e+\text{Au}$~collisions in the EIC kinematics, for cylinder and sphere  gold nucleus shape. The $t$ has three value (a) $-t=0.5$ GeV$^2$(b)  $-t=1.5$ GeV$^2$(c)(b)  $-t=5$ GeV$^2$. The $t$ dependent model assume a sphere shape of the gold nucleus. The EIC kinematics are $\sqrt{s}_{eN}=89$ GeV, $x=0.008$, $y_l=2.41$ with $\beta=0.94, x_{\text{IP}}<0.01, R=0.4$. }  
 \label{fig:EIC_tDep_Au}
\end{figure*}

We plot the diffractive harmonics for the cylinder and sphere proton(nucleus) in both the HERA and the EIC kinematics. Figure~\ref{fig:HERA_tDep} displays the $e+p$ results in the HERA kinematics. For EIC kinematics, we predict both  $e+p$  and $e+\text{Au}$~results, in  Fig.~\ref{fig:EIC_tDep} and Fig.~\ref{fig:EIC_tDep_Au}, respectively. For the proton, we compute with $-t=0.5$ GeV$^2$ and $-t=1.5$ GeV$^2$. Regarding the gold nucleus, we select $-t=0.5$ GeV$^2$, $-t=1.5$ GeV$^2$ and $-t=5$ GeV$^2$.
The sizable difference between cylinder and sphere proton(nucleus) suggests harmonics as new probes for the density profile of the target. Various density profiles can be tested in diffractive harmonics in the future study, such as Gaussian~\cite{Mantysaari:2019csc} or a more flexible parameterization~\cite{Lappi:2023frf}. The notable sharp peaks of diffractive harmonics for sphere shape Au at $-t=0.5$ GeV$^2$ in Fig.~\ref{fig:EIC_tDep_Au} originate from the diffractive nature of this process. We will explain this behavior in the following discussion.

\begin{figure}[htpb]
\centering
\includegraphics[scale=0.9]{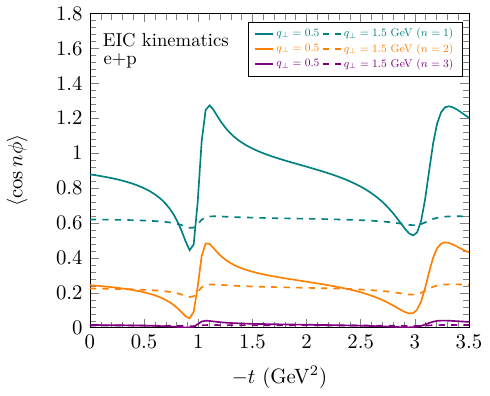}
 \caption{The $t$ distribution of harmonics for diffractive lepton-jet production in $e+p$ collisions in the EIC kinematics, with the spherical density profile. Two different imbalanced momentum $q_\perp =0.5$ GeV and $q_\perp =1.5$ GeV have been chosen. The EIC kinematics are $\sqrt{s}_{eN}=89$ GeV, $x=0.008$, $y_l=2.41$ with $\beta=0.94, x_{\text{IP}}<0.01, R=0.4$.}  
 \label{fig:EIC_ep_tDes}
\end{figure}

\begin{figure}[htpb]
\centering
\includegraphics[scale=0.9]{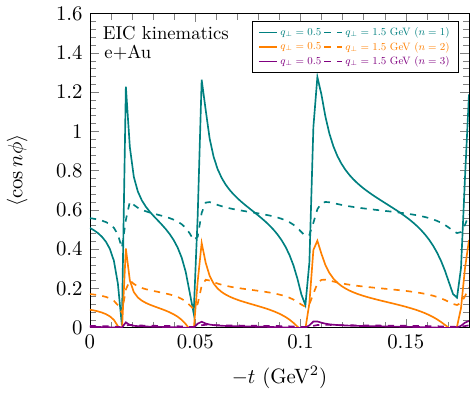}
 \caption{The $t$ distribution of harmonics for diffractive lepton-jet production in $e+\text{Au}$~collisions in the EIC kinematics, with the spherical density profile. Two different imbalanced momentum $q_\perp =0.5$ GeV and $q_\perp =1.5$ GeV have been chosen. The EIC kinematics are $\sqrt{s}_{eN}=89$ GeV, $x=0.008$, $y_l=2.41$ with $\beta=0.94, x_{\text{IP}}<0.01, R=0.4$.}  
 \label{fig:EIC_eA_tDes}
\end{figure}

The $t$-distribution of diffractive scattering cross-sections in nuclear and hadronic physics always show a pulse shape\cite{Caldwell:2010zza,Barone:2002cv,Kovchegov:2012mbw}, resembling the diffraction pattern in optics. We present the $t$-distribution of diffractive harmonics of $e+p$ collisions with a spherical proton in Fig.~\ref{fig:EIC_ep_tDes}, with $q_\perp = 0.5$ GeV and $q_\perp = 1.5$ GeV. Since the sphere-shaped proton is circular in the transverse plane, the Fourier transform of a circle is the Bessel function of the first kind $J_1(r_p \Delta_\perp)$. The positions of the minima can be determined by zeros of the bessel function $J_1(r_p \Delta_\perp)$ at $r_p \Delta_\perp = 3.8, 7.0, 10.2, ...$~. The first minina of Fig.~\ref{fig:EIC_ep_tDes} can be calculated as $-t = [3.8/r_p]^2 \approx 0.9$ GeV$^2$. Furthermore, we calculate the $t$-distribution of diffractive harmonics for $e+\text{Au}$  collisions with spherical gold nucleus in Fig.~\ref{fig:EIC_eA_tDes}, with $q_\perp = 0.5$ GeV and $q_\perp = 1.5$ GeV. The positions of minima are the same as those of the t-distribution for $J/\psi$~photoproduction\cite{Caldwell:2010zza}, with first minima as $-t = [3.8/r_A]^2 \approx 0.014$ GeV$^2$. In Fig.~\ref{fig:EIC_tDep_Au}, the sharp peaks of $q_\perp$ distribution of $e$+Au diffractive harmonics with $-t=0.5$ GeV$^2$  arise from the divergent behavior at $-t=0.5$ GeV$^2$, which coincides with one of the minima. In contrast, if the density profile were cylinder-like, the harmonics in Fig.~\ref{fig:EIC_ep_tDes} and Fig.~\ref{fig:EIC_eA_tDes} would be constant as $t$ varies.


The above calculation provides quantitative predictions to the future experimental studies on diffractive lepton-jet production at HERA and EIC.

\section{Conclusion}  
In this paper, we propose novel observables for studying gluon saturation: harmonics and its nuclear modification factors of inclusive and diffractive lepton-jet correlation. 

Using the small-$x$ factorization, the detailed derivation of the azimuthal angle dependent lepton-jet correlation in small-$x$ framework is presented. We obtain analytical expressions for the harmonics, which predicts the suppression of the harmonics with increasing of saturation scale $Q_s$. This behavior is confirmed by numerical calculation. Furthermore, we find that the impact of QED radiation corrections on the harmonics are sizable, while negligible for nuclear modification factor. These can be seen both from expressions and numerical calculation. The striking difference in the nuclear modification factor for non-saturation and saturation framework make it a robust observable to distinguish these two frameworks.

In addition, the parallel study on the diffractive lepton-jet production is carried out. Numerical calculation demonstrate that the diffractive harmonics are twice the value of the inclusive harmonics, while the nuclear modification factors are almost the same. These findings suggest that the diffractive harmonics and their nuclear modification factors serve as sensitive observables for probing gluon saturation phenomenon. In particular, the t-dependent diffractive harmonics can distinguish different nuclear density profiles.

\par 

\section*{Acknowledgments} 
We thank Feng Yuan and Heikki Mäntysaari for discussions. This work is supported by the CUHK-Shenzhen university development fund under the grant No. UDF01001859. Xuan-Bo Tong is also supported by the Research Council of Finland, the Centre of Excellence in Quark Matter and under the European Union’s Horizon 2020 research and innovation programme by the European Research Council (ERC, grant agreement No. ERC-2018-ADG-835105 YoctoLHC) and by the STRONG-2020 project (grant agreement No. 824093). The content of this article does not reflect the official opinion of the European Union and responsibility for the information and views expressed therein lies entirely with the authors.

\par

\bibliographystyle{apsrev4-1}
\bibliography{lepjet_long_v2}

\begin{thebibliography}{200}%
\makeatletter
\providecommand \@ifxundefined [1]{%
 \@ifx{#1\undefined}
}%
\providecommand \@ifnum [1]{%
 \ifnum #1\expandafter \@firstoftwo
 \else \expandafter \@secondoftwo
 \fi
}%
\providecommand \@ifx [1]{%
 \ifx #1\expandafter \@firstoftwo
 \else \expandafter \@secondoftwo
 \fi
}%
\providecommand \natexlab [1]{#1}%
\providecommand \enquote  [1]{``#1''}%
\providecommand \bibnamefont  [1]{#1}%
\providecommand \bibfnamefont [1]{#1}%
\providecommand \citenamefont [1]{#1}%
\providecommand \href@noop [0]{\@secondoftwo}%
\providecommand \href [0]{\begingroup \@sanitize@url \@href}%
\providecommand \@href[1]{\@@startlink{#1}\@@href}%
\providecommand \@@href[1]{\endgroup#1\@@endlink}%
\providecommand \@sanitize@url [0]{\catcode `\\12\catcode `\$12\catcode
  `\&12\catcode `\#12\catcode `\^12\catcode `\_12\catcode `\%12\relax}%
\providecommand \@@startlink[1]{}%
\providecommand \@@endlink[0]{}%
\providecommand \url  [0]{\begingroup\@sanitize@url \@url }%
\providecommand \@url [1]{\endgroup\@href {#1}{\urlprefix }}%
\providecommand \urlprefix  [0]{URL }%
\providecommand \Eprint [0]{\href }%
\providecommand \doibase [0]{http://dx.doi.org/}%
\providecommand \selectlanguage [0]{\@gobble}%
\providecommand \bibinfo  [0]{\@secondoftwo}%
\providecommand \bibfield  [0]{\@secondoftwo}%
\providecommand \translation [1]{[#1]}%
\providecommand \BibitemOpen [0]{}%
\providecommand \bibitemStop [0]{}%
\providecommand \bibitemNoStop [0]{.\EOS\space}%
\providecommand \EOS [0]{\spacefactor3000\relax}%
\providecommand \BibitemShut  [1]{\csname bibitem#1\endcsname}%
\let\auto@bib@innerbib\@empty
\bibitem [{\citenamefont {Tong}\ \emph {et~al.}(2023)\citenamefont {Tong},
  \citenamefont {Xiao},\ and\ \citenamefont {Zhang}}]{Tong:2022zwp}%
  \BibitemOpen
  \bibfield  {author} {\bibinfo {author} {\bibfnamefont {X.-B.}\ \bibnamefont
  {Tong}}, \bibinfo {author} {\bibfnamefont {B.-W.}\ \bibnamefont {Xiao}}, \
  and\ \bibinfo {author} {\bibfnamefont {Y.-Y.}\ \bibnamefont {Zhang}},\ }\href
  {\doibase 10.1103/PhysRevLett.130.151902} {\bibfield  {journal} {\bibinfo
  {journal} {Phys. Rev. Lett.}\ }\textbf {\bibinfo {volume} {130}},\ \bibinfo
  {pages} {151902} (\bibinfo {year} {2023})},\ \Eprint
  {http://arxiv.org/abs/2211.01647} {arXiv:2211.01647 [hep-ph]} \BibitemShut
  {NoStop}%
\bibitem [{\citenamefont {Gribov}\ \emph {et~al.}(1983)\citenamefont {Gribov},
  \citenamefont {Levin},\ and\ \citenamefont {Ryskin}}]{Gribov:1983ivg}%
  \BibitemOpen
  \bibfield  {author} {\bibinfo {author} {\bibfnamefont {L.~V.}\ \bibnamefont
  {Gribov}}, \bibinfo {author} {\bibfnamefont {E.~M.}\ \bibnamefont {Levin}}, \
  and\ \bibinfo {author} {\bibfnamefont {M.~G.}\ \bibnamefont {Ryskin}},\
  }\href {\doibase 10.1016/0370-1573(83)90022-4} {\bibfield  {journal}
  {\bibinfo  {journal} {Phys. Rept.}\ }\textbf {\bibinfo {volume} {100}},\
  \bibinfo {pages} {1} (\bibinfo {year} {1983})}\BibitemShut {NoStop}%
\bibitem [{\citenamefont {Mueller}\ and\ \citenamefont
  {Qiu}(1986)}]{Mueller:1985wy}%
  \BibitemOpen
  \bibfield  {author} {\bibinfo {author} {\bibfnamefont {A.~H.}\ \bibnamefont
  {Mueller}}\ and\ \bibinfo {author} {\bibfnamefont {J.-w.}\ \bibnamefont
  {Qiu}},\ }\href {\doibase 10.1016/0550-3213(86)90164-1} {\bibfield  {journal}
  {\bibinfo  {journal} {Nucl. Phys. B}\ }\textbf {\bibinfo {volume} {268}},\
  \bibinfo {pages} {427} (\bibinfo {year} {1986})}\BibitemShut {NoStop}%
\bibitem [{\citenamefont {Mueller}(1990)}]{Mueller:1989st}%
  \BibitemOpen
  \bibfield  {author} {\bibinfo {author} {\bibfnamefont {A.~H.}\ \bibnamefont
  {Mueller}},\ }\href {\doibase 10.1016/0550-3213(90)90173-B} {\bibfield
  {journal} {\bibinfo  {journal} {Nucl. Phys. B}\ }\textbf {\bibinfo {volume}
  {335}},\ \bibinfo {pages} {115} (\bibinfo {year} {1990})}\BibitemShut
  {NoStop}%
\bibitem [{\citenamefont {McLerran}\ and\ \citenamefont
  {Venugopalan}(1994{\natexlab{a}})}]{McLerran:1993ni}%
  \BibitemOpen
  \bibfield  {author} {\bibinfo {author} {\bibfnamefont {L.~D.}\ \bibnamefont
  {McLerran}}\ and\ \bibinfo {author} {\bibfnamefont {R.}~\bibnamefont
  {Venugopalan}},\ }\href {\doibase 10.1103/PhysRevD.49.2233} {\bibfield
  {journal} {\bibinfo  {journal} {Phys. Rev. D}\ }\textbf {\bibinfo {volume}
  {49}},\ \bibinfo {pages} {2233} (\bibinfo {year} {1994}{\natexlab{a}})},\
  \Eprint {http://arxiv.org/abs/hep-ph/9309289} {arXiv:hep-ph/9309289}
  \BibitemShut {NoStop}%
\bibitem [{\citenamefont {McLerran}\ and\ \citenamefont
  {Venugopalan}(1994{\natexlab{b}})}]{McLerran:1993ka}%
  \BibitemOpen
  \bibfield  {author} {\bibinfo {author} {\bibfnamefont {L.~D.}\ \bibnamefont
  {McLerran}}\ and\ \bibinfo {author} {\bibfnamefont {R.}~\bibnamefont
  {Venugopalan}},\ }\href {\doibase 10.1103/PhysRevD.49.3352} {\bibfield
  {journal} {\bibinfo  {journal} {Phys. Rev. D}\ }\textbf {\bibinfo {volume}
  {49}},\ \bibinfo {pages} {3352} (\bibinfo {year} {1994}{\natexlab{b}})},\
  \Eprint {http://arxiv.org/abs/hep-ph/9311205} {arXiv:hep-ph/9311205}
  \BibitemShut {NoStop}%
\bibitem [{\citenamefont {McLerran}\ and\ \citenamefont
  {Venugopalan}(1994{\natexlab{c}})}]{McLerran:1994vd}%
  \BibitemOpen
  \bibfield  {author} {\bibinfo {author} {\bibfnamefont {L.~D.}\ \bibnamefont
  {McLerran}}\ and\ \bibinfo {author} {\bibfnamefont {R.}~\bibnamefont
  {Venugopalan}},\ }\href {\doibase 10.1103/PhysRevD.50.2225} {\bibfield
  {journal} {\bibinfo  {journal} {Phys. Rev. D}\ }\textbf {\bibinfo {volume}
  {50}},\ \bibinfo {pages} {2225} (\bibinfo {year} {1994}{\natexlab{c}})},\
  \Eprint {http://arxiv.org/abs/hep-ph/9402335} {arXiv:hep-ph/9402335}
  \BibitemShut {NoStop}%
\bibitem [{\citenamefont {Morreale}\ and\ \citenamefont
  {Salazar}(2021)}]{Morreale:2021pnn}%
  \BibitemOpen
  \bibfield  {author} {\bibinfo {author} {\bibfnamefont {A.}~\bibnamefont
  {Morreale}}\ and\ \bibinfo {author} {\bibfnamefont {F.}~\bibnamefont
  {Salazar}},\ }\href {\doibase 10.3390/universe7080312} {\bibfield  {journal}
  {\bibinfo  {journal} {Universe}\ }\textbf {\bibinfo {volume} {7}},\ \bibinfo
  {pages} {312} (\bibinfo {year} {2021})},\ \Eprint
  {http://arxiv.org/abs/2108.08254} {arXiv:2108.08254 [hep-ph]} \BibitemShut
  {NoStop}%
\bibitem [{\citenamefont {Gelis}\ \emph {et~al.}(2010)\citenamefont {Gelis},
  \citenamefont {Iancu}, \citenamefont {Jalilian-Marian},\ and\ \citenamefont
  {Venugopalan}}]{Gelis:2010nm}%
  \BibitemOpen
  \bibfield  {author} {\bibinfo {author} {\bibfnamefont {F.}~\bibnamefont
  {Gelis}}, \bibinfo {author} {\bibfnamefont {E.}~\bibnamefont {Iancu}},
  \bibinfo {author} {\bibfnamefont {J.}~\bibnamefont {Jalilian-Marian}}, \ and\
  \bibinfo {author} {\bibfnamefont {R.}~\bibnamefont {Venugopalan}},\ }\href
  {\doibase 10.1146/annurev.nucl.010909.083629} {\bibfield  {journal} {\bibinfo
   {journal} {Ann. Rev. Nucl. Part. Sci.}\ }\textbf {\bibinfo {volume} {60}},\
  \bibinfo {pages} {463} (\bibinfo {year} {2010})},\ \Eprint
  {http://arxiv.org/abs/1002.0333} {arXiv:1002.0333 [hep-ph]} \BibitemShut
  {NoStop}%
\bibitem [{\citenamefont {Iancu}\ and\ \citenamefont
  {Venugopalan}(2003)}]{Iancu:2003xm}%
  \BibitemOpen
  \bibfield  {author} {\bibinfo {author} {\bibfnamefont {E.}~\bibnamefont
  {Iancu}}\ and\ \bibinfo {author} {\bibfnamefont {R.}~\bibnamefont
  {Venugopalan}},\ }\enquote {\bibinfo {title} {{The Color glass condensate and
  high-energy scattering in QCD}},}\ in\ \href {\doibase
  10.1142/9789812795533_0005} {\emph {\bibinfo {booktitle} {{Quark-gluon plasma
  4}}}},\ \bibinfo {editor} {edited by\ \bibinfo {editor} {\bibfnamefont
  {R.~C.}\ \bibnamefont {Hwa}}\ and\ \bibinfo {editor} {\bibfnamefont {X.-N.}\
  \bibnamefont {Wang}}}\ (\bibinfo {year} {2003})\ pp.\ \bibinfo {pages}
  {249--3363},\ \Eprint {http://arxiv.org/abs/hep-ph/0303204}
  {arXiv:hep-ph/0303204} \BibitemShut {NoStop}%
\bibitem [{\citenamefont {Kovchegov}\ and\ \citenamefont
  {Levin}(2013)}]{Kovchegov:2012mbw}%
  \BibitemOpen
  \bibfield  {author} {\bibinfo {author} {\bibfnamefont {Y.~V.}\ \bibnamefont
  {Kovchegov}}\ and\ \bibinfo {author} {\bibfnamefont {E.}~\bibnamefont
  {Levin}},\ }\href {\doibase 10.1017/9781009291446} {\emph {\bibinfo {title}
  {{Quantum Chromodynamics at High Energy}}}},\ Vol.~\bibinfo {volume} {33}\
  (\bibinfo  {publisher} {Oxford University Press},\ \bibinfo {year}
  {2013})\BibitemShut {NoStop}%
\bibitem [{\citenamefont {Albacete}\ and\ \citenamefont
  {Marquet}(2014)}]{Albacete:2014fwa}%
  \BibitemOpen
  \bibfield  {author} {\bibinfo {author} {\bibfnamefont {J.~L.}\ \bibnamefont
  {Albacete}}\ and\ \bibinfo {author} {\bibfnamefont {C.}~\bibnamefont
  {Marquet}},\ }\href {\doibase 10.1016/j.ppnp.2014.01.004} {\bibfield
  {journal} {\bibinfo  {journal} {Prog. Part. Nucl. Phys.}\ }\textbf {\bibinfo
  {volume} {76}},\ \bibinfo {pages} {1} (\bibinfo {year} {2014})},\ \Eprint
  {http://arxiv.org/abs/1401.4866} {arXiv:1401.4866 [hep-ph]} \BibitemShut
  {NoStop}%
\bibitem [{\citenamefont {Blaizot}(2017)}]{Blaizot:2016qgz}%
  \BibitemOpen
  \bibfield  {author} {\bibinfo {author} {\bibfnamefont {J.-P.}\ \bibnamefont
  {Blaizot}},\ }\href {\doibase 10.1088/1361-6633/aa5435} {\bibfield  {journal}
  {\bibinfo  {journal} {Rept. Prog. Phys.}\ }\textbf {\bibinfo {volume} {80}},\
  \bibinfo {pages} {032301} (\bibinfo {year} {2017})},\ \Eprint
  {http://arxiv.org/abs/1607.04448} {arXiv:1607.04448 [hep-ph]} \BibitemShut
  {NoStop}%
\bibitem [{\citenamefont {Gelis}\ and\ \citenamefont
  {Jalilian-Marian}(2003)}]{Gelis:2002nn}%
  \BibitemOpen
  \bibfield  {author} {\bibinfo {author} {\bibfnamefont {F.}~\bibnamefont
  {Gelis}}\ and\ \bibinfo {author} {\bibfnamefont {J.}~\bibnamefont
  {Jalilian-Marian}},\ }\href {\doibase 10.1103/PhysRevD.67.074019} {\bibfield
  {journal} {\bibinfo  {journal} {Phys. Rev. D}\ }\textbf {\bibinfo {volume}
  {67}},\ \bibinfo {pages} {074019} (\bibinfo {year} {2003})},\ \Eprint
  {http://arxiv.org/abs/hep-ph/0211363} {arXiv:hep-ph/0211363} \BibitemShut
  {NoStop}%
\bibitem [{\citenamefont {Baier}\ \emph {et~al.}(2005)\citenamefont {Baier},
  \citenamefont {Kovner}, \citenamefont {Nardi},\ and\ \citenamefont
  {Wiedemann}}]{Baier:2005dv}%
  \BibitemOpen
  \bibfield  {author} {\bibinfo {author} {\bibfnamefont {R.}~\bibnamefont
  {Baier}}, \bibinfo {author} {\bibfnamefont {A.}~\bibnamefont {Kovner}},
  \bibinfo {author} {\bibfnamefont {M.}~\bibnamefont {Nardi}}, \ and\ \bibinfo
  {author} {\bibfnamefont {U.~A.}\ \bibnamefont {Wiedemann}},\ }\href {\doibase
  10.1103/PhysRevD.72.094013} {\bibfield  {journal} {\bibinfo  {journal} {Phys.
  Rev. D}\ }\textbf {\bibinfo {volume} {72}},\ \bibinfo {pages} {094013}
  (\bibinfo {year} {2005})},\ \Eprint {http://arxiv.org/abs/hep-ph/0506126}
  {arXiv:hep-ph/0506126} \BibitemShut {NoStop}%
\bibitem [{\citenamefont {Dominguez}\ \emph
  {et~al.}(2011{\natexlab{a}})\citenamefont {Dominguez}, \citenamefont {Xiao},\
  and\ \citenamefont {Yuan}}]{Dominguez:2010xd}%
  \BibitemOpen
  \bibfield  {author} {\bibinfo {author} {\bibfnamefont {F.}~\bibnamefont
  {Dominguez}}, \bibinfo {author} {\bibfnamefont {B.-W.}\ \bibnamefont {Xiao}},
  \ and\ \bibinfo {author} {\bibfnamefont {F.}~\bibnamefont {Yuan}},\ }\href
  {\doibase 10.1103/PhysRevLett.106.022301} {\bibfield  {journal} {\bibinfo
  {journal} {Phys. Rev. Lett.}\ }\textbf {\bibinfo {volume} {106}},\ \bibinfo
  {pages} {022301} (\bibinfo {year} {2011}{\natexlab{a}})},\ \Eprint
  {http://arxiv.org/abs/1009.2141} {arXiv:1009.2141 [hep-ph]} \BibitemShut
  {NoStop}%
\bibitem [{\citenamefont {Dominguez}\ \emph
  {et~al.}(2011{\natexlab{b}})\citenamefont {Dominguez}, \citenamefont
  {Marquet}, \citenamefont {Xiao},\ and\ \citenamefont
  {Yuan}}]{Dominguez:2011wm}%
  \BibitemOpen
  \bibfield  {author} {\bibinfo {author} {\bibfnamefont {F.}~\bibnamefont
  {Dominguez}}, \bibinfo {author} {\bibfnamefont {C.}~\bibnamefont {Marquet}},
  \bibinfo {author} {\bibfnamefont {B.-W.}\ \bibnamefont {Xiao}}, \ and\
  \bibinfo {author} {\bibfnamefont {F.}~\bibnamefont {Yuan}},\ }\href {\doibase
  10.1103/PhysRevD.83.105005} {\bibfield  {journal} {\bibinfo  {journal} {Phys.
  Rev. D}\ }\textbf {\bibinfo {volume} {83}},\ \bibinfo {pages} {105005}
  (\bibinfo {year} {2011}{\natexlab{b}})},\ \Eprint
  {http://arxiv.org/abs/1101.0715} {arXiv:1101.0715 [hep-ph]} \BibitemShut
  {NoStop}%
\bibitem [{\citenamefont {Mueller}\ \emph
  {et~al.}(2013{\natexlab{a}})\citenamefont {Mueller}, \citenamefont {Xiao},\
  and\ \citenamefont {Yuan}}]{Mueller:2013wwa}%
  \BibitemOpen
  \bibfield  {author} {\bibinfo {author} {\bibfnamefont {A.~H.}\ \bibnamefont
  {Mueller}}, \bibinfo {author} {\bibfnamefont {B.-W.}\ \bibnamefont {Xiao}}, \
  and\ \bibinfo {author} {\bibfnamefont {F.}~\bibnamefont {Yuan}},\ }\href
  {\doibase 10.1103/PhysRevD.88.114010} {\bibfield  {journal} {\bibinfo
  {journal} {Phys. Rev. D}\ }\textbf {\bibinfo {volume} {88}},\ \bibinfo
  {pages} {114010} (\bibinfo {year} {2013}{\natexlab{a}})},\ \Eprint
  {http://arxiv.org/abs/1308.2993} {arXiv:1308.2993 [hep-ph]} \BibitemShut
  {NoStop}%
\bibitem [{\citenamefont {Dumitru}\ \emph {et~al.}(2015)\citenamefont
  {Dumitru}, \citenamefont {Lappi},\ and\ \citenamefont
  {Skokov}}]{Dumitru:2015gaa}%
  \BibitemOpen
  \bibfield  {author} {\bibinfo {author} {\bibfnamefont {A.}~\bibnamefont
  {Dumitru}}, \bibinfo {author} {\bibfnamefont {T.}~\bibnamefont {Lappi}}, \
  and\ \bibinfo {author} {\bibfnamefont {V.}~\bibnamefont {Skokov}},\ }\href
  {\doibase 10.1103/PhysRevLett.115.252301} {\bibfield  {journal} {\bibinfo
  {journal} {Phys. Rev. Lett.}\ }\textbf {\bibinfo {volume} {115}},\ \bibinfo
  {pages} {252301} (\bibinfo {year} {2015})},\ \Eprint
  {http://arxiv.org/abs/1508.04438} {arXiv:1508.04438 [hep-ph]} \BibitemShut
  {NoStop}%
\bibitem [{\citenamefont {Dumitru}\ and\ \citenamefont
  {Skokov}(2016)}]{Dumitru:2016jku}%
  \BibitemOpen
  \bibfield  {author} {\bibinfo {author} {\bibfnamefont {A.}~\bibnamefont
  {Dumitru}}\ and\ \bibinfo {author} {\bibfnamefont {V.}~\bibnamefont
  {Skokov}},\ }\href {\doibase 10.1103/PhysRevD.94.014030} {\bibfield
  {journal} {\bibinfo  {journal} {Phys. Rev. D}\ }\textbf {\bibinfo {volume}
  {94}},\ \bibinfo {pages} {014030} (\bibinfo {year} {2016})},\ \Eprint
  {http://arxiv.org/abs/1605.02739} {arXiv:1605.02739 [hep-ph]} \BibitemShut
  {NoStop}%
\bibitem [{\citenamefont {Boer}\ \emph {et~al.}(2016)\citenamefont {Boer},
  \citenamefont {Mulders}, \citenamefont {Pisano},\ and\ \citenamefont
  {Zhou}}]{Boer:2016fqd}%
  \BibitemOpen
  \bibfield  {author} {\bibinfo {author} {\bibfnamefont {D.}~\bibnamefont
  {Boer}}, \bibinfo {author} {\bibfnamefont {P.~J.}\ \bibnamefont {Mulders}},
  \bibinfo {author} {\bibfnamefont {C.}~\bibnamefont {Pisano}}, \ and\ \bibinfo
  {author} {\bibfnamefont {J.}~\bibnamefont {Zhou}},\ }\href {\doibase
  10.1007/JHEP08(2016)001} {\bibfield  {journal} {\bibinfo  {journal} {JHEP}\
  }\textbf {\bibinfo {volume} {08}},\ \bibinfo {pages} {001} (\bibinfo {year}
  {2016})},\ \Eprint {http://arxiv.org/abs/1605.07934} {arXiv:1605.07934
  [hep-ph]} \BibitemShut {NoStop}%
\bibitem [{\citenamefont {Dumitru}\ \emph {et~al.}(2019)\citenamefont
  {Dumitru}, \citenamefont {Skokov},\ and\ \citenamefont
  {Ullrich}}]{Dumitru:2018kuw}%
  \BibitemOpen
  \bibfield  {author} {\bibinfo {author} {\bibfnamefont {A.}~\bibnamefont
  {Dumitru}}, \bibinfo {author} {\bibfnamefont {V.}~\bibnamefont {Skokov}}, \
  and\ \bibinfo {author} {\bibfnamefont {T.}~\bibnamefont {Ullrich}},\ }\href
  {\doibase 10.1103/PhysRevC.99.015204} {\bibfield  {journal} {\bibinfo
  {journal} {Phys. Rev. C}\ }\textbf {\bibinfo {volume} {99}},\ \bibinfo
  {pages} {015204} (\bibinfo {year} {2019})},\ \Eprint
  {http://arxiv.org/abs/1809.02615} {arXiv:1809.02615 [hep-ph]} \BibitemShut
  {NoStop}%
\bibitem [{\citenamefont {Zhao}\ \emph {et~al.}(2021)\citenamefont {Zhao},
  \citenamefont {Xu}, \citenamefont {Chen}, \citenamefont {Zhang},\ and\
  \citenamefont {Wu}}]{Zhao:2021kae}%
  \BibitemOpen
  \bibfield  {author} {\bibinfo {author} {\bibfnamefont {Y.-Y.}\ \bibnamefont
  {Zhao}}, \bibinfo {author} {\bibfnamefont {M.-M.}\ \bibnamefont {Xu}},
  \bibinfo {author} {\bibfnamefont {L.-Z.}\ \bibnamefont {Chen}}, \bibinfo
  {author} {\bibfnamefont {D.-H.}\ \bibnamefont {Zhang}}, \ and\ \bibinfo
  {author} {\bibfnamefont {Y.-F.}\ \bibnamefont {Wu}},\ }\href {\doibase
  10.1103/PhysRevD.104.114032} {\bibfield  {journal} {\bibinfo  {journal}
  {Phys. Rev. D}\ }\textbf {\bibinfo {volume} {104}},\ \bibinfo {pages}
  {114032} (\bibinfo {year} {2021})},\ \Eprint
  {http://arxiv.org/abs/2105.08818} {arXiv:2105.08818 [hep-ph]} \BibitemShut
  {NoStop}%
\bibitem [{\citenamefont {Boussarie}\ \emph {et~al.}(2021)\citenamefont
  {Boussarie}, \citenamefont {M\"antysaari}, \citenamefont {Salazar},\ and\
  \citenamefont {Schenke}}]{Boussarie:2021ybe}%
  \BibitemOpen
  \bibfield  {author} {\bibinfo {author} {\bibfnamefont {R.}~\bibnamefont
  {Boussarie}}, \bibinfo {author} {\bibfnamefont {H.}~\bibnamefont
  {M\"antysaari}}, \bibinfo {author} {\bibfnamefont {F.}~\bibnamefont
  {Salazar}}, \ and\ \bibinfo {author} {\bibfnamefont {B.}~\bibnamefont
  {Schenke}},\ }\href {\doibase 10.1007/JHEP09(2021)178} {\bibfield  {journal}
  {\bibinfo  {journal} {JHEP}\ }\textbf {\bibinfo {volume} {09}},\ \bibinfo
  {pages} {178} (\bibinfo {year} {2021})},\ \Eprint
  {http://arxiv.org/abs/2106.11301} {arXiv:2106.11301 [hep-ph]} \BibitemShut
  {NoStop}%
\bibitem [{\citenamefont {Taels}\ \emph {et~al.}(2022)\citenamefont {Taels},
  \citenamefont {Altinoluk}, \citenamefont {Beuf},\ and\ \citenamefont
  {Marquet}}]{Taels:2022tza}%
  \BibitemOpen
  \bibfield  {author} {\bibinfo {author} {\bibfnamefont {P.}~\bibnamefont
  {Taels}}, \bibinfo {author} {\bibfnamefont {T.}~\bibnamefont {Altinoluk}},
  \bibinfo {author} {\bibfnamefont {G.}~\bibnamefont {Beuf}}, \ and\ \bibinfo
  {author} {\bibfnamefont {C.}~\bibnamefont {Marquet}},\ }\href {\doibase
  10.1007/JHEP10(2022)184} {\bibfield  {journal} {\bibinfo  {journal} {JHEP}\
  }\textbf {\bibinfo {volume} {10}},\ \bibinfo {pages} {184} (\bibinfo {year}
  {2022})},\ \Eprint {http://arxiv.org/abs/2204.11650} {arXiv:2204.11650
  [hep-ph]} \BibitemShut {NoStop}%
\bibitem [{\citenamefont {Caucal}\ \emph {et~al.}(2022)\citenamefont {Caucal},
  \citenamefont {Salazar}, \citenamefont {Schenke},\ and\ \citenamefont
  {Venugopalan}}]{Caucal:2022ulg}%
  \BibitemOpen
  \bibfield  {author} {\bibinfo {author} {\bibfnamefont {P.}~\bibnamefont
  {Caucal}}, \bibinfo {author} {\bibfnamefont {F.}~\bibnamefont {Salazar}},
  \bibinfo {author} {\bibfnamefont {B.}~\bibnamefont {Schenke}}, \ and\
  \bibinfo {author} {\bibfnamefont {R.}~\bibnamefont {Venugopalan}},\ }\href
  {\doibase 10.1007/JHEP11(2022)169} {\bibfield  {journal} {\bibinfo  {journal}
  {JHEP}\ }\textbf {\bibinfo {volume} {11}},\ \bibinfo {pages} {169} (\bibinfo
  {year} {2022})},\ \Eprint {http://arxiv.org/abs/2208.13872} {arXiv:2208.13872
  [hep-ph]} \BibitemShut {NoStop}%
\bibitem [{\citenamefont {Caucal}\ \emph
  {et~al.}(2023{\natexlab{a}})\citenamefont {Caucal}, \citenamefont {Salazar},
  \citenamefont {Schenke}, \citenamefont {Stebel},\ and\ \citenamefont
  {Venugopalan}}]{Caucal:2023nci}%
  \BibitemOpen
  \bibfield  {author} {\bibinfo {author} {\bibfnamefont {P.}~\bibnamefont
  {Caucal}}, \bibinfo {author} {\bibfnamefont {F.}~\bibnamefont {Salazar}},
  \bibinfo {author} {\bibfnamefont {B.}~\bibnamefont {Schenke}}, \bibinfo
  {author} {\bibfnamefont {T.}~\bibnamefont {Stebel}}, \ and\ \bibinfo {author}
  {\bibfnamefont {R.}~\bibnamefont {Venugopalan}},\ }\href {\doibase
  10.1007/JHEP08(2023)062} {\bibfield  {journal} {\bibinfo  {journal} {JHEP}\
  }\textbf {\bibinfo {volume} {08}},\ \bibinfo {pages} {062} (\bibinfo {year}
  {2023}{\natexlab{a}})},\ \Eprint {http://arxiv.org/abs/2304.03304}
  {arXiv:2304.03304 [hep-ph]} \BibitemShut {NoStop}%
\bibitem [{\citenamefont {Caucal}\ \emph
  {et~al.}(2023{\natexlab{b}})\citenamefont {Caucal}, \citenamefont {Salazar},
  \citenamefont {Schenke}, \citenamefont {Stebel},\ and\ \citenamefont
  {Venugopalan}}]{Caucal:2023fsf}%
  \BibitemOpen
  \bibfield  {author} {\bibinfo {author} {\bibfnamefont {P.}~\bibnamefont
  {Caucal}}, \bibinfo {author} {\bibfnamefont {F.}~\bibnamefont {Salazar}},
  \bibinfo {author} {\bibfnamefont {B.}~\bibnamefont {Schenke}}, \bibinfo
  {author} {\bibfnamefont {T.}~\bibnamefont {Stebel}}, \ and\ \bibinfo {author}
  {\bibfnamefont {R.}~\bibnamefont {Venugopalan}},\ }\href@noop {} {\
  (\bibinfo {year} {2023}{\natexlab{b}})},\ \Eprint
  {http://arxiv.org/abs/2308.00022} {arXiv:2308.00022 [hep-ph]} \BibitemShut
  {NoStop}%
\bibitem [{\citenamefont {Caucal}\ \emph {et~al.}(2021)\citenamefont {Caucal},
  \citenamefont {Salazar},\ and\ \citenamefont {Venugopalan}}]{Caucal:2021ent}%
  \BibitemOpen
  \bibfield  {author} {\bibinfo {author} {\bibfnamefont {P.}~\bibnamefont
  {Caucal}}, \bibinfo {author} {\bibfnamefont {F.}~\bibnamefont {Salazar}}, \
  and\ \bibinfo {author} {\bibfnamefont {R.}~\bibnamefont {Venugopalan}},\
  }\href {\doibase 10.1007/JHEP11(2021)222} {\bibfield  {journal} {\bibinfo
  {journal} {JHEP}\ }\textbf {\bibinfo {volume} {11}},\ \bibinfo {pages} {222}
  (\bibinfo {year} {2021})},\ \Eprint {http://arxiv.org/abs/2108.06347}
  {arXiv:2108.06347 [hep-ph]} \BibitemShut {NoStop}%
\bibitem [{\citenamefont {Zhang}\ and\ \citenamefont
  {Wang}(2022)}]{Zhang:2021tcc}%
  \BibitemOpen
  \bibfield  {author} {\bibinfo {author} {\bibfnamefont {Y.-Y.}\ \bibnamefont
  {Zhang}}\ and\ \bibinfo {author} {\bibfnamefont {X.-N.}\ \bibnamefont
  {Wang}},\ }\href {\doibase 10.1103/PhysRevD.105.034015} {\bibfield  {journal}
  {\bibinfo  {journal} {Phys. Rev. D}\ }\textbf {\bibinfo {volume} {105}},\
  \bibinfo {pages} {034015} (\bibinfo {year} {2022})},\ \Eprint
  {http://arxiv.org/abs/2104.04520} {arXiv:2104.04520 [hep-ph]} \BibitemShut
  {NoStop}%
\bibitem [{\citenamefont {Metz}\ and\ \citenamefont
  {Zhou}(2011)}]{Metz:2011wb}%
  \BibitemOpen
  \bibfield  {author} {\bibinfo {author} {\bibfnamefont {A.}~\bibnamefont
  {Metz}}\ and\ \bibinfo {author} {\bibfnamefont {J.}~\bibnamefont {Zhou}},\
  }\href {\doibase 10.1103/PhysRevD.84.051503} {\bibfield  {journal} {\bibinfo
  {journal} {Phys. Rev. D}\ }\textbf {\bibinfo {volume} {84}},\ \bibinfo
  {pages} {051503} (\bibinfo {year} {2011})},\ \Eprint
  {http://arxiv.org/abs/1105.1991} {arXiv:1105.1991 [hep-ph]} \BibitemShut
  {NoStop}%
\bibitem [{\citenamefont {Dominguez}\ \emph {et~al.}(2012)\citenamefont
  {Dominguez}, \citenamefont {Qiu}, \citenamefont {Xiao},\ and\ \citenamefont
  {Yuan}}]{Dominguez:2011br}%
  \BibitemOpen
  \bibfield  {author} {\bibinfo {author} {\bibfnamefont {F.}~\bibnamefont
  {Dominguez}}, \bibinfo {author} {\bibfnamefont {J.-W.}\ \bibnamefont {Qiu}},
  \bibinfo {author} {\bibfnamefont {B.-W.}\ \bibnamefont {Xiao}}, \ and\
  \bibinfo {author} {\bibfnamefont {F.}~\bibnamefont {Yuan}},\ }\href {\doibase
  10.1103/PhysRevD.85.045003} {\bibfield  {journal} {\bibinfo  {journal} {Phys.
  Rev. D}\ }\textbf {\bibinfo {volume} {85}},\ \bibinfo {pages} {045003}
  (\bibinfo {year} {2012})},\ \Eprint {http://arxiv.org/abs/1109.6293}
  {arXiv:1109.6293 [hep-ph]} \BibitemShut {NoStop}%
\bibitem [{\citenamefont {Boussarie}\ \emph {et~al.}(2014)\citenamefont
  {Boussarie}, \citenamefont {Grabovsky}, \citenamefont {Szymanowski},\ and\
  \citenamefont {Wallon}}]{Boussarie:2014lxa}%
  \BibitemOpen
  \bibfield  {author} {\bibinfo {author} {\bibfnamefont {R.}~\bibnamefont
  {Boussarie}}, \bibinfo {author} {\bibfnamefont {A.~V.}\ \bibnamefont
  {Grabovsky}}, \bibinfo {author} {\bibfnamefont {L.}~\bibnamefont
  {Szymanowski}}, \ and\ \bibinfo {author} {\bibfnamefont {S.}~\bibnamefont
  {Wallon}},\ }\href {\doibase 10.1007/JHEP09(2014)026} {\bibfield  {journal}
  {\bibinfo  {journal} {JHEP}\ }\textbf {\bibinfo {volume} {09}},\ \bibinfo
  {pages} {026} (\bibinfo {year} {2014})},\ \Eprint
  {http://arxiv.org/abs/1405.7676} {arXiv:1405.7676 [hep-ph]} \BibitemShut
  {NoStop}%
\bibitem [{\citenamefont {Boussarie}\ \emph {et~al.}(2016)\citenamefont
  {Boussarie}, \citenamefont {Grabovsky}, \citenamefont {Szymanowski},\ and\
  \citenamefont {Wallon}}]{Boussarie:2016ogo}%
  \BibitemOpen
  \bibfield  {author} {\bibinfo {author} {\bibfnamefont {R.}~\bibnamefont
  {Boussarie}}, \bibinfo {author} {\bibfnamefont {A.~V.}\ \bibnamefont
  {Grabovsky}}, \bibinfo {author} {\bibfnamefont {L.}~\bibnamefont
  {Szymanowski}}, \ and\ \bibinfo {author} {\bibfnamefont {S.}~\bibnamefont
  {Wallon}},\ }\href {\doibase 10.1007/JHEP11(2016)149} {\bibfield  {journal}
  {\bibinfo  {journal} {JHEP}\ }\textbf {\bibinfo {volume} {11}},\ \bibinfo
  {pages} {149} (\bibinfo {year} {2016})},\ \Eprint
  {http://arxiv.org/abs/1606.00419} {arXiv:1606.00419 [hep-ph]} \BibitemShut
  {NoStop}%
\bibitem [{\citenamefont {Salazar}\ and\ \citenamefont
  {Schenke}(2019)}]{Salazar:2019ncp}%
  \BibitemOpen
  \bibfield  {author} {\bibinfo {author} {\bibfnamefont {F.}~\bibnamefont
  {Salazar}}\ and\ \bibinfo {author} {\bibfnamefont {B.}~\bibnamefont
  {Schenke}},\ }\href {\doibase 10.1103/PhysRevD.100.034007} {\bibfield
  {journal} {\bibinfo  {journal} {Phys. Rev. D}\ }\textbf {\bibinfo {volume}
  {100}},\ \bibinfo {pages} {034007} (\bibinfo {year} {2019})},\ \Eprint
  {http://arxiv.org/abs/1905.03763} {arXiv:1905.03763 [hep-ph]} \BibitemShut
  {NoStop}%
\bibitem [{\citenamefont {Boussarie}\ \emph {et~al.}(2019)\citenamefont
  {Boussarie}, \citenamefont {Grabovsky}, \citenamefont {Szymanowski},\ and\
  \citenamefont {Wallon}}]{Boussarie:2019ero}%
  \BibitemOpen
  \bibfield  {author} {\bibinfo {author} {\bibfnamefont {R.}~\bibnamefont
  {Boussarie}}, \bibinfo {author} {\bibfnamefont {A.~V.}\ \bibnamefont
  {Grabovsky}}, \bibinfo {author} {\bibfnamefont {L.}~\bibnamefont
  {Szymanowski}}, \ and\ \bibinfo {author} {\bibfnamefont {S.}~\bibnamefont
  {Wallon}},\ }\href {\doibase 10.1103/PhysRevD.100.074020} {\bibfield
  {journal} {\bibinfo  {journal} {Phys. Rev. D}\ }\textbf {\bibinfo {volume}
  {100}},\ \bibinfo {pages} {074020} (\bibinfo {year} {2019})},\ \Eprint
  {http://arxiv.org/abs/1905.07371} {arXiv:1905.07371 [hep-ph]} \BibitemShut
  {NoStop}%
\bibitem [{\citenamefont {M\"antysaari}\ \emph {et~al.}(2020)\citenamefont
  {M\"antysaari}, \citenamefont {Mueller}, \citenamefont {Salazar},\ and\
  \citenamefont {Schenke}}]{Mantysaari:2019hkq}%
  \BibitemOpen
  \bibfield  {author} {\bibinfo {author} {\bibfnamefont {H.}~\bibnamefont
  {M\"antysaari}}, \bibinfo {author} {\bibfnamefont {N.}~\bibnamefont
  {Mueller}}, \bibinfo {author} {\bibfnamefont {F.}~\bibnamefont {Salazar}}, \
  and\ \bibinfo {author} {\bibfnamefont {B.}~\bibnamefont {Schenke}},\ }\href
  {\doibase 10.1103/PhysRevLett.124.112301} {\bibfield  {journal} {\bibinfo
  {journal} {Phys. Rev. Lett.}\ }\textbf {\bibinfo {volume} {124}},\ \bibinfo
  {pages} {112301} (\bibinfo {year} {2020})},\ \Eprint
  {http://arxiv.org/abs/1912.05586} {arXiv:1912.05586 [nucl-th]} \BibitemShut
  {NoStop}%
\bibitem [{\citenamefont {Boer}\ and\ \citenamefont
  {Setyadi}(2021)}]{Boer:2021upt}%
  \BibitemOpen
  \bibfield  {author} {\bibinfo {author} {\bibfnamefont {D.}~\bibnamefont
  {Boer}}\ and\ \bibinfo {author} {\bibfnamefont {C.}~\bibnamefont {Setyadi}},\
  }\href {\doibase 10.1103/PhysRevD.104.074006} {\bibfield  {journal} {\bibinfo
   {journal} {Phys. Rev. D}\ }\textbf {\bibinfo {volume} {104}},\ \bibinfo
  {pages} {074006} (\bibinfo {year} {2021})},\ \Eprint
  {http://arxiv.org/abs/2106.15148} {arXiv:2106.15148 [hep-ph]} \BibitemShut
  {NoStop}%
\bibitem [{\citenamefont {Iancu}\ \emph
  {et~al.}(2022{\natexlab{a}})\citenamefont {Iancu}, \citenamefont {Mueller},\
  and\ \citenamefont {Triantafyllopoulos}}]{Iancu:2021rup}%
  \BibitemOpen
  \bibfield  {author} {\bibinfo {author} {\bibfnamefont {E.}~\bibnamefont
  {Iancu}}, \bibinfo {author} {\bibfnamefont {A.~H.}\ \bibnamefont {Mueller}},
  \ and\ \bibinfo {author} {\bibfnamefont {D.~N.}\ \bibnamefont
  {Triantafyllopoulos}},\ }\href {\doibase 10.1103/PhysRevLett.128.202001}
  {\bibfield  {journal} {\bibinfo  {journal} {Phys. Rev. Lett.}\ }\textbf
  {\bibinfo {volume} {128}},\ \bibinfo {pages} {202001} (\bibinfo {year}
  {2022}{\natexlab{a}})},\ \Eprint {http://arxiv.org/abs/2112.06353}
  {arXiv:2112.06353 [hep-ph]} \BibitemShut {NoStop}%
\bibitem [{\citenamefont {Iancu}\ \emph
  {et~al.}(2022{\natexlab{b}})\citenamefont {Iancu}, \citenamefont {Mueller},
  \citenamefont {Triantafyllopoulos},\ and\ \citenamefont
  {Wei}}]{Iancu:2022lcw}%
  \BibitemOpen
  \bibfield  {author} {\bibinfo {author} {\bibfnamefont {E.}~\bibnamefont
  {Iancu}}, \bibinfo {author} {\bibfnamefont {A.~H.}\ \bibnamefont {Mueller}},
  \bibinfo {author} {\bibfnamefont {D.~N.}\ \bibnamefont {Triantafyllopoulos}},
  \ and\ \bibinfo {author} {\bibfnamefont {S.~Y.}\ \bibnamefont {Wei}},\ }\href
  {\doibase 10.1007/JHEP10(2022)103} {\bibfield  {journal} {\bibinfo  {journal}
  {JHEP}\ }\textbf {\bibinfo {volume} {10}},\ \bibinfo {pages} {103} (\bibinfo
  {year} {2022}{\natexlab{b}})},\ \Eprint {http://arxiv.org/abs/2207.06268}
  {arXiv:2207.06268 [hep-ph]} \BibitemShut {NoStop}%
\bibitem [{\citenamefont {Hatta}\ \emph {et~al.}(2016)\citenamefont {Hatta},
  \citenamefont {Xiao},\ and\ \citenamefont {Yuan}}]{Hatta:2016dxp}%
  \BibitemOpen
  \bibfield  {author} {\bibinfo {author} {\bibfnamefont {Y.}~\bibnamefont
  {Hatta}}, \bibinfo {author} {\bibfnamefont {B.-W.}\ \bibnamefont {Xiao}}, \
  and\ \bibinfo {author} {\bibfnamefont {F.}~\bibnamefont {Yuan}},\ }\href
  {\doibase 10.1103/PhysRevLett.116.202301} {\bibfield  {journal} {\bibinfo
  {journal} {Phys. Rev. Lett.}\ }\textbf {\bibinfo {volume} {116}},\ \bibinfo
  {pages} {202301} (\bibinfo {year} {2016})},\ \Eprint
  {http://arxiv.org/abs/1601.01585} {arXiv:1601.01585 [hep-ph]} \BibitemShut
  {NoStop}%
\bibitem [{\citenamefont {Altinoluk}\ \emph {et~al.}(2016)\citenamefont
  {Altinoluk}, \citenamefont {Armesto}, \citenamefont {Beuf},\ and\
  \citenamefont {Rezaeian}}]{Altinoluk:2015dpi}%
  \BibitemOpen
  \bibfield  {author} {\bibinfo {author} {\bibfnamefont {T.}~\bibnamefont
  {Altinoluk}}, \bibinfo {author} {\bibfnamefont {N.}~\bibnamefont {Armesto}},
  \bibinfo {author} {\bibfnamefont {G.}~\bibnamefont {Beuf}}, \ and\ \bibinfo
  {author} {\bibfnamefont {A.~H.}\ \bibnamefont {Rezaeian}},\ }\href {\doibase
  10.1016/j.physletb.2016.05.032} {\bibfield  {journal} {\bibinfo  {journal}
  {Phys. Lett. B}\ }\textbf {\bibinfo {volume} {758}},\ \bibinfo {pages} {373}
  (\bibinfo {year} {2016})},\ \Eprint {http://arxiv.org/abs/1511.07452}
  {arXiv:1511.07452 [hep-ph]} \BibitemShut {NoStop}%
\bibitem [{\citenamefont {M\"antysaari}\ \emph {et~al.}(2019)\citenamefont
  {M\"antysaari}, \citenamefont {Mueller},\ and\ \citenamefont
  {Schenke}}]{Mantysaari:2019csc}%
  \BibitemOpen
  \bibfield  {author} {\bibinfo {author} {\bibfnamefont {H.}~\bibnamefont
  {M\"antysaari}}, \bibinfo {author} {\bibfnamefont {N.}~\bibnamefont
  {Mueller}}, \ and\ \bibinfo {author} {\bibfnamefont {B.}~\bibnamefont
  {Schenke}},\ }\href {\doibase 10.1103/PhysRevD.99.074004} {\bibfield
  {journal} {\bibinfo  {journal} {Phys. Rev. D}\ }\textbf {\bibinfo {volume}
  {99}},\ \bibinfo {pages} {074004} (\bibinfo {year} {2019})},\ \Eprint
  {http://arxiv.org/abs/1902.05087} {arXiv:1902.05087 [hep-ph]} \BibitemShut
  {NoStop}%
\bibitem [{\citenamefont {Fucilla}\ \emph {et~al.}(2023)\citenamefont
  {Fucilla}, \citenamefont {Grabovsky}, \citenamefont {Li}, \citenamefont
  {Szymanowski},\ and\ \citenamefont {Wallon}}]{Fucilla:2022wcg}%
  \BibitemOpen
  \bibfield  {author} {\bibinfo {author} {\bibfnamefont {M.}~\bibnamefont
  {Fucilla}}, \bibinfo {author} {\bibfnamefont {A.~V.}\ \bibnamefont
  {Grabovsky}}, \bibinfo {author} {\bibfnamefont {E.}~\bibnamefont {Li}},
  \bibinfo {author} {\bibfnamefont {L.}~\bibnamefont {Szymanowski}}, \ and\
  \bibinfo {author} {\bibfnamefont {S.}~\bibnamefont {Wallon}},\ }\href
  {\doibase 10.1007/JHEP03(2023)159} {\bibfield  {journal} {\bibinfo  {journal}
  {JHEP}\ }\textbf {\bibinfo {volume} {03}},\ \bibinfo {pages} {159} (\bibinfo
  {year} {2023})},\ \Eprint {http://arxiv.org/abs/2211.05774} {arXiv:2211.05774
  [hep-ph]} \BibitemShut {NoStop}%
\bibitem [{\citenamefont {Rodriguez-Aguilar}\ \emph {et~al.}(2023)\citenamefont
  {Rodriguez-Aguilar}, \citenamefont {Triantafyllopoulos},\ and\ \citenamefont
  {Wei}}]{Rodriguez-Aguilar:2023ihz}%
  \BibitemOpen
  \bibfield  {author} {\bibinfo {author} {\bibfnamefont {B.}~\bibnamefont
  {Rodriguez-Aguilar}}, \bibinfo {author} {\bibfnamefont {D.~N.}\ \bibnamefont
  {Triantafyllopoulos}}, \ and\ \bibinfo {author} {\bibfnamefont {S.~Y.}\
  \bibnamefont {Wei}},\ }\href {\doibase 10.1103/PhysRevD.107.114007}
  {\bibfield  {journal} {\bibinfo  {journal} {Phys. Rev. D}\ }\textbf {\bibinfo
  {volume} {107}},\ \bibinfo {pages} {114007} (\bibinfo {year} {2023})},\
  \Eprint {http://arxiv.org/abs/2302.01106} {arXiv:2302.01106 [hep-ph]}
  \BibitemShut {NoStop}%
\bibitem [{\citenamefont {Altinoluk}\ \emph
  {et~al.}(2023{\natexlab{a}})\citenamefont {Altinoluk}, \citenamefont {Beuf},
  \citenamefont {Czajka},\ and\ \citenamefont {Tymowska}}]{Altinoluk:2022jkk}%
  \BibitemOpen
  \bibfield  {author} {\bibinfo {author} {\bibfnamefont {T.}~\bibnamefont
  {Altinoluk}}, \bibinfo {author} {\bibfnamefont {G.}~\bibnamefont {Beuf}},
  \bibinfo {author} {\bibfnamefont {A.}~\bibnamefont {Czajka}}, \ and\ \bibinfo
  {author} {\bibfnamefont {A.}~\bibnamefont {Tymowska}},\ }\href {\doibase
  10.1103/PhysRevD.107.074016} {\bibfield  {journal} {\bibinfo  {journal}
  {Phys. Rev. D}\ }\textbf {\bibinfo {volume} {107}},\ \bibinfo {pages}
  {074016} (\bibinfo {year} {2023}{\natexlab{a}})},\ \Eprint
  {http://arxiv.org/abs/2212.10484} {arXiv:2212.10484 [hep-ph]} \BibitemShut
  {NoStop}%
\bibitem [{\citenamefont {Altinoluk}\ \emph
  {et~al.}(2023{\natexlab{b}})\citenamefont {Altinoluk}, \citenamefont
  {Armesto},\ and\ \citenamefont {Beuf}}]{Altinoluk:2023qfr}%
  \BibitemOpen
  \bibfield  {author} {\bibinfo {author} {\bibfnamefont {T.}~\bibnamefont
  {Altinoluk}}, \bibinfo {author} {\bibfnamefont {N.}~\bibnamefont {Armesto}},
  \ and\ \bibinfo {author} {\bibfnamefont {G.}~\bibnamefont {Beuf}},\
  }\href@noop {} {\  (\bibinfo {year} {2023}{\natexlab{b}})},\ \Eprint
  {http://arxiv.org/abs/2303.12691} {arXiv:2303.12691 [hep-ph]} \BibitemShut
  {NoStop}%
\bibitem [{\citenamefont {Hagiwara}\ \emph {et~al.}(2021)\citenamefont
  {Hagiwara}, \citenamefont {Zhang}, \citenamefont {Zhou},\ and\ \citenamefont
  {Zhou}}]{Hagiwara:2021xkf}%
  \BibitemOpen
  \bibfield  {author} {\bibinfo {author} {\bibfnamefont {Y.}~\bibnamefont
  {Hagiwara}}, \bibinfo {author} {\bibfnamefont {C.}~\bibnamefont {Zhang}},
  \bibinfo {author} {\bibfnamefont {J.}~\bibnamefont {Zhou}}, \ and\ \bibinfo
  {author} {\bibfnamefont {Y.-j.}\ \bibnamefont {Zhou}},\ }\href {\doibase
  10.1103/PhysRevD.104.094021} {\bibfield  {journal} {\bibinfo  {journal}
  {Phys. Rev. D}\ }\textbf {\bibinfo {volume} {104}},\ \bibinfo {pages}
  {094021} (\bibinfo {year} {2021})},\ \Eprint
  {http://arxiv.org/abs/2106.13466} {arXiv:2106.13466 [hep-ph]} \BibitemShut
  {NoStop}%
\bibitem [{\citenamefont {Zheng}\ \emph {et~al.}(2014)\citenamefont {Zheng},
  \citenamefont {Aschenauer}, \citenamefont {Lee},\ and\ \citenamefont
  {Xiao}}]{Zheng:2014vka}%
  \BibitemOpen
  \bibfield  {author} {\bibinfo {author} {\bibfnamefont {L.}~\bibnamefont
  {Zheng}}, \bibinfo {author} {\bibfnamefont {E.~C.}\ \bibnamefont
  {Aschenauer}}, \bibinfo {author} {\bibfnamefont {J.~H.}\ \bibnamefont {Lee}},
  \ and\ \bibinfo {author} {\bibfnamefont {B.-W.}\ \bibnamefont {Xiao}},\
  }\href {\doibase 10.1103/PhysRevD.89.074037} {\bibfield  {journal} {\bibinfo
  {journal} {Phys. Rev. D}\ }\textbf {\bibinfo {volume} {89}},\ \bibinfo
  {pages} {074037} (\bibinfo {year} {2014})},\ \Eprint
  {http://arxiv.org/abs/1403.2413} {arXiv:1403.2413 [hep-ph]} \BibitemShut
  {NoStop}%
\bibitem [{\citenamefont {Bergabo}\ and\ \citenamefont
  {Jalilian-Marian}(2022{\natexlab{a}})}]{Bergabo:2021woe}%
  \BibitemOpen
  \bibfield  {author} {\bibinfo {author} {\bibfnamefont {F.}~\bibnamefont
  {Bergabo}}\ and\ \bibinfo {author} {\bibfnamefont {J.}~\bibnamefont
  {Jalilian-Marian}},\ }\href {\doibase 10.1016/j.nuclphysa.2021.122358}
  {\bibfield  {journal} {\bibinfo  {journal} {Nucl. Phys. A}\ }\textbf
  {\bibinfo {volume} {1018}},\ \bibinfo {pages} {122358} (\bibinfo {year}
  {2022}{\natexlab{a}})},\ \Eprint {http://arxiv.org/abs/2108.10428}
  {arXiv:2108.10428 [hep-ph]} \BibitemShut {NoStop}%
\bibitem [{\citenamefont {Bergabo}\ and\ \citenamefont
  {Jalilian-Marian}(2022{\natexlab{b}})}]{Bergabo:2022tcu}%
  \BibitemOpen
  \bibfield  {author} {\bibinfo {author} {\bibfnamefont {F.}~\bibnamefont
  {Bergabo}}\ and\ \bibinfo {author} {\bibfnamefont {J.}~\bibnamefont
  {Jalilian-Marian}},\ }\href {\doibase 10.1103/PhysRevD.106.054035} {\bibfield
   {journal} {\bibinfo  {journal} {Phys. Rev. D}\ }\textbf {\bibinfo {volume}
  {106}},\ \bibinfo {pages} {054035} (\bibinfo {year} {2022}{\natexlab{b}})},\
  \Eprint {http://arxiv.org/abs/2207.03606} {arXiv:2207.03606 [hep-ph]}
  \BibitemShut {NoStop}%
\bibitem [{\citenamefont {Iancu}\ and\ \citenamefont
  {Mulian}(2023)}]{Iancu:2022gpw}%
  \BibitemOpen
  \bibfield  {author} {\bibinfo {author} {\bibfnamefont {E.}~\bibnamefont
  {Iancu}}\ and\ \bibinfo {author} {\bibfnamefont {Y.}~\bibnamefont {Mulian}},\
  }\href {\doibase 10.1007/JHEP07(2023)121} {\bibfield  {journal} {\bibinfo
  {journal} {JHEP}\ }\textbf {\bibinfo {volume} {07}},\ \bibinfo {pages} {121}
  (\bibinfo {year} {2023})},\ \Eprint {http://arxiv.org/abs/2211.04837}
  {arXiv:2211.04837 [hep-ph]} \BibitemShut {NoStop}%
\bibitem [{\citenamefont {Bergabo}\ and\ \citenamefont
  {Jalilian-Marian}(2023)}]{Bergabo:2023wed}%
  \BibitemOpen
  \bibfield  {author} {\bibinfo {author} {\bibfnamefont {F.}~\bibnamefont
  {Bergabo}}\ and\ \bibinfo {author} {\bibfnamefont {J.}~\bibnamefont
  {Jalilian-Marian}},\ }\href {\doibase 10.1103/PhysRevD.107.054036} {\bibfield
   {journal} {\bibinfo  {journal} {Phys. Rev. D}\ }\textbf {\bibinfo {volume}
  {107}},\ \bibinfo {pages} {054036} (\bibinfo {year} {2023})},\ \Eprint
  {http://arxiv.org/abs/2301.03117} {arXiv:2301.03117 [hep-ph]} \BibitemShut
  {NoStop}%
\bibitem [{\citenamefont {Kolb\'e}\ \emph {et~al.}(2021)\citenamefont
  {Kolb\'e}, \citenamefont {Roy}, \citenamefont {Salazar}, \citenamefont
  {Schenke},\ and\ \citenamefont {Venugopalan}}]{Kolbe:2020tlq}%
  \BibitemOpen
  \bibfield  {author} {\bibinfo {author} {\bibfnamefont {I.}~\bibnamefont
  {Kolb\'e}}, \bibinfo {author} {\bibfnamefont {K.}~\bibnamefont {Roy}},
  \bibinfo {author} {\bibfnamefont {F.}~\bibnamefont {Salazar}}, \bibinfo
  {author} {\bibfnamefont {B.}~\bibnamefont {Schenke}}, \ and\ \bibinfo
  {author} {\bibfnamefont {R.}~\bibnamefont {Venugopalan}},\ }\href {\doibase
  10.1007/JHEP01(2021)052} {\bibfield  {journal} {\bibinfo  {journal} {JHEP}\
  }\textbf {\bibinfo {volume} {01}},\ \bibinfo {pages} {052} (\bibinfo {year}
  {2021})},\ \Eprint {http://arxiv.org/abs/2008.04372} {arXiv:2008.04372
  [hep-ph]} \BibitemShut {NoStop}%
\bibitem [{\citenamefont {Boer}\ \emph {et~al.}(2011)\citenamefont {Boer} \emph
  {et~al.}}]{Boer:2011fh}%
  \BibitemOpen
  \bibfield  {author} {\bibinfo {author} {\bibfnamefont {D.}~\bibnamefont
  {Boer}} \emph {et~al.},\ }\href@noop {} {\  (\bibinfo {year} {2011})},\
  \Eprint {http://arxiv.org/abs/1108.1713} {arXiv:1108.1713 [nucl-th]}
  \BibitemShut {NoStop}%
\bibitem [{\citenamefont {Accardi}\ \emph {et~al.}(2016)\citenamefont {Accardi}
  \emph {et~al.}}]{Accardi:2012qut}%
  \BibitemOpen
  \bibfield  {author} {\bibinfo {author} {\bibfnamefont {A.}~\bibnamefont
  {Accardi}} \emph {et~al.},\ }\href {\doibase 10.1140/epja/i2016-16268-9}
  {\bibfield  {journal} {\bibinfo  {journal} {Eur. Phys. J. A}\ }\textbf
  {\bibinfo {volume} {52}},\ \bibinfo {pages} {268} (\bibinfo {year} {2016})},\
  \Eprint {http://arxiv.org/abs/1212.1701} {arXiv:1212.1701 [nucl-ex]}
  \BibitemShut {NoStop}%
\bibitem [{\citenamefont {Abdul~Khalek}\ \emph
  {et~al.}(2022{\natexlab{a}})\citenamefont {Abdul~Khalek} \emph
  {et~al.}}]{AbdulKhalek:2021gbh}%
  \BibitemOpen
  \bibfield  {author} {\bibinfo {author} {\bibfnamefont {R.}~\bibnamefont
  {Abdul~Khalek}} \emph {et~al.},\ }\href {\doibase
  10.1016/j.nuclphysa.2022.122447} {\bibfield  {journal} {\bibinfo  {journal}
  {Nucl. Phys. A}\ }\textbf {\bibinfo {volume} {1026}},\ \bibinfo {pages}
  {122447} (\bibinfo {year} {2022}{\natexlab{a}})},\ \Eprint
  {http://arxiv.org/abs/2103.05419} {arXiv:2103.05419 [physics.ins-det]}
  \BibitemShut {NoStop}%
\bibitem [{\citenamefont {Abdul~Khalek}\ \emph
  {et~al.}(2022{\natexlab{b}})\citenamefont {Abdul~Khalek} \emph
  {et~al.}}]{AbdulKhalek:2022hcn}%
  \BibitemOpen
  \bibfield  {author} {\bibinfo {author} {\bibfnamefont {R.}~\bibnamefont
  {Abdul~Khalek}} \emph {et~al.},\ }\href@noop {} {\  (\bibinfo {year}
  {2022}{\natexlab{b}})},\ \Eprint {http://arxiv.org/abs/2203.13199}
  {arXiv:2203.13199 [hep-ph]} \BibitemShut {NoStop}%
\bibitem [{\citenamefont {Abir}\ \emph {et~al.}(2023)\citenamefont {Abir} \emph
  {et~al.}}]{Abir:2023fpo}%
  \BibitemOpen
  \bibfield  {author} {\bibinfo {author} {\bibfnamefont {R.}~\bibnamefont
  {Abir}} \emph {et~al.},\ }\href@noop {} {\  (\bibinfo {year} {2023})},\
  \Eprint {http://arxiv.org/abs/2305.14572} {arXiv:2305.14572 [hep-ph]}
  \BibitemShut {NoStop}%
\bibitem [{\citenamefont {Mueller}\ \emph
  {et~al.}(2013{\natexlab{b}})\citenamefont {Mueller}, \citenamefont {Xiao},\
  and\ \citenamefont {Yuan}}]{Mueller:2012uf}%
  \BibitemOpen
  \bibfield  {author} {\bibinfo {author} {\bibfnamefont {A.~H.}\ \bibnamefont
  {Mueller}}, \bibinfo {author} {\bibfnamefont {B.-W.}\ \bibnamefont {Xiao}}, \
  and\ \bibinfo {author} {\bibfnamefont {F.}~\bibnamefont {Yuan}},\ }\href
  {\doibase 10.1103/PhysRevLett.110.082301} {\bibfield  {journal} {\bibinfo
  {journal} {Phys. Rev. Lett.}\ }\textbf {\bibinfo {volume} {110}},\ \bibinfo
  {pages} {082301} (\bibinfo {year} {2013}{\natexlab{b}})},\ \Eprint
  {http://arxiv.org/abs/1210.5792} {arXiv:1210.5792 [hep-ph]} \BibitemShut
  {NoStop}%
\bibitem [{\citenamefont {Jalilian-Marian}\ and\ \citenamefont
  {Kovchegov}(2004)}]{Jalilian-Marian:2004vhw}%
  \BibitemOpen
  \bibfield  {author} {\bibinfo {author} {\bibfnamefont {J.}~\bibnamefont
  {Jalilian-Marian}}\ and\ \bibinfo {author} {\bibfnamefont {Y.~V.}\
  \bibnamefont {Kovchegov}},\ }\href {\doibase 10.1103/PhysRevD.71.079901}
  {\bibfield  {journal} {\bibinfo  {journal} {Phys. Rev. D}\ }\textbf {\bibinfo
  {volume} {70}},\ \bibinfo {pages} {114017} (\bibinfo {year} {2004})},\
  \bibinfo {note} {[Erratum: Phys.Rev.D 71, 079901 (2005)]},\ \Eprint
  {http://arxiv.org/abs/hep-ph/0405266} {arXiv:hep-ph/0405266} \BibitemShut
  {NoStop}%
\bibitem [{\citenamefont {Kharzeev}\ \emph {et~al.}(2005)\citenamefont
  {Kharzeev}, \citenamefont {Levin},\ and\ \citenamefont
  {McLerran}}]{Kharzeev:2004bw}%
  \BibitemOpen
  \bibfield  {author} {\bibinfo {author} {\bibfnamefont {D.}~\bibnamefont
  {Kharzeev}}, \bibinfo {author} {\bibfnamefont {E.}~\bibnamefont {Levin}}, \
  and\ \bibinfo {author} {\bibfnamefont {L.}~\bibnamefont {McLerran}},\ }\href
  {\doibase 10.1016/j.nuclphysa.2004.10.031} {\bibfield  {journal} {\bibinfo
  {journal} {Nucl. Phys. A}\ }\textbf {\bibinfo {volume} {748}},\ \bibinfo
  {pages} {627} (\bibinfo {year} {2005})},\ \Eprint
  {http://arxiv.org/abs/hep-ph/0403271} {arXiv:hep-ph/0403271} \BibitemShut
  {NoStop}%
\bibitem [{\citenamefont {Marquet}(2007{\natexlab{a}})}]{Marquet:2007vb}%
  \BibitemOpen
  \bibfield  {author} {\bibinfo {author} {\bibfnamefont {C.}~\bibnamefont
  {Marquet}},\ }\href {\doibase 10.1016/j.nuclphysa.2007.09.001} {\bibfield
  {journal} {\bibinfo  {journal} {Nucl. Phys. A}\ }\textbf {\bibinfo {volume}
  {796}},\ \bibinfo {pages} {41} (\bibinfo {year} {2007}{\natexlab{a}})},\
  \Eprint {http://arxiv.org/abs/0708.0231} {arXiv:0708.0231 [hep-ph]}
  \BibitemShut {NoStop}%
\bibitem [{\citenamefont {Tuchin}(2010)}]{Tuchin:2009nf}%
  \BibitemOpen
  \bibfield  {author} {\bibinfo {author} {\bibfnamefont {K.}~\bibnamefont
  {Tuchin}},\ }\href {\doibase 10.1016/j.nuclphysa.2010.06.001} {\bibfield
  {journal} {\bibinfo  {journal} {Nucl. Phys. A}\ }\textbf {\bibinfo {volume}
  {846}},\ \bibinfo {pages} {83} (\bibinfo {year} {2010})},\ \Eprint
  {http://arxiv.org/abs/0912.5479} {arXiv:0912.5479 [hep-ph]} \BibitemShut
  {NoStop}%
\bibitem [{\citenamefont {Dumitru}\ and\ \citenamefont
  {Jalilian-Marian}(2010)}]{Dumitru:2010ak}%
  \BibitemOpen
  \bibfield  {author} {\bibinfo {author} {\bibfnamefont {A.}~\bibnamefont
  {Dumitru}}\ and\ \bibinfo {author} {\bibfnamefont {J.}~\bibnamefont
  {Jalilian-Marian}},\ }\href {\doibase 10.1103/PhysRevD.82.074023} {\bibfield
  {journal} {\bibinfo  {journal} {Phys. Rev. D}\ }\textbf {\bibinfo {volume}
  {82}},\ \bibinfo {pages} {074023} (\bibinfo {year} {2010})},\ \Eprint
  {http://arxiv.org/abs/1008.0480} {arXiv:1008.0480 [hep-ph]} \BibitemShut
  {NoStop}%
\bibitem [{\citenamefont {Kutak}\ and\ \citenamefont
  {Sapeta}(2012)}]{Kutak:2012rf}%
  \BibitemOpen
  \bibfield  {author} {\bibinfo {author} {\bibfnamefont {K.}~\bibnamefont
  {Kutak}}\ and\ \bibinfo {author} {\bibfnamefont {S.}~\bibnamefont {Sapeta}},\
  }\href {\doibase 10.1103/PhysRevD.86.094043} {\bibfield  {journal} {\bibinfo
  {journal} {Phys. Rev. D}\ }\textbf {\bibinfo {volume} {86}},\ \bibinfo
  {pages} {094043} (\bibinfo {year} {2012})},\ \Eprint
  {http://arxiv.org/abs/1205.5035} {arXiv:1205.5035 [hep-ph]} \BibitemShut
  {NoStop}%
\bibitem [{\citenamefont {van Hameren}\ \emph {et~al.}(2014)\citenamefont {van
  Hameren}, \citenamefont {Kotko}, \citenamefont {Kutak},\ and\ \citenamefont
  {Sapeta}}]{vanHameren:2014ala}%
  \BibitemOpen
  \bibfield  {author} {\bibinfo {author} {\bibfnamefont {A.}~\bibnamefont {van
  Hameren}}, \bibinfo {author} {\bibfnamefont {P.}~\bibnamefont {Kotko}},
  \bibinfo {author} {\bibfnamefont {K.}~\bibnamefont {Kutak}}, \ and\ \bibinfo
  {author} {\bibfnamefont {S.}~\bibnamefont {Sapeta}},\ }\href {\doibase
  10.1016/j.physletb.2014.09.005} {\bibfield  {journal} {\bibinfo  {journal}
  {Phys. Lett. B}\ }\textbf {\bibinfo {volume} {737}},\ \bibinfo {pages} {335}
  (\bibinfo {year} {2014})},\ \Eprint {http://arxiv.org/abs/1404.6204}
  {arXiv:1404.6204 [hep-ph]} \BibitemShut {NoStop}%
\bibitem [{\citenamefont {Kotko}\ \emph {et~al.}(2015)\citenamefont {Kotko},
  \citenamefont {Kutak}, \citenamefont {Marquet}, \citenamefont {Petreska},
  \citenamefont {Sapeta},\ and\ \citenamefont {van Hameren}}]{Kotko:2015ura}%
  \BibitemOpen
  \bibfield  {author} {\bibinfo {author} {\bibfnamefont {P.}~\bibnamefont
  {Kotko}}, \bibinfo {author} {\bibfnamefont {K.}~\bibnamefont {Kutak}},
  \bibinfo {author} {\bibfnamefont {C.}~\bibnamefont {Marquet}}, \bibinfo
  {author} {\bibfnamefont {E.}~\bibnamefont {Petreska}}, \bibinfo {author}
  {\bibfnamefont {S.}~\bibnamefont {Sapeta}}, \ and\ \bibinfo {author}
  {\bibfnamefont {A.}~\bibnamefont {van Hameren}},\ }\href {\doibase
  10.1007/JHEP09(2015)106} {\bibfield  {journal} {\bibinfo  {journal} {JHEP}\
  }\textbf {\bibinfo {volume} {09}},\ \bibinfo {pages} {106} (\bibinfo {year}
  {2015})},\ \Eprint {http://arxiv.org/abs/1503.03421} {arXiv:1503.03421
  [hep-ph]} \BibitemShut {NoStop}%
\bibitem [{\citenamefont {van Hameren}\ \emph {et~al.}(2016)\citenamefont {van
  Hameren}, \citenamefont {Kotko}, \citenamefont {Kutak}, \citenamefont
  {Marquet}, \citenamefont {Petreska},\ and\ \citenamefont
  {Sapeta}}]{vanHameren:2016ftb}%
  \BibitemOpen
  \bibfield  {author} {\bibinfo {author} {\bibfnamefont {A.}~\bibnamefont {van
  Hameren}}, \bibinfo {author} {\bibfnamefont {P.}~\bibnamefont {Kotko}},
  \bibinfo {author} {\bibfnamefont {K.}~\bibnamefont {Kutak}}, \bibinfo
  {author} {\bibfnamefont {C.}~\bibnamefont {Marquet}}, \bibinfo {author}
  {\bibfnamefont {E.}~\bibnamefont {Petreska}}, \ and\ \bibinfo {author}
  {\bibfnamefont {S.}~\bibnamefont {Sapeta}},\ }\href {\doibase
  10.1007/JHEP12(2016)034} {\bibfield  {journal} {\bibinfo  {journal} {JHEP}\
  }\textbf {\bibinfo {volume} {12}},\ \bibinfo {pages} {034} (\bibinfo {year}
  {2016})},\ \bibinfo {note} {[Erratum: JHEP 02, 158 (2019)]},\ \Eprint
  {http://arxiv.org/abs/1607.03121} {arXiv:1607.03121 [hep-ph]} \BibitemShut
  {NoStop}%
\bibitem [{\citenamefont {van Hameren}\ \emph {et~al.}(2019)\citenamefont {van
  Hameren}, \citenamefont {Kotko}, \citenamefont {Kutak},\ and\ \citenamefont
  {Sapeta}}]{vanHameren:2019ysa}%
  \BibitemOpen
  \bibfield  {author} {\bibinfo {author} {\bibfnamefont {A.}~\bibnamefont {van
  Hameren}}, \bibinfo {author} {\bibfnamefont {P.}~\bibnamefont {Kotko}},
  \bibinfo {author} {\bibfnamefont {K.}~\bibnamefont {Kutak}}, \ and\ \bibinfo
  {author} {\bibfnamefont {S.}~\bibnamefont {Sapeta}},\ }\href {\doibase
  10.1016/j.physletb.2019.06.055} {\bibfield  {journal} {\bibinfo  {journal}
  {Phys. Lett. B}\ }\textbf {\bibinfo {volume} {795}},\ \bibinfo {pages} {511}
  (\bibinfo {year} {2019})},\ \Eprint {http://arxiv.org/abs/1903.01361}
  {arXiv:1903.01361 [hep-ph]} \BibitemShut {NoStop}%
\bibitem [{\citenamefont {van Hameren}\ \emph {et~al.}(2021)\citenamefont {van
  Hameren}, \citenamefont {Kotko}, \citenamefont {Kutak},\ and\ \citenamefont
  {Sapeta}}]{vanHameren:2020rqt}%
  \BibitemOpen
  \bibfield  {author} {\bibinfo {author} {\bibfnamefont {A.}~\bibnamefont {van
  Hameren}}, \bibinfo {author} {\bibfnamefont {P.}~\bibnamefont {Kotko}},
  \bibinfo {author} {\bibfnamefont {K.}~\bibnamefont {Kutak}}, \ and\ \bibinfo
  {author} {\bibfnamefont {S.}~\bibnamefont {Sapeta}},\ }\href {\doibase
  10.1016/j.physletb.2021.136078} {\bibfield  {journal} {\bibinfo  {journal}
  {Phys. Lett. B}\ }\textbf {\bibinfo {volume} {814}},\ \bibinfo {pages}
  {136078} (\bibinfo {year} {2021})},\ \Eprint
  {http://arxiv.org/abs/2010.13066} {arXiv:2010.13066 [hep-ph]} \BibitemShut
  {NoStop}%
\bibitem [{\citenamefont {van Hameren}\ \emph {et~al.}(2023)\citenamefont {van
  Hameren}, \citenamefont {Kakkad}, \citenamefont {Kotko}, \citenamefont
  {Kutak},\ and\ \citenamefont {Sapeta}}]{vanHameren:2023oiq}%
  \BibitemOpen
  \bibfield  {author} {\bibinfo {author} {\bibfnamefont {A.}~\bibnamefont {van
  Hameren}}, \bibinfo {author} {\bibfnamefont {H.}~\bibnamefont {Kakkad}},
  \bibinfo {author} {\bibfnamefont {P.}~\bibnamefont {Kotko}}, \bibinfo
  {author} {\bibfnamefont {K.}~\bibnamefont {Kutak}}, \ and\ \bibinfo {author}
  {\bibfnamefont {S.}~\bibnamefont {Sapeta}},\ }\href@noop {} {\  (\bibinfo
  {year} {2023})},\ \Eprint {http://arxiv.org/abs/2306.17513} {arXiv:2306.17513
  [hep-ph]} \BibitemShut {NoStop}%
\bibitem [{\citenamefont {Marquet}\ \emph {et~al.}(2016)\citenamefont
  {Marquet}, \citenamefont {Petreska},\ and\ \citenamefont
  {Roiesnel}}]{Marquet:2016cgx}%
  \BibitemOpen
  \bibfield  {author} {\bibinfo {author} {\bibfnamefont {C.}~\bibnamefont
  {Marquet}}, \bibinfo {author} {\bibfnamefont {E.}~\bibnamefont {Petreska}}, \
  and\ \bibinfo {author} {\bibfnamefont {C.}~\bibnamefont {Roiesnel}},\ }\href
  {\doibase 10.1007/JHEP10(2016)065} {\bibfield  {journal} {\bibinfo  {journal}
  {JHEP}\ }\textbf {\bibinfo {volume} {10}},\ \bibinfo {pages} {065} (\bibinfo
  {year} {2016})},\ \Eprint {http://arxiv.org/abs/1608.02577} {arXiv:1608.02577
  [hep-ph]} \BibitemShut {NoStop}%
\bibitem [{\citenamefont {Klein}\ and\ \citenamefont
  {M\"antysaari}(2019)}]{Klein:2019qfb}%
  \BibitemOpen
  \bibfield  {author} {\bibinfo {author} {\bibfnamefont {S.~R.}\ \bibnamefont
  {Klein}}\ and\ \bibinfo {author} {\bibfnamefont {H.}~\bibnamefont
  {M\"antysaari}},\ }\href {\doibase 10.1038/s42254-019-0107-6} {\bibfield
  {journal} {\bibinfo  {journal} {Nature Rev. Phys.}\ }\textbf {\bibinfo
  {volume} {1}},\ \bibinfo {pages} {662} (\bibinfo {year} {2019})},\ \Eprint
  {http://arxiv.org/abs/1910.10858} {arXiv:1910.10858 [hep-ex]} \BibitemShut
  {NoStop}%
\bibitem [{\citenamefont {Iancu}\ and\ \citenamefont
  {Mulian}(2021)}]{Iancu:2020mos}%
  \BibitemOpen
  \bibfield  {author} {\bibinfo {author} {\bibfnamefont {E.}~\bibnamefont
  {Iancu}}\ and\ \bibinfo {author} {\bibfnamefont {Y.}~\bibnamefont {Mulian}},\
  }\href {\doibase 10.1007/JHEP03(2021)005} {\bibfield  {journal} {\bibinfo
  {journal} {JHEP}\ }\textbf {\bibinfo {volume} {03}},\ \bibinfo {pages} {005}
  (\bibinfo {year} {2021})},\ \Eprint {http://arxiv.org/abs/2009.11930}
  {arXiv:2009.11930 [hep-ph]} \BibitemShut {NoStop}%
\bibitem [{\citenamefont {Bolognino}\ \emph {et~al.}(2021)\citenamefont
  {Bolognino}, \citenamefont {Celiberto}, \citenamefont {Fucilla},
  \citenamefont {Ivanov},\ and\ \citenamefont {Papa}}]{Bolognino:2021mrc}%
  \BibitemOpen
  \bibfield  {author} {\bibinfo {author} {\bibfnamefont {A.~D.}\ \bibnamefont
  {Bolognino}}, \bibinfo {author} {\bibfnamefont {F.~G.}\ \bibnamefont
  {Celiberto}}, \bibinfo {author} {\bibfnamefont {M.}~\bibnamefont {Fucilla}},
  \bibinfo {author} {\bibfnamefont {D.~Y.}\ \bibnamefont {Ivanov}}, \ and\
  \bibinfo {author} {\bibfnamefont {A.}~\bibnamefont {Papa}},\ }\href {\doibase
  10.1103/PhysRevD.103.094004} {\bibfield  {journal} {\bibinfo  {journal}
  {Phys. Rev. D}\ }\textbf {\bibinfo {volume} {103}},\ \bibinfo {pages}
  {094004} (\bibinfo {year} {2021})},\ \Eprint
  {http://arxiv.org/abs/2103.07396} {arXiv:2103.07396 [hep-ph]} \BibitemShut
  {NoStop}%
\bibitem [{\citenamefont {Al-Mashad}\ \emph {et~al.}(2022)\citenamefont
  {Al-Mashad}, \citenamefont {van Hameren}, \citenamefont {Kakkad},
  \citenamefont {Kotko}, \citenamefont {Kutak}, \citenamefont {van Mechelen},\
  and\ \citenamefont {Sapeta}}]{Al-Mashad:2022zbq}%
  \BibitemOpen
  \bibfield  {author} {\bibinfo {author} {\bibfnamefont {M.~A.}\ \bibnamefont
  {Al-Mashad}}, \bibinfo {author} {\bibfnamefont {A.}~\bibnamefont {van
  Hameren}}, \bibinfo {author} {\bibfnamefont {H.}~\bibnamefont {Kakkad}},
  \bibinfo {author} {\bibfnamefont {P.}~\bibnamefont {Kotko}}, \bibinfo
  {author} {\bibfnamefont {K.}~\bibnamefont {Kutak}}, \bibinfo {author}
  {\bibfnamefont {P.}~\bibnamefont {van Mechelen}}, \ and\ \bibinfo {author}
  {\bibfnamefont {S.}~\bibnamefont {Sapeta}},\ }\href {\doibase
  10.1007/JHEP12(2022)131} {\bibfield  {journal} {\bibinfo  {journal} {JHEP}\
  }\textbf {\bibinfo {volume} {12}},\ \bibinfo {pages} {131} (\bibinfo {year}
  {2022})},\ \Eprint {http://arxiv.org/abs/2210.06613} {arXiv:2210.06613
  [hep-ph]} \BibitemShut {NoStop}%
\bibitem [{\citenamefont {Agostini}\ \emph {et~al.}(2023)\citenamefont
  {Agostini}, \citenamefont {Altinoluk},\ and\ \citenamefont
  {Armesto}}]{Agostini:2022oge}%
  \BibitemOpen
  \bibfield  {author} {\bibinfo {author} {\bibfnamefont {P.}~\bibnamefont
  {Agostini}}, \bibinfo {author} {\bibfnamefont {T.}~\bibnamefont {Altinoluk}},
  \ and\ \bibinfo {author} {\bibfnamefont {N.}~\bibnamefont {Armesto}},\ }\href
  {\doibase 10.1016/j.physletb.2023.137892} {\bibfield  {journal} {\bibinfo
  {journal} {Phys. Lett. B}\ }\textbf {\bibinfo {volume} {840}},\ \bibinfo
  {pages} {137892} (\bibinfo {year} {2023})},\ \Eprint
  {http://arxiv.org/abs/2212.03633} {arXiv:2212.03633 [hep-ph]} \BibitemShut
  {NoStop}%
\bibitem [{\citenamefont {Iancu}\ \emph {et~al.}(2023)\citenamefont {Iancu},
  \citenamefont {Mueller}, \citenamefont {Triantafyllopoulos},\ and\
  \citenamefont {Wei}}]{Iancu:2023lel}%
  \BibitemOpen
  \bibfield  {author} {\bibinfo {author} {\bibfnamefont {E.}~\bibnamefont
  {Iancu}}, \bibinfo {author} {\bibfnamefont {A.~H.}\ \bibnamefont {Mueller}},
  \bibinfo {author} {\bibfnamefont {D.~N.}\ \bibnamefont {Triantafyllopoulos}},
  \ and\ \bibinfo {author} {\bibfnamefont {S.~Y.}\ \bibnamefont {Wei}},\
  }\href@noop {} {\  (\bibinfo {year} {2023})},\ \Eprint
  {http://arxiv.org/abs/2304.12401} {arXiv:2304.12401 [hep-ph]} \BibitemShut
  {NoStop}%
\bibitem [{\citenamefont {Hagiwara}\ \emph {et~al.}(2017)\citenamefont
  {Hagiwara}, \citenamefont {Hatta}, \citenamefont {Pasechnik}, \citenamefont
  {Tasevsky},\ and\ \citenamefont {Teryaev}}]{Hagiwara:2017fye}%
  \BibitemOpen
  \bibfield  {author} {\bibinfo {author} {\bibfnamefont {Y.}~\bibnamefont
  {Hagiwara}}, \bibinfo {author} {\bibfnamefont {Y.}~\bibnamefont {Hatta}},
  \bibinfo {author} {\bibfnamefont {R.}~\bibnamefont {Pasechnik}}, \bibinfo
  {author} {\bibfnamefont {M.}~\bibnamefont {Tasevsky}}, \ and\ \bibinfo
  {author} {\bibfnamefont {O.}~\bibnamefont {Teryaev}},\ }\href {\doibase
  10.1103/PhysRevD.96.034009} {\bibfield  {journal} {\bibinfo  {journal} {Phys.
  Rev. D}\ }\textbf {\bibinfo {volume} {96}},\ \bibinfo {pages} {034009}
  (\bibinfo {year} {2017})},\ \Eprint {http://arxiv.org/abs/1706.01765}
  {arXiv:1706.01765 [hep-ph]} \BibitemShut {NoStop}%
\bibitem [{\citenamefont {Kotko}\ \emph {et~al.}(2017)\citenamefont {Kotko},
  \citenamefont {Kutak}, \citenamefont {Sapeta}, \citenamefont {Stasto},\ and\
  \citenamefont {Strikman}}]{Kotko:2017oxg}%
  \BibitemOpen
  \bibfield  {author} {\bibinfo {author} {\bibfnamefont {P.}~\bibnamefont
  {Kotko}}, \bibinfo {author} {\bibfnamefont {K.}~\bibnamefont {Kutak}},
  \bibinfo {author} {\bibfnamefont {S.}~\bibnamefont {Sapeta}}, \bibinfo
  {author} {\bibfnamefont {A.~M.}\ \bibnamefont {Stasto}}, \ and\ \bibinfo
  {author} {\bibfnamefont {M.}~\bibnamefont {Strikman}},\ }\href {\doibase
  10.1140/epjc/s10052-017-4906-6} {\bibfield  {journal} {\bibinfo  {journal}
  {Eur. Phys. J. C}\ }\textbf {\bibinfo {volume} {77}},\ \bibinfo {pages} {353}
  (\bibinfo {year} {2017})},\ \Eprint {http://arxiv.org/abs/1702.03063}
  {arXiv:1702.03063 [hep-ph]} \BibitemShut {NoStop}%
\bibitem [{\citenamefont {Bhattacharya}\ \emph {et~al.}(2022)\citenamefont
  {Bhattacharya}, \citenamefont {Metz}, \citenamefont {Ojha}, \citenamefont
  {Tsai},\ and\ \citenamefont {Zhou}}]{Bhattacharya:2018lgm}%
  \BibitemOpen
  \bibfield  {author} {\bibinfo {author} {\bibfnamefont {S.}~\bibnamefont
  {Bhattacharya}}, \bibinfo {author} {\bibfnamefont {A.}~\bibnamefont {Metz}},
  \bibinfo {author} {\bibfnamefont {V.~K.}\ \bibnamefont {Ojha}}, \bibinfo
  {author} {\bibfnamefont {J.-Y.}\ \bibnamefont {Tsai}}, \ and\ \bibinfo
  {author} {\bibfnamefont {J.}~\bibnamefont {Zhou}},\ }\href {\doibase
  10.1016/j.physletb.2022.137383} {\bibfield  {journal} {\bibinfo  {journal}
  {Phys. Lett. B}\ }\textbf {\bibinfo {volume} {833}},\ \bibinfo {pages}
  {137383} (\bibinfo {year} {2022})},\ \Eprint
  {http://arxiv.org/abs/1802.10550} {arXiv:1802.10550 [hep-ph]} \BibitemShut
  {NoStop}%
\bibitem [{\citenamefont {Boussarie}\ \emph {et~al.}(2018)\citenamefont
  {Boussarie}, \citenamefont {Hatta}, \citenamefont {Xiao},\ and\ \citenamefont
  {Yuan}}]{Boussarie:2018zwg}%
  \BibitemOpen
  \bibfield  {author} {\bibinfo {author} {\bibfnamefont {R.}~\bibnamefont
  {Boussarie}}, \bibinfo {author} {\bibfnamefont {Y.}~\bibnamefont {Hatta}},
  \bibinfo {author} {\bibfnamefont {B.-W.}\ \bibnamefont {Xiao}}, \ and\
  \bibinfo {author} {\bibfnamefont {F.}~\bibnamefont {Yuan}},\ }\href {\doibase
  10.1103/PhysRevD.98.074015} {\bibfield  {journal} {\bibinfo  {journal} {Phys.
  Rev. D}\ }\textbf {\bibinfo {volume} {98}},\ \bibinfo {pages} {074015}
  (\bibinfo {year} {2018})},\ \Eprint {http://arxiv.org/abs/1807.08697}
  {arXiv:1807.08697 [hep-ph]} \BibitemShut {NoStop}%
\bibitem [{\citenamefont {Albacete}\ and\ \citenamefont
  {Marquet}(2010)}]{Albacete:2010pg}%
  \BibitemOpen
  \bibfield  {author} {\bibinfo {author} {\bibfnamefont {J.~L.}\ \bibnamefont
  {Albacete}}\ and\ \bibinfo {author} {\bibfnamefont {C.}~\bibnamefont
  {Marquet}},\ }\href {\doibase 10.1103/PhysRevLett.105.162301} {\bibfield
  {journal} {\bibinfo  {journal} {Phys. Rev. Lett.}\ }\textbf {\bibinfo
  {volume} {105}},\ \bibinfo {pages} {162301} (\bibinfo {year} {2010})},\
  \Eprint {http://arxiv.org/abs/1005.4065} {arXiv:1005.4065 [hep-ph]}
  \BibitemShut {NoStop}%
\bibitem [{\citenamefont {Stasto}\ \emph
  {et~al.}(2012{\natexlab{a}})\citenamefont {Stasto}, \citenamefont {Xiao},\
  and\ \citenamefont {Yuan}}]{Stasto:2011ru}%
  \BibitemOpen
  \bibfield  {author} {\bibinfo {author} {\bibfnamefont {A.}~\bibnamefont
  {Stasto}}, \bibinfo {author} {\bibfnamefont {B.-W.}\ \bibnamefont {Xiao}}, \
  and\ \bibinfo {author} {\bibfnamefont {F.}~\bibnamefont {Yuan}},\ }\href
  {\doibase 10.1016/j.physletb.2012.08.044} {\bibfield  {journal} {\bibinfo
  {journal} {Phys. Lett. B}\ }\textbf {\bibinfo {volume} {716}},\ \bibinfo
  {pages} {430} (\bibinfo {year} {2012}{\natexlab{a}})},\ \Eprint
  {http://arxiv.org/abs/1109.1817} {arXiv:1109.1817 [hep-ph]} \BibitemShut
  {NoStop}%
\bibitem [{\citenamefont {Lappi}\ and\ \citenamefont
  {Mantysaari}(2013)}]{Lappi:2012nh}%
  \BibitemOpen
  \bibfield  {author} {\bibinfo {author} {\bibfnamefont {T.}~\bibnamefont
  {Lappi}}\ and\ \bibinfo {author} {\bibfnamefont {H.}~\bibnamefont
  {Mantysaari}},\ }\href {\doibase 10.1016/j.nuclphysa.2013.03.017} {\bibfield
  {journal} {\bibinfo  {journal} {Nucl. Phys. A}\ }\textbf {\bibinfo {volume}
  {908}},\ \bibinfo {pages} {51} (\bibinfo {year} {2013})},\ \Eprint
  {http://arxiv.org/abs/1209.2853} {arXiv:1209.2853 [hep-ph]} \BibitemShut
  {NoStop}%
\bibitem [{\citenamefont {Iancu}\ and\ \citenamefont
  {Laidet}(2013)}]{Iancu:2013dta}%
  \BibitemOpen
  \bibfield  {author} {\bibinfo {author} {\bibfnamefont {E.}~\bibnamefont
  {Iancu}}\ and\ \bibinfo {author} {\bibfnamefont {J.}~\bibnamefont {Laidet}},\
  }\href {\doibase 10.1016/j.nuclphysa.2013.07.012} {\bibfield  {journal}
  {\bibinfo  {journal} {Nucl. Phys. A}\ }\textbf {\bibinfo {volume} {916}},\
  \bibinfo {pages} {48} (\bibinfo {year} {2013})},\ \Eprint
  {http://arxiv.org/abs/1305.5926} {arXiv:1305.5926 [hep-ph]} \BibitemShut
  {NoStop}%
\bibitem [{\citenamefont {Albacete}\ \emph {et~al.}(2019)\citenamefont
  {Albacete}, \citenamefont {Giacalone}, \citenamefont {Marquet},\ and\
  \citenamefont {Matas}}]{Albacete:2018ruq}%
  \BibitemOpen
  \bibfield  {author} {\bibinfo {author} {\bibfnamefont {J.~L.}\ \bibnamefont
  {Albacete}}, \bibinfo {author} {\bibfnamefont {G.}~\bibnamefont {Giacalone}},
  \bibinfo {author} {\bibfnamefont {C.}~\bibnamefont {Marquet}}, \ and\
  \bibinfo {author} {\bibfnamefont {M.}~\bibnamefont {Matas}},\ }\href
  {\doibase 10.1103/PhysRevD.99.014002} {\bibfield  {journal} {\bibinfo
  {journal} {Phys. Rev. D}\ }\textbf {\bibinfo {volume} {99}},\ \bibinfo
  {pages} {014002} (\bibinfo {year} {2019})},\ \Eprint
  {http://arxiv.org/abs/1805.05711} {arXiv:1805.05711 [hep-ph]} \BibitemShut
  {NoStop}%
\bibitem [{\citenamefont {Stasto}\ \emph {et~al.}(2018)\citenamefont {Stasto},
  \citenamefont {Wei}, \citenamefont {Xiao},\ and\ \citenamefont
  {Yuan}}]{Stasto:2018rci}%
  \BibitemOpen
  \bibfield  {author} {\bibinfo {author} {\bibfnamefont {A.}~\bibnamefont
  {Stasto}}, \bibinfo {author} {\bibfnamefont {S.-Y.}\ \bibnamefont {Wei}},
  \bibinfo {author} {\bibfnamefont {B.-W.}\ \bibnamefont {Xiao}}, \ and\
  \bibinfo {author} {\bibfnamefont {F.}~\bibnamefont {Yuan}},\ }\href {\doibase
  10.1016/j.physletb.2018.08.011} {\bibfield  {journal} {\bibinfo  {journal}
  {Phys. Lett. B}\ }\textbf {\bibinfo {volume} {784}},\ \bibinfo {pages} {301}
  (\bibinfo {year} {2018})},\ \Eprint {http://arxiv.org/abs/1805.05712}
  {arXiv:1805.05712 [hep-ph]} \BibitemShut {NoStop}%
\bibitem [{\citenamefont {Jalilian-Marian}(2006)}]{Jalilian-Marian:2005qbq}%
  \BibitemOpen
  \bibfield  {author} {\bibinfo {author} {\bibfnamefont {J.}~\bibnamefont
  {Jalilian-Marian}},\ }\href {\doibase 10.1016/j.nuclphysa.2006.02.013}
  {\bibfield  {journal} {\bibinfo  {journal} {Nucl. Phys. A}\ }\textbf
  {\bibinfo {volume} {770}},\ \bibinfo {pages} {210} (\bibinfo {year}
  {2006})},\ \Eprint {http://arxiv.org/abs/hep-ph/0509338}
  {arXiv:hep-ph/0509338} \BibitemShut {NoStop}%
\bibitem [{\citenamefont {Jalilian-Marian}\ and\ \citenamefont
  {Rezaeian}(2012)}]{Jalilian-Marian:2012wwi}%
  \BibitemOpen
  \bibfield  {author} {\bibinfo {author} {\bibfnamefont {J.}~\bibnamefont
  {Jalilian-Marian}}\ and\ \bibinfo {author} {\bibfnamefont {A.~H.}\
  \bibnamefont {Rezaeian}},\ }\href {\doibase 10.1103/PhysRevD.86.034016}
  {\bibfield  {journal} {\bibinfo  {journal} {Phys. Rev. D}\ }\textbf {\bibinfo
  {volume} {86}},\ \bibinfo {pages} {034016} (\bibinfo {year} {2012})},\
  \Eprint {http://arxiv.org/abs/1204.1319} {arXiv:1204.1319 [hep-ph]}
  \BibitemShut {NoStop}%
\bibitem [{\citenamefont {Stasto}\ \emph
  {et~al.}(2012{\natexlab{b}})\citenamefont {Stasto}, \citenamefont {Xiao},\
  and\ \citenamefont {Zaslavsky}}]{Stasto:2012ru}%
  \BibitemOpen
  \bibfield  {author} {\bibinfo {author} {\bibfnamefont {A.}~\bibnamefont
  {Stasto}}, \bibinfo {author} {\bibfnamefont {B.-W.}\ \bibnamefont {Xiao}}, \
  and\ \bibinfo {author} {\bibfnamefont {D.}~\bibnamefont {Zaslavsky}},\ }\href
  {\doibase 10.1103/PhysRevD.86.014009} {\bibfield  {journal} {\bibinfo
  {journal} {Phys. Rev. D}\ }\textbf {\bibinfo {volume} {86}},\ \bibinfo
  {pages} {014009} (\bibinfo {year} {2012}{\natexlab{b}})},\ \Eprint
  {http://arxiv.org/abs/1204.4861} {arXiv:1204.4861 [hep-ph]} \BibitemShut
  {NoStop}%
\bibitem [{\citenamefont {Rezaeian}(2012)}]{Rezaeian:2012wa}%
  \BibitemOpen
  \bibfield  {author} {\bibinfo {author} {\bibfnamefont {A.~H.}\ \bibnamefont
  {Rezaeian}},\ }\href {\doibase 10.1103/PhysRevD.86.094016} {\bibfield
  {journal} {\bibinfo  {journal} {Phys. Rev. D}\ }\textbf {\bibinfo {volume}
  {86}},\ \bibinfo {pages} {094016} (\bibinfo {year} {2012})},\ \Eprint
  {http://arxiv.org/abs/1209.0478} {arXiv:1209.0478 [hep-ph]} \BibitemShut
  {NoStop}%
\bibitem [{\citenamefont {Basso}\ \emph
  {et~al.}(2016{\natexlab{a}})\citenamefont {Basso}, \citenamefont {Goncalves},
  \citenamefont {Nemchik}, \citenamefont {Pasechnik},\ and\ \citenamefont
  {Sumbera}}]{Basso:2015pba}%
  \BibitemOpen
  \bibfield  {author} {\bibinfo {author} {\bibfnamefont {E.}~\bibnamefont
  {Basso}}, \bibinfo {author} {\bibfnamefont {V.~P.}\ \bibnamefont
  {Goncalves}}, \bibinfo {author} {\bibfnamefont {J.}~\bibnamefont {Nemchik}},
  \bibinfo {author} {\bibfnamefont {R.}~\bibnamefont {Pasechnik}}, \ and\
  \bibinfo {author} {\bibfnamefont {M.}~\bibnamefont {Sumbera}},\ }\href
  {\doibase 10.1103/PhysRevD.93.034023} {\bibfield  {journal} {\bibinfo
  {journal} {Phys. Rev. D}\ }\textbf {\bibinfo {volume} {93}},\ \bibinfo
  {pages} {034023} (\bibinfo {year} {2016}{\natexlab{a}})},\ \Eprint
  {http://arxiv.org/abs/1510.00650} {arXiv:1510.00650 [hep-ph]} \BibitemShut
  {NoStop}%
\bibitem [{\citenamefont {Rezaeian}(2016)}]{Rezaeian:2016szi}%
  \BibitemOpen
  \bibfield  {author} {\bibinfo {author} {\bibfnamefont {A.~H.}\ \bibnamefont
  {Rezaeian}},\ }\href {\doibase 10.1103/PhysRevD.93.094030} {\bibfield
  {journal} {\bibinfo  {journal} {Phys. Rev. D}\ }\textbf {\bibinfo {volume}
  {93}},\ \bibinfo {pages} {094030} (\bibinfo {year} {2016})},\ \Eprint
  {http://arxiv.org/abs/1603.07354} {arXiv:1603.07354 [hep-ph]} \BibitemShut
  {NoStop}%
\bibitem [{\citenamefont {Basso}\ \emph
  {et~al.}(2016{\natexlab{b}})\citenamefont {Basso}, \citenamefont {Goncalves},
  \citenamefont {Krelina}, \citenamefont {Nemchik},\ and\ \citenamefont
  {Pasechnik}}]{Basso:2016ulb}%
  \BibitemOpen
  \bibfield  {author} {\bibinfo {author} {\bibfnamefont {E.}~\bibnamefont
  {Basso}}, \bibinfo {author} {\bibfnamefont {V.~P.}\ \bibnamefont
  {Goncalves}}, \bibinfo {author} {\bibfnamefont {M.}~\bibnamefont {Krelina}},
  \bibinfo {author} {\bibfnamefont {J.}~\bibnamefont {Nemchik}}, \ and\
  \bibinfo {author} {\bibfnamefont {R.}~\bibnamefont {Pasechnik}},\ }\href
  {\doibase 10.1103/PhysRevD.93.094027} {\bibfield  {journal} {\bibinfo
  {journal} {Phys. Rev. D}\ }\textbf {\bibinfo {volume} {93}},\ \bibinfo
  {pages} {094027} (\bibinfo {year} {2016}{\natexlab{b}})},\ \Eprint
  {http://arxiv.org/abs/1603.01893} {arXiv:1603.01893 [hep-ph]} \BibitemShut
  {NoStop}%
\bibitem [{\citenamefont {Boer}\ \emph {et~al.}(2017)\citenamefont {Boer},
  \citenamefont {Mulders}, \citenamefont {Zhou},\ and\ \citenamefont
  {Zhou}}]{Boer:2017xpy}%
  \BibitemOpen
  \bibfield  {author} {\bibinfo {author} {\bibfnamefont {D.}~\bibnamefont
  {Boer}}, \bibinfo {author} {\bibfnamefont {P.~J.}\ \bibnamefont {Mulders}},
  \bibinfo {author} {\bibfnamefont {J.}~\bibnamefont {Zhou}}, \ and\ \bibinfo
  {author} {\bibfnamefont {Y.-j.}\ \bibnamefont {Zhou}},\ }\href {\doibase
  10.1007/JHEP10(2017)196} {\bibfield  {journal} {\bibinfo  {journal} {JHEP}\
  }\textbf {\bibinfo {volume} {10}},\ \bibinfo {pages} {196} (\bibinfo {year}
  {2017})},\ \Eprint {http://arxiv.org/abs/1702.08195} {arXiv:1702.08195
  [hep-ph]} \BibitemShut {NoStop}%
\bibitem [{\citenamefont {Beni\'c}\ and\ \citenamefont
  {Dumitru}(2018)}]{Benic:2017znu}%
  \BibitemOpen
  \bibfield  {author} {\bibinfo {author} {\bibfnamefont {S.}~\bibnamefont
  {Beni\'c}}\ and\ \bibinfo {author} {\bibfnamefont {A.}~\bibnamefont
  {Dumitru}},\ }\href {\doibase 10.1103/PhysRevD.97.014012} {\bibfield
  {journal} {\bibinfo  {journal} {Phys. Rev. D}\ }\textbf {\bibinfo {volume}
  {97}},\ \bibinfo {pages} {014012} (\bibinfo {year} {2018})},\ \Eprint
  {http://arxiv.org/abs/1710.01991} {arXiv:1710.01991 [hep-ph]} \BibitemShut
  {NoStop}%
\bibitem [{\citenamefont {Goncalves}\ \emph {et~al.}(2020)\citenamefont
  {Goncalves}, \citenamefont {Lima}, \citenamefont {Pasechnik},\ and\
  \citenamefont {\v{S}umbera}}]{Goncalves:2020tvh}%
  \BibitemOpen
  \bibfield  {author} {\bibinfo {author} {\bibfnamefont {V.~P.}\ \bibnamefont
  {Goncalves}}, \bibinfo {author} {\bibfnamefont {Y.}~\bibnamefont {Lima}},
  \bibinfo {author} {\bibfnamefont {R.}~\bibnamefont {Pasechnik}}, \ and\
  \bibinfo {author} {\bibfnamefont {M.}~\bibnamefont {\v{S}umbera}},\ }\href
  {\doibase 10.1103/PhysRevD.101.094019} {\bibfield  {journal} {\bibinfo
  {journal} {Phys. Rev. D}\ }\textbf {\bibinfo {volume} {101}},\ \bibinfo
  {pages} {094019} (\bibinfo {year} {2020})},\ \Eprint
  {http://arxiv.org/abs/2003.02555} {arXiv:2003.02555 [hep-ph]} \BibitemShut
  {NoStop}%
\bibitem [{\citenamefont {Beni\'c}\ \emph {et~al.}(2022)\citenamefont
  {Beni\'c}, \citenamefont {Garcia-Montero},\ and\ \citenamefont
  {Perkov}}]{Benic:2022ixp}%
  \BibitemOpen
  \bibfield  {author} {\bibinfo {author} {\bibfnamefont {S.}~\bibnamefont
  {Beni\'c}}, \bibinfo {author} {\bibfnamefont {O.}~\bibnamefont
  {Garcia-Montero}}, \ and\ \bibinfo {author} {\bibfnamefont {A.}~\bibnamefont
  {Perkov}},\ }\href {\doibase 10.1103/PhysRevD.105.114052} {\bibfield
  {journal} {\bibinfo  {journal} {Phys. Rev. D}\ }\textbf {\bibinfo {volume}
  {105}},\ \bibinfo {pages} {114052} (\bibinfo {year} {2022})},\ \Eprint
  {http://arxiv.org/abs/2203.01685} {arXiv:2203.01685 [hep-ph]} \BibitemShut
  {NoStop}%
\bibitem [{\citenamefont {Boer}\ \emph {et~al.}(2022)\citenamefont {Boer},
  \citenamefont {Hagiwara}, \citenamefont {Zhou},\ and\ \citenamefont
  {Zhou}}]{Boer:2022njw}%
  \BibitemOpen
  \bibfield  {author} {\bibinfo {author} {\bibfnamefont {D.}~\bibnamefont
  {Boer}}, \bibinfo {author} {\bibfnamefont {Y.}~\bibnamefont {Hagiwara}},
  \bibinfo {author} {\bibfnamefont {J.}~\bibnamefont {Zhou}}, \ and\ \bibinfo
  {author} {\bibfnamefont {Y.-j.}\ \bibnamefont {Zhou}},\ }\href {\doibase
  10.1103/PhysRevD.105.096017} {\bibfield  {journal} {\bibinfo  {journal}
  {Phys. Rev. D}\ }\textbf {\bibinfo {volume} {105}},\ \bibinfo {pages}
  {096017} (\bibinfo {year} {2022})},\ \Eprint
  {http://arxiv.org/abs/2203.00267} {arXiv:2203.00267 [hep-ph]} \BibitemShut
  {NoStop}%
\bibitem [{\citenamefont {Gelis}\ and\ \citenamefont
  {Jalilian-Marian}(2002)}]{Gelis:2002fw}%
  \BibitemOpen
  \bibfield  {author} {\bibinfo {author} {\bibfnamefont {F.}~\bibnamefont
  {Gelis}}\ and\ \bibinfo {author} {\bibfnamefont {J.}~\bibnamefont
  {Jalilian-Marian}},\ }\href {\doibase 10.1103/PhysRevD.66.094014} {\bibfield
  {journal} {\bibinfo  {journal} {Phys. Rev. D}\ }\textbf {\bibinfo {volume}
  {66}},\ \bibinfo {pages} {094014} (\bibinfo {year} {2002})},\ \Eprint
  {http://arxiv.org/abs/hep-ph/0208141} {arXiv:hep-ph/0208141} \BibitemShut
  {NoStop}%
\bibitem [{\citenamefont {Kovner}\ and\ \citenamefont
  {Rezaeian}(2014)}]{Kovner:2014qea}%
  \BibitemOpen
  \bibfield  {author} {\bibinfo {author} {\bibfnamefont {A.}~\bibnamefont
  {Kovner}}\ and\ \bibinfo {author} {\bibfnamefont {A.~H.}\ \bibnamefont
  {Rezaeian}},\ }\href {\doibase 10.1103/PhysRevD.90.014031} {\bibfield
  {journal} {\bibinfo  {journal} {Phys. Rev. D}\ }\textbf {\bibinfo {volume}
  {90}},\ \bibinfo {pages} {014031} (\bibinfo {year} {2014})},\ \Eprint
  {http://arxiv.org/abs/1404.5632} {arXiv:1404.5632 [hep-ph]} \BibitemShut
  {NoStop}%
\bibitem [{\citenamefont {Kovner}\ and\ \citenamefont
  {Rezaeian}(2015)}]{Kovner:2015rna}%
  \BibitemOpen
  \bibfield  {author} {\bibinfo {author} {\bibfnamefont {A.}~\bibnamefont
  {Kovner}}\ and\ \bibinfo {author} {\bibfnamefont {A.~H.}\ \bibnamefont
  {Rezaeian}},\ }\href {\doibase 10.1103/PhysRevD.92.074045} {\bibfield
  {journal} {\bibinfo  {journal} {Phys. Rev. D}\ }\textbf {\bibinfo {volume}
  {92}},\ \bibinfo {pages} {074045} (\bibinfo {year} {2015})},\ \Eprint
  {http://arxiv.org/abs/1508.02412} {arXiv:1508.02412 [hep-ph]} \BibitemShut
  {NoStop}%
\bibitem [{\citenamefont {Marquet}\ \emph {et~al.}(2020)\citenamefont
  {Marquet}, \citenamefont {Wei},\ and\ \citenamefont
  {Xiao}}]{Marquet:2019ltn}%
  \BibitemOpen
  \bibfield  {author} {\bibinfo {author} {\bibfnamefont {C.}~\bibnamefont
  {Marquet}}, \bibinfo {author} {\bibfnamefont {S.-Y.}\ \bibnamefont {Wei}}, \
  and\ \bibinfo {author} {\bibfnamefont {B.-W.}\ \bibnamefont {Xiao}},\ }\href
  {\doibase 10.1016/j.physletb.2020.135253} {\bibfield  {journal} {\bibinfo
  {journal} {Phys. Lett. B}\ }\textbf {\bibinfo {volume} {802}},\ \bibinfo
  {pages} {135253} (\bibinfo {year} {2020})},\ \Eprint
  {http://arxiv.org/abs/1909.08572} {arXiv:1909.08572 [hep-ph]} \BibitemShut
  {NoStop}%
\bibitem [{\citenamefont {Akcakaya}\ \emph {et~al.}(2013)\citenamefont
  {Akcakaya}, \citenamefont {Sch\"afer},\ and\ \citenamefont
  {Zhou}}]{Akcakaya:2012si}%
  \BibitemOpen
  \bibfield  {author} {\bibinfo {author} {\bibfnamefont {E.}~\bibnamefont
  {Akcakaya}}, \bibinfo {author} {\bibfnamefont {A.}~\bibnamefont {Sch\"afer}},
  \ and\ \bibinfo {author} {\bibfnamefont {J.}~\bibnamefont {Zhou}},\ }\href
  {\doibase 10.1103/PhysRevD.87.054010} {\bibfield  {journal} {\bibinfo
  {journal} {Phys. Rev. D}\ }\textbf {\bibinfo {volume} {87}},\ \bibinfo
  {pages} {054010} (\bibinfo {year} {2013})},\ \Eprint
  {http://arxiv.org/abs/1208.4965} {arXiv:1208.4965 [hep-ph]} \BibitemShut
  {NoStop}%
\bibitem [{\citenamefont {Marquet}\ \emph {et~al.}(2018)\citenamefont
  {Marquet}, \citenamefont {Roiesnel},\ and\ \citenamefont
  {Taels}}]{Marquet:2017xwy}%
  \BibitemOpen
  \bibfield  {author} {\bibinfo {author} {\bibfnamefont {C.}~\bibnamefont
  {Marquet}}, \bibinfo {author} {\bibfnamefont {C.}~\bibnamefont {Roiesnel}}, \
  and\ \bibinfo {author} {\bibfnamefont {P.}~\bibnamefont {Taels}},\ }\href
  {\doibase 10.1103/PhysRevD.97.014004} {\bibfield  {journal} {\bibinfo
  {journal} {Phys. Rev. D}\ }\textbf {\bibinfo {volume} {97}},\ \bibinfo
  {pages} {014004} (\bibinfo {year} {2018})},\ \Eprint
  {http://arxiv.org/abs/1710.05698} {arXiv:1710.05698 [hep-ph]} \BibitemShut
  {NoStop}%
\bibitem [{\citenamefont {Ju}\ and\ \citenamefont
  {Sch\"onherr}(2023)}]{Ju:2022wia}%
  \BibitemOpen
  \bibfield  {author} {\bibinfo {author} {\bibfnamefont {W.-L.}\ \bibnamefont
  {Ju}}\ and\ \bibinfo {author} {\bibfnamefont {M.}~\bibnamefont
  {Sch\"onherr}},\ }\href {\doibase 10.1007/JHEP02(2023)075} {\bibfield
  {journal} {\bibinfo  {journal} {JHEP}\ }\textbf {\bibinfo {volume} {02}},\
  \bibinfo {pages} {075} (\bibinfo {year} {2023})},\ \Eprint
  {http://arxiv.org/abs/2210.09272} {arXiv:2210.09272 [hep-ph]} \BibitemShut
  {NoStop}%
\bibitem [{\citenamefont {Ganguli}\ \emph {et~al.}(2023)\citenamefont
  {Ganguli}, \citenamefont {van Hameren}, \citenamefont {Kotko},\ and\
  \citenamefont {Kutak}}]{Ganguli:2023joy}%
  \BibitemOpen
  \bibfield  {author} {\bibinfo {author} {\bibfnamefont {I.}~\bibnamefont
  {Ganguli}}, \bibinfo {author} {\bibfnamefont {A.}~\bibnamefont {van
  Hameren}}, \bibinfo {author} {\bibfnamefont {P.}~\bibnamefont {Kotko}}, \
  and\ \bibinfo {author} {\bibfnamefont {K.}~\bibnamefont {Kutak}},\ }\href
  {\doibase 10.1140/epjc/s10052-023-12043-3} {\bibfield  {journal} {\bibinfo
  {journal} {Eur. Phys. J. C}\ }\textbf {\bibinfo {volume} {83}},\ \bibinfo
  {pages} {868} (\bibinfo {year} {2023})},\ \Eprint
  {http://arxiv.org/abs/2306.04706} {arXiv:2306.04706 [hep-ph]} \BibitemShut
  {NoStop}%
\bibitem [{\citenamefont {Liu}\ \emph {et~al.}(2019)\citenamefont {Liu},
  \citenamefont {Ringer}, \citenamefont {Vogelsang},\ and\ \citenamefont
  {Yuan}}]{Liu:2018trl}%
  \BibitemOpen
  \bibfield  {author} {\bibinfo {author} {\bibfnamefont {X.}~\bibnamefont
  {Liu}}, \bibinfo {author} {\bibfnamefont {F.}~\bibnamefont {Ringer}},
  \bibinfo {author} {\bibfnamefont {W.}~\bibnamefont {Vogelsang}}, \ and\
  \bibinfo {author} {\bibfnamefont {F.}~\bibnamefont {Yuan}},\ }\href {\doibase
  10.1103/PhysRevLett.122.192003} {\bibfield  {journal} {\bibinfo  {journal}
  {Phys. Rev. Lett.}\ }\textbf {\bibinfo {volume} {122}},\ \bibinfo {pages}
  {192003} (\bibinfo {year} {2019})},\ \Eprint
  {http://arxiv.org/abs/1812.08077} {arXiv:1812.08077 [hep-ph]} \BibitemShut
  {NoStop}%
\bibitem [{\citenamefont {Liu}\ \emph {et~al.}(2020)\citenamefont {Liu},
  \citenamefont {Ringer}, \citenamefont {Vogelsang},\ and\ \citenamefont
  {Yuan}}]{Liu:2020dct}%
  \BibitemOpen
  \bibfield  {author} {\bibinfo {author} {\bibfnamefont {X.}~\bibnamefont
  {Liu}}, \bibinfo {author} {\bibfnamefont {F.}~\bibnamefont {Ringer}},
  \bibinfo {author} {\bibfnamefont {W.}~\bibnamefont {Vogelsang}}, \ and\
  \bibinfo {author} {\bibfnamefont {F.}~\bibnamefont {Yuan}},\ }\href {\doibase
  10.1103/PhysRevD.102.094022} {\bibfield  {journal} {\bibinfo  {journal}
  {Phys. Rev. D}\ }\textbf {\bibinfo {volume} {102}},\ \bibinfo {pages}
  {094022} (\bibinfo {year} {2020})},\ \Eprint
  {http://arxiv.org/abs/2007.12866} {arXiv:2007.12866 [hep-ph]} \BibitemShut
  {NoStop}%
\bibitem [{\citenamefont {Arratia}\ \emph
  {et~al.}(2020{\natexlab{a}})\citenamefont {Arratia}, \citenamefont {Kang},
  \citenamefont {Prokudin},\ and\ \citenamefont {Ringer}}]{Arratia:2020nxw}%
  \BibitemOpen
  \bibfield  {author} {\bibinfo {author} {\bibfnamefont {M.}~\bibnamefont
  {Arratia}}, \bibinfo {author} {\bibfnamefont {Z.-B.}\ \bibnamefont {Kang}},
  \bibinfo {author} {\bibfnamefont {A.}~\bibnamefont {Prokudin}}, \ and\
  \bibinfo {author} {\bibfnamefont {F.}~\bibnamefont {Ringer}},\ }\href
  {\doibase 10.1103/PhysRevD.102.074015} {\bibfield  {journal} {\bibinfo
  {journal} {Phys. Rev. D}\ }\textbf {\bibinfo {volume} {102}},\ \bibinfo
  {pages} {074015} (\bibinfo {year} {2020}{\natexlab{a}})},\ \Eprint
  {http://arxiv.org/abs/2007.07281} {arXiv:2007.07281 [hep-ph]} \BibitemShut
  {NoStop}%
\bibitem [{\citenamefont {Kang}\ \emph {et~al.}(2021)\citenamefont {Kang},
  \citenamefont {Lee}, \citenamefont {Shao},\ and\ \citenamefont
  {Zhao}}]{Kang:2021ffh}%
  \BibitemOpen
  \bibfield  {author} {\bibinfo {author} {\bibfnamefont {Z.-B.}\ \bibnamefont
  {Kang}}, \bibinfo {author} {\bibfnamefont {K.}~\bibnamefont {Lee}}, \bibinfo
  {author} {\bibfnamefont {D.~Y.}\ \bibnamefont {Shao}}, \ and\ \bibinfo
  {author} {\bibfnamefont {F.}~\bibnamefont {Zhao}},\ }\href {\doibase
  10.1007/JHEP11(2021)005} {\bibfield  {journal} {\bibinfo  {journal} {JHEP}\
  }\textbf {\bibinfo {volume} {11}},\ \bibinfo {pages} {005} (\bibinfo {year}
  {2021})},\ \Eprint {http://arxiv.org/abs/2106.15624} {arXiv:2106.15624
  [hep-ph]} \BibitemShut {NoStop}%
\bibitem [{\citenamefont {Arratia}\ \emph {et~al.}(2023)\citenamefont
  {Arratia}, \citenamefont {Kang}, \citenamefont {Paul}, \citenamefont
  {Prokudin}, \citenamefont {Ringer},\ and\ \citenamefont
  {Zhao}}]{Arratia:2022oxd}%
  \BibitemOpen
  \bibfield  {author} {\bibinfo {author} {\bibfnamefont {M.}~\bibnamefont
  {Arratia}}, \bibinfo {author} {\bibfnamefont {Z.-B.}\ \bibnamefont {Kang}},
  \bibinfo {author} {\bibfnamefont {S.~J.}\ \bibnamefont {Paul}}, \bibinfo
  {author} {\bibfnamefont {A.}~\bibnamefont {Prokudin}}, \bibinfo {author}
  {\bibfnamefont {F.}~\bibnamefont {Ringer}}, \ and\ \bibinfo {author}
  {\bibfnamefont {F.}~\bibnamefont {Zhao}},\ }\href {\doibase
  10.1103/PhysRevD.107.094036} {\bibfield  {journal} {\bibinfo  {journal}
  {Phys. Rev. D}\ }\textbf {\bibinfo {volume} {107}},\ \bibinfo {pages}
  {094036} (\bibinfo {year} {2023})},\ \Eprint
  {http://arxiv.org/abs/2212.02432} {arXiv:2212.02432 [hep-ph]} \BibitemShut
  {NoStop}%
\bibitem [{\citenamefont {Kang}\ \emph {et~al.}(2020)\citenamefont {Kang},
  \citenamefont {Liu}, \citenamefont {Mantry},\ and\ \citenamefont
  {Shao}}]{Kang:2020fka}%
  \BibitemOpen
  \bibfield  {author} {\bibinfo {author} {\bibfnamefont {Z.-B.}\ \bibnamefont
  {Kang}}, \bibinfo {author} {\bibfnamefont {X.}~\bibnamefont {Liu}}, \bibinfo
  {author} {\bibfnamefont {S.}~\bibnamefont {Mantry}}, \ and\ \bibinfo {author}
  {\bibfnamefont {D.~Y.}\ \bibnamefont {Shao}},\ }\href {\doibase
  10.1103/PhysRevLett.125.242003} {\bibfield  {journal} {\bibinfo  {journal}
  {Phys. Rev. Lett.}\ }\textbf {\bibinfo {volume} {125}},\ \bibinfo {pages}
  {242003} (\bibinfo {year} {2020})},\ \Eprint
  {http://arxiv.org/abs/2008.00655} {arXiv:2008.00655 [hep-ph]} \BibitemShut
  {NoStop}%
\bibitem [{\citenamefont {Yang}\ and\ \citenamefont
  {Yang}(2023)}]{Yang:2023vyv}%
  \BibitemOpen
  \bibfield  {author} {\bibinfo {author} {\bibfnamefont {W.}~\bibnamefont
  {Yang}}\ and\ \bibinfo {author} {\bibfnamefont {X.}~\bibnamefont {Yang}},\
  }\href {\doibase 10.1016/j.nuclphysb.2023.116181} {\bibfield  {journal}
  {\bibinfo  {journal} {Nucl. Phys. B}\ }\textbf {\bibinfo {volume} {990}},\
  \bibinfo {pages} {116181} (\bibinfo {year} {2023})}\BibitemShut {NoStop}%
\bibitem [{\citenamefont {Yang}(2023)}]{Yang:2023zod}%
  \BibitemOpen
  \bibfield  {author} {\bibinfo {author} {\bibfnamefont {W.}~\bibnamefont
  {Yang}},\ }\href {\doibase 10.1103/PhysRevD.108.056022} {\bibfield  {journal}
  {\bibinfo  {journal} {Phys. Rev. D}\ }\textbf {\bibinfo {volume} {108}},\
  \bibinfo {pages} {056022} (\bibinfo {year} {2023})},\ \Eprint
  {http://arxiv.org/abs/2306.12632} {arXiv:2306.12632 [hep-ph]} \BibitemShut
  {NoStop}%
\bibitem [{\citenamefont {Abramowicz}\ and\ \citenamefont
  {Caldwell}(1999)}]{Abramowicz:1998ii}%
  \BibitemOpen
  \bibfield  {author} {\bibinfo {author} {\bibfnamefont {H.}~\bibnamefont
  {Abramowicz}}\ and\ \bibinfo {author} {\bibfnamefont {A.}~\bibnamefont
  {Caldwell}},\ }\href {\doibase 10.1103/RevModPhys.71.1275} {\bibfield
  {journal} {\bibinfo  {journal} {Rev. Mod. Phys.}\ }\textbf {\bibinfo {volume}
  {71}},\ \bibinfo {pages} {1275} (\bibinfo {year} {1999})},\ \Eprint
  {http://arxiv.org/abs/hep-ex/9903037} {arXiv:hep-ex/9903037} \BibitemShut
  {NoStop}%
\bibitem [{\citenamefont {Andreev}\ \emph {et~al.}(2022)\citenamefont {Andreev}
  \emph {et~al.}}]{H1:2021wkz}%
  \BibitemOpen
  \bibfield  {author} {\bibinfo {author} {\bibfnamefont {V.}~\bibnamefont
  {Andreev}} \emph {et~al.} (\bibinfo {collaboration} {H1}),\ }\href {\doibase
  10.1103/PhysRevLett.128.132002} {\bibfield  {journal} {\bibinfo  {journal}
  {Phys. Rev. Lett.}\ }\textbf {\bibinfo {volume} {128}},\ \bibinfo {pages}
  {132002} (\bibinfo {year} {2022})},\ \Eprint
  {http://arxiv.org/abs/2108.12376} {arXiv:2108.12376 [hep-ex]} \BibitemShut
  {NoStop}%
\bibitem [{\citenamefont {Arratia}\ \emph
  {et~al.}(2020{\natexlab{b}})\citenamefont {Arratia}, \citenamefont {Song},
  \citenamefont {Ringer},\ and\ \citenamefont {Jacak}}]{Arratia:2019vju}%
  \BibitemOpen
  \bibfield  {author} {\bibinfo {author} {\bibfnamefont {M.}~\bibnamefont
  {Arratia}}, \bibinfo {author} {\bibfnamefont {Y.}~\bibnamefont {Song}},
  \bibinfo {author} {\bibfnamefont {F.}~\bibnamefont {Ringer}}, \ and\ \bibinfo
  {author} {\bibfnamefont {B.~V.}\ \bibnamefont {Jacak}},\ }\href {\doibase
  10.1103/PhysRevC.101.065204} {\bibfield  {journal} {\bibinfo  {journal}
  {Phys. Rev. C}\ }\textbf {\bibinfo {volume} {101}},\ \bibinfo {pages}
  {065204} (\bibinfo {year} {2020}{\natexlab{b}})},\ \Eprint
  {http://arxiv.org/abs/1912.05931} {arXiv:1912.05931 [nucl-ex]} \BibitemShut
  {NoStop}%
\bibitem [{\citenamefont {Hatta}\ \emph
  {et~al.}(2021{\natexlab{a}})\citenamefont {Hatta}, \citenamefont {Xiao},
  \citenamefont {Yuan},\ and\ \citenamefont {Zhou}}]{Hatta:2021jcd}%
  \BibitemOpen
  \bibfield  {author} {\bibinfo {author} {\bibfnamefont {Y.}~\bibnamefont
  {Hatta}}, \bibinfo {author} {\bibfnamefont {B.-W.}\ \bibnamefont {Xiao}},
  \bibinfo {author} {\bibfnamefont {F.}~\bibnamefont {Yuan}}, \ and\ \bibinfo
  {author} {\bibfnamefont {J.}~\bibnamefont {Zhou}},\ }\href {\doibase
  10.1103/PhysRevD.104.054037} {\bibfield  {journal} {\bibinfo  {journal}
  {Phys. Rev. D}\ }\textbf {\bibinfo {volume} {104}},\ \bibinfo {pages}
  {054037} (\bibinfo {year} {2021}{\natexlab{a}})},\ \Eprint
  {http://arxiv.org/abs/2106.05307} {arXiv:2106.05307 [hep-ph]} \BibitemShut
  {NoStop}%
\bibitem [{\citenamefont {Hatta}\ \emph
  {et~al.}(2021{\natexlab{b}})\citenamefont {Hatta}, \citenamefont {Xiao},
  \citenamefont {Yuan},\ and\ \citenamefont {Zhou}}]{Hatta:2020bgy}%
  \BibitemOpen
  \bibfield  {author} {\bibinfo {author} {\bibfnamefont {Y.}~\bibnamefont
  {Hatta}}, \bibinfo {author} {\bibfnamefont {B.-W.}\ \bibnamefont {Xiao}},
  \bibinfo {author} {\bibfnamefont {F.}~\bibnamefont {Yuan}}, \ and\ \bibinfo
  {author} {\bibfnamefont {J.}~\bibnamefont {Zhou}},\ }\href {\doibase
  10.1103/PhysRevLett.126.142001} {\bibfield  {journal} {\bibinfo  {journal}
  {Phys. Rev. Lett.}\ }\textbf {\bibinfo {volume} {126}},\ \bibinfo {pages}
  {142001} (\bibinfo {year} {2021}{\natexlab{b}})},\ \Eprint
  {http://arxiv.org/abs/2010.10774} {arXiv:2010.10774 [hep-ph]} \BibitemShut
  {NoStop}%
\bibitem [{\citenamefont {Marquet}\ \emph {et~al.}(2009)\citenamefont
  {Marquet}, \citenamefont {Xiao},\ and\ \citenamefont
  {Yuan}}]{Marquet:2009ca}%
  \BibitemOpen
  \bibfield  {author} {\bibinfo {author} {\bibfnamefont {C.}~\bibnamefont
  {Marquet}}, \bibinfo {author} {\bibfnamefont {B.-W.}\ \bibnamefont {Xiao}}, \
  and\ \bibinfo {author} {\bibfnamefont {F.}~\bibnamefont {Yuan}},\ }\href
  {\doibase 10.1016/j.physletb.2009.10.099} {\bibfield  {journal} {\bibinfo
  {journal} {Phys. Lett. B}\ }\textbf {\bibinfo {volume} {682}},\ \bibinfo
  {pages} {207} (\bibinfo {year} {2009})},\ \Eprint
  {http://arxiv.org/abs/0906.1454} {arXiv:0906.1454 [hep-ph]} \BibitemShut
  {NoStop}%
\bibitem [{\citenamefont {Xiao}\ and\ \citenamefont
  {Yuan}(2010{\natexlab{a}})}]{Xiao:2010sa}%
  \BibitemOpen
  \bibfield  {author} {\bibinfo {author} {\bibfnamefont {B.-W.}\ \bibnamefont
  {Xiao}}\ and\ \bibinfo {author} {\bibfnamefont {F.}~\bibnamefont {Yuan}},\
  }\href {\doibase 10.1103/PhysRevD.82.114009} {\bibfield  {journal} {\bibinfo
  {journal} {Phys. Rev. D}\ }\textbf {\bibinfo {volume} {82}},\ \bibinfo
  {pages} {114009} (\bibinfo {year} {2010}{\natexlab{a}})},\ \Eprint
  {http://arxiv.org/abs/1008.4432} {arXiv:1008.4432 [hep-ph]} \BibitemShut
  {NoStop}%
\bibitem [{\citenamefont {Xiao}\ and\ \citenamefont
  {Yuan}(2010{\natexlab{b}})}]{Xiao:2010sp}%
  \BibitemOpen
  \bibfield  {author} {\bibinfo {author} {\bibfnamefont {B.-W.}\ \bibnamefont
  {Xiao}}\ and\ \bibinfo {author} {\bibfnamefont {F.}~\bibnamefont {Yuan}},\
  }\href {\doibase 10.1103/PhysRevLett.105.062001} {\bibfield  {journal}
  {\bibinfo  {journal} {Phys. Rev. Lett.}\ }\textbf {\bibinfo {volume} {105}},\
  \bibinfo {pages} {062001} (\bibinfo {year} {2010}{\natexlab{b}})},\ \Eprint
  {http://arxiv.org/abs/1003.0482} {arXiv:1003.0482 [hep-ph]} \BibitemShut
  {NoStop}%
\bibitem [{\citenamefont {Catani}\ \emph {et~al.}(2014)\citenamefont {Catani},
  \citenamefont {Grazzini},\ and\ \citenamefont {Torre}}]{Catani:2014qha}%
  \BibitemOpen
  \bibfield  {author} {\bibinfo {author} {\bibfnamefont {S.}~\bibnamefont
  {Catani}}, \bibinfo {author} {\bibfnamefont {M.}~\bibnamefont {Grazzini}}, \
  and\ \bibinfo {author} {\bibfnamefont {A.}~\bibnamefont {Torre}},\ }\href
  {\doibase 10.1016/j.nuclphysb.2014.11.019} {\bibfield  {journal} {\bibinfo
  {journal} {Nucl. Phys. B}\ }\textbf {\bibinfo {volume} {890}},\ \bibinfo
  {pages} {518} (\bibinfo {year} {2014})},\ \Eprint
  {http://arxiv.org/abs/1408.4564} {arXiv:1408.4564 [hep-ph]} \BibitemShut
  {NoStop}%
\bibitem [{\citenamefont {Catani}\ \emph {et~al.}(2017)\citenamefont {Catani},
  \citenamefont {Grazzini},\ and\ \citenamefont {Sargsyan}}]{Catani:2017tuc}%
  \BibitemOpen
  \bibfield  {author} {\bibinfo {author} {\bibfnamefont {S.}~\bibnamefont
  {Catani}}, \bibinfo {author} {\bibfnamefont {M.}~\bibnamefont {Grazzini}}, \
  and\ \bibinfo {author} {\bibfnamefont {H.}~\bibnamefont {Sargsyan}},\ }\href
  {\doibase 10.1007/JHEP06(2017)017} {\bibfield  {journal} {\bibinfo  {journal}
  {JHEP}\ }\textbf {\bibinfo {volume} {06}},\ \bibinfo {pages} {017} (\bibinfo
  {year} {2017})},\ \Eprint {http://arxiv.org/abs/1703.08468} {arXiv:1703.08468
  [hep-ph]} \BibitemShut {NoStop}%
\bibitem [{\citenamefont {Hatta}\ \emph {et~al.}(2022)\citenamefont {Hatta},
  \citenamefont {Xiao},\ and\ \citenamefont {Yuan}}]{Hatta:2022lzj}%
  \BibitemOpen
  \bibfield  {author} {\bibinfo {author} {\bibfnamefont {Y.}~\bibnamefont
  {Hatta}}, \bibinfo {author} {\bibfnamefont {B.-W.}\ \bibnamefont {Xiao}}, \
  and\ \bibinfo {author} {\bibfnamefont {F.}~\bibnamefont {Yuan}},\ }\href
  {\doibase 10.1103/PhysRevD.106.094015} {\bibfield  {journal} {\bibinfo
  {journal} {Phys. Rev. D}\ }\textbf {\bibinfo {volume} {106}},\ \bibinfo
  {pages} {094015} (\bibinfo {year} {2022})},\ \Eprint
  {http://arxiv.org/abs/2205.08060} {arXiv:2205.08060 [hep-ph]} \BibitemShut
  {NoStop}%
\bibitem [{\citenamefont {Berera}\ and\ \citenamefont
  {Soper}(1996)}]{Berera:1995fj}%
  \BibitemOpen
  \bibfield  {author} {\bibinfo {author} {\bibfnamefont {A.}~\bibnamefont
  {Berera}}\ and\ \bibinfo {author} {\bibfnamefont {D.~E.}\ \bibnamefont
  {Soper}},\ }\href {\doibase 10.1103/PhysRevD.53.6162} {\bibfield  {journal}
  {\bibinfo  {journal} {Phys. Rev. D}\ }\textbf {\bibinfo {volume} {53}},\
  \bibinfo {pages} {6162} (\bibinfo {year} {1996})},\ \Eprint
  {http://arxiv.org/abs/hep-ph/9509239} {arXiv:hep-ph/9509239} \BibitemShut
  {NoStop}%
\bibitem [{\citenamefont {Hebecker}(1997)}]{Hebecker:1997gp}%
  \BibitemOpen
  \bibfield  {author} {\bibinfo {author} {\bibfnamefont {A.}~\bibnamefont
  {Hebecker}},\ }\href {\doibase 10.1016/S0550-3213(97)00512-9} {\bibfield
  {journal} {\bibinfo  {journal} {Nucl. Phys. B}\ }\textbf {\bibinfo {volume}
  {505}},\ \bibinfo {pages} {349} (\bibinfo {year} {1997})},\ \Eprint
  {http://arxiv.org/abs/hep-ph/9702373} {arXiv:hep-ph/9702373} \BibitemShut
  {NoStop}%
\bibitem [{\citenamefont {Buchmuller}\ \emph {et~al.}(1999)\citenamefont
  {Buchmuller}, \citenamefont {Gehrmann},\ and\ \citenamefont
  {Hebecker}}]{Buchmuller:1998jv}%
  \BibitemOpen
  \bibfield  {author} {\bibinfo {author} {\bibfnamefont {W.}~\bibnamefont
  {Buchmuller}}, \bibinfo {author} {\bibfnamefont {T.}~\bibnamefont
  {Gehrmann}}, \ and\ \bibinfo {author} {\bibfnamefont {A.}~\bibnamefont
  {Hebecker}},\ }\href {\doibase 10.1016/S0550-3213(98)00682-8} {\bibfield
  {journal} {\bibinfo  {journal} {Nucl. Phys. B}\ }\textbf {\bibinfo {volume}
  {537}},\ \bibinfo {pages} {477} (\bibinfo {year} {1999})},\ \Eprint
  {http://arxiv.org/abs/hep-ph/9808454} {arXiv:hep-ph/9808454} \BibitemShut
  {NoStop}%
\bibitem [{\citenamefont {Golec-Biernat}\ and\ \citenamefont
  {Wusthoff}(1999)}]{Golec-Biernat:1999qor}%
  \BibitemOpen
  \bibfield  {author} {\bibinfo {author} {\bibfnamefont {K.~J.}\ \bibnamefont
  {Golec-Biernat}}\ and\ \bibinfo {author} {\bibfnamefont {M.}~\bibnamefont
  {Wusthoff}},\ }\href {\doibase 10.1103/PhysRevD.60.114023} {\bibfield
  {journal} {\bibinfo  {journal} {Phys. Rev. D}\ }\textbf {\bibinfo {volume}
  {60}},\ \bibinfo {pages} {114023} (\bibinfo {year} {1999})},\ \Eprint
  {http://arxiv.org/abs/hep-ph/9903358} {arXiv:hep-ph/9903358} \BibitemShut
  {NoStop}%
\bibitem [{\citenamefont {Hautmann}\ \emph {et~al.}(1999)\citenamefont
  {Hautmann}, \citenamefont {Kunszt},\ and\ \citenamefont
  {Soper}}]{Hautmann:1999ui}%
  \BibitemOpen
  \bibfield  {author} {\bibinfo {author} {\bibfnamefont {F.}~\bibnamefont
  {Hautmann}}, \bibinfo {author} {\bibfnamefont {Z.}~\bibnamefont {Kunszt}}, \
  and\ \bibinfo {author} {\bibfnamefont {D.~E.}\ \bibnamefont {Soper}},\ }\href
  {\doibase 10.1016/S0550-3213(99)00568-4} {\bibfield  {journal} {\bibinfo
  {journal} {Nucl. Phys. B}\ }\textbf {\bibinfo {volume} {563}},\ \bibinfo
  {pages} {153} (\bibinfo {year} {1999})},\ \Eprint
  {http://arxiv.org/abs/hep-ph/9906284} {arXiv:hep-ph/9906284} \BibitemShut
  {NoStop}%
\bibitem [{\citenamefont {Hautmann}\ and\ \citenamefont
  {Soper}(2001)}]{Hautmann:2000pw}%
  \BibitemOpen
  \bibfield  {author} {\bibinfo {author} {\bibfnamefont {F.}~\bibnamefont
  {Hautmann}}\ and\ \bibinfo {author} {\bibfnamefont {D.~E.}\ \bibnamefont
  {Soper}},\ }\href {\doibase 10.1103/PhysRevD.63.011501} {\bibfield  {journal}
  {\bibinfo  {journal} {Phys. Rev. D}\ }\textbf {\bibinfo {volume} {63}},\
  \bibinfo {pages} {011501} (\bibinfo {year} {2001})},\ \Eprint
  {http://arxiv.org/abs/hep-ph/0008224} {arXiv:hep-ph/0008224} \BibitemShut
  {NoStop}%
\bibitem [{\citenamefont {Golec-Biernat}\ and\ \citenamefont
  {Wusthoff}(2001)}]{Golec-Biernat:2001gyl}%
  \BibitemOpen
  \bibfield  {author} {\bibinfo {author} {\bibfnamefont {K.~J.}\ \bibnamefont
  {Golec-Biernat}}\ and\ \bibinfo {author} {\bibfnamefont {M.}~\bibnamefont
  {Wusthoff}},\ }\href {\doibase 10.1007/s100520100661} {\bibfield  {journal}
  {\bibinfo  {journal} {Eur. Phys. J. C}\ }\textbf {\bibinfo {volume} {20}},\
  \bibinfo {pages} {313} (\bibinfo {year} {2001})},\ \Eprint
  {http://arxiv.org/abs/hep-ph/0102093} {arXiv:hep-ph/0102093} \BibitemShut
  {NoStop}%
\bibitem [{\citenamefont {M\"antysaari}(2020)}]{Mantysaari:2020axf}%
  \BibitemOpen
  \bibfield  {author} {\bibinfo {author} {\bibfnamefont {H.}~\bibnamefont
  {M\"antysaari}},\ }\href {\doibase 10.1088/1361-6633/aba347} {\bibfield
  {journal} {\bibinfo  {journal} {Rept. Prog. Phys.}\ }\textbf {\bibinfo
  {volume} {83}},\ \bibinfo {pages} {082201} (\bibinfo {year} {2020})},\
  \Eprint {http://arxiv.org/abs/2001.10705} {arXiv:2001.10705 [hep-ph]}
  \BibitemShut {NoStop}%
\bibitem [{\citenamefont {Frankfurt}\ \emph {et~al.}(2022)\citenamefont
  {Frankfurt}, \citenamefont {Guzey}, \citenamefont {Stasto},\ and\
  \citenamefont {Strikman}}]{Frankfurt:2022jns}%
  \BibitemOpen
  \bibfield  {author} {\bibinfo {author} {\bibfnamefont {L.}~\bibnamefont
  {Frankfurt}}, \bibinfo {author} {\bibfnamefont {V.}~\bibnamefont {Guzey}},
  \bibinfo {author} {\bibfnamefont {A.}~\bibnamefont {Stasto}}, \ and\ \bibinfo
  {author} {\bibfnamefont {M.}~\bibnamefont {Strikman}},\ }\href {\doibase
  10.1088/1361-6633/ac8228} {\bibfield  {journal} {\bibinfo  {journal} {Rept.
  Prog. Phys.}\ }\textbf {\bibinfo {volume} {85}},\ \bibinfo {pages} {126301}
  (\bibinfo {year} {2022})},\ \Eprint {http://arxiv.org/abs/2203.12289}
  {arXiv:2203.12289 [hep-ph]} \BibitemShut {NoStop}%
\bibitem [{\citenamefont {Munier}\ and\ \citenamefont
  {Shoshi}(2004)}]{Munier:2003zb}%
  \BibitemOpen
  \bibfield  {author} {\bibinfo {author} {\bibfnamefont {S.}~\bibnamefont
  {Munier}}\ and\ \bibinfo {author} {\bibfnamefont {A.}~\bibnamefont
  {Shoshi}},\ }\href {\doibase 10.1103/PhysRevD.69.074022} {\bibfield
  {journal} {\bibinfo  {journal} {Phys. Rev. D}\ }\textbf {\bibinfo {volume}
  {69}},\ \bibinfo {pages} {074022} (\bibinfo {year} {2004})},\ \Eprint
  {http://arxiv.org/abs/hep-ph/0312022} {arXiv:hep-ph/0312022} \BibitemShut
  {NoStop}%
\bibitem [{\citenamefont {Marquet}(2007{\natexlab{b}})}]{Marquet:2007nf}%
  \BibitemOpen
  \bibfield  {author} {\bibinfo {author} {\bibfnamefont {C.}~\bibnamefont
  {Marquet}},\ }\href {\doibase 10.1103/PhysRevD.76.094017} {\bibfield
  {journal} {\bibinfo  {journal} {Phys. Rev. D}\ }\textbf {\bibinfo {volume}
  {76}},\ \bibinfo {pages} {094017} (\bibinfo {year} {2007}{\natexlab{b}})},\
  \Eprint {http://arxiv.org/abs/0706.2682} {arXiv:0706.2682 [hep-ph]}
  \BibitemShut {NoStop}%
\bibitem [{\citenamefont {Kowalski}\ \emph {et~al.}(2008)\citenamefont
  {Kowalski}, \citenamefont {Lappi}, \citenamefont {Marquet},\ and\
  \citenamefont {Venugopalan}}]{Kowalski:2008sa}%
  \BibitemOpen
  \bibfield  {author} {\bibinfo {author} {\bibfnamefont {H.}~\bibnamefont
  {Kowalski}}, \bibinfo {author} {\bibfnamefont {T.}~\bibnamefont {Lappi}},
  \bibinfo {author} {\bibfnamefont {C.}~\bibnamefont {Marquet}}, \ and\
  \bibinfo {author} {\bibfnamefont {R.}~\bibnamefont {Venugopalan}},\ }\href
  {\doibase 10.1103/PhysRevC.78.045201} {\bibfield  {journal} {\bibinfo
  {journal} {Phys. Rev. C}\ }\textbf {\bibinfo {volume} {78}},\ \bibinfo
  {pages} {045201} (\bibinfo {year} {2008})},\ \Eprint
  {http://arxiv.org/abs/0805.4071} {arXiv:0805.4071 [hep-ph]} \BibitemShut
  {NoStop}%
\bibitem [{\citenamefont {Kugeratski}\ \emph {et~al.}(2006)\citenamefont
  {Kugeratski}, \citenamefont {Goncalves},\ and\ \citenamefont
  {Navarra}}]{Kugeratski:2005ck}%
  \BibitemOpen
  \bibfield  {author} {\bibinfo {author} {\bibfnamefont {M.~S.}\ \bibnamefont
  {Kugeratski}}, \bibinfo {author} {\bibfnamefont {V.~P.}\ \bibnamefont
  {Goncalves}}, \ and\ \bibinfo {author} {\bibfnamefont {F.~S.}\ \bibnamefont
  {Navarra}},\ }\href {\doibase 10.1140/epjc/s2006-02517-7} {\bibfield
  {journal} {\bibinfo  {journal} {Eur. Phys. J. C}\ }\textbf {\bibinfo {volume}
  {46}},\ \bibinfo {pages} {413} (\bibinfo {year} {2006})},\ \Eprint
  {http://arxiv.org/abs/hep-ph/0511224} {arXiv:hep-ph/0511224} \BibitemShut
  {NoStop}%
\bibitem [{\citenamefont {Cazaroto}\ \emph {et~al.}(2009)\citenamefont
  {Cazaroto}, \citenamefont {Carvalho}, \citenamefont {Goncalves},\ and\
  \citenamefont {Navarra}}]{Cazaroto:2008iy}%
  \BibitemOpen
  \bibfield  {author} {\bibinfo {author} {\bibfnamefont {E.~R.}\ \bibnamefont
  {Cazaroto}}, \bibinfo {author} {\bibfnamefont {F.}~\bibnamefont {Carvalho}},
  \bibinfo {author} {\bibfnamefont {V.~P.}\ \bibnamefont {Goncalves}}, \ and\
  \bibinfo {author} {\bibfnamefont {F.~S.}\ \bibnamefont {Navarra}},\ }\href
  {\doibase 10.1016/j.physletb.2008.12.036} {\bibfield  {journal} {\bibinfo
  {journal} {Phys. Lett. B}\ }\textbf {\bibinfo {volume} {671}},\ \bibinfo
  {pages} {233} (\bibinfo {year} {2009})},\ \Eprint
  {http://arxiv.org/abs/0805.1255} {arXiv:0805.1255 [hep-ph]} \BibitemShut
  {NoStop}%
\bibitem [{\citenamefont {Kovchegov}\ and\ \citenamefont
  {Levin}(2000)}]{Kovchegov:1999ji}%
  \BibitemOpen
  \bibfield  {author} {\bibinfo {author} {\bibfnamefont {Y.~V.}\ \bibnamefont
  {Kovchegov}}\ and\ \bibinfo {author} {\bibfnamefont {E.}~\bibnamefont
  {Levin}},\ }\href {\doibase 10.1016/S0550-3213(00)00125-5} {\bibfield
  {journal} {\bibinfo  {journal} {Nucl. Phys. B}\ }\textbf {\bibinfo {volume}
  {577}},\ \bibinfo {pages} {221} (\bibinfo {year} {2000})},\ \Eprint
  {http://arxiv.org/abs/hep-ph/9911523} {arXiv:hep-ph/9911523} \BibitemShut
  {NoStop}%
\bibitem [{\citenamefont {Kovchegov}(2012)}]{Kovchegov:2011aa}%
  \BibitemOpen
  \bibfield  {author} {\bibinfo {author} {\bibfnamefont {Y.~V.}\ \bibnamefont
  {Kovchegov}},\ }\href {\doibase 10.1016/j.physletb.2012.02.073} {\bibfield
  {journal} {\bibinfo  {journal} {Phys. Lett. B}\ }\textbf {\bibinfo {volume}
  {710}},\ \bibinfo {pages} {192} (\bibinfo {year} {2012})},\ \Eprint
  {http://arxiv.org/abs/1112.2598} {arXiv:1112.2598 [hep-ph]} \BibitemShut
  {NoStop}%
\bibitem [{\citenamefont {Kovner}\ \emph {et~al.}(2006)\citenamefont {Kovner},
  \citenamefont {Lublinsky},\ and\ \citenamefont {Weigert}}]{Kovner:2006ge}%
  \BibitemOpen
  \bibfield  {author} {\bibinfo {author} {\bibfnamefont {A.}~\bibnamefont
  {Kovner}}, \bibinfo {author} {\bibfnamefont {M.}~\bibnamefont {Lublinsky}}, \
  and\ \bibinfo {author} {\bibfnamefont {H.}~\bibnamefont {Weigert}},\ }\href
  {\doibase 10.1103/PhysRevD.74.114023} {\bibfield  {journal} {\bibinfo
  {journal} {Phys. Rev. D}\ }\textbf {\bibinfo {volume} {74}},\ \bibinfo
  {pages} {114023} (\bibinfo {year} {2006})},\ \Eprint
  {http://arxiv.org/abs/hep-ph/0608258} {arXiv:hep-ph/0608258} \BibitemShut
  {NoStop}%
\bibitem [{\citenamefont {Lublinsky}(2014)}]{Lublinsky:2014bma}%
  \BibitemOpen
  \bibfield  {author} {\bibinfo {author} {\bibfnamefont {M.}~\bibnamefont
  {Lublinsky}},\ }\href {\doibase 10.1016/j.physletb.2014.06.037} {\bibfield
  {journal} {\bibinfo  {journal} {Phys. Lett. B}\ }\textbf {\bibinfo {volume}
  {735}},\ \bibinfo {pages} {200} (\bibinfo {year} {2014})},\ \Eprint
  {http://arxiv.org/abs/1404.0369} {arXiv:1404.0369 [hep-ph]} \BibitemShut
  {NoStop}%
\bibitem [{\citenamefont {Levin}\ and\ \citenamefont
  {Lublinsky}(2002{\natexlab{a}})}]{Levin:2001pr}%
  \BibitemOpen
  \bibfield  {author} {\bibinfo {author} {\bibfnamefont {E.}~\bibnamefont
  {Levin}}\ and\ \bibinfo {author} {\bibfnamefont {M.}~\bibnamefont
  {Lublinsky}},\ }\href {\doibase 10.1007/s100520100839} {\bibfield  {journal}
  {\bibinfo  {journal} {Eur. Phys. J. C}\ }\textbf {\bibinfo {volume} {22}},\
  \bibinfo {pages} {647} (\bibinfo {year} {2002}{\natexlab{a}})},\ \Eprint
  {http://arxiv.org/abs/hep-ph/0108239} {arXiv:hep-ph/0108239} \BibitemShut
  {NoStop}%
\bibitem [{\citenamefont {Levin}\ and\ \citenamefont
  {Lublinsky}(2001)}]{Levin:2001yv}%
  \BibitemOpen
  \bibfield  {author} {\bibinfo {author} {\bibfnamefont {E.}~\bibnamefont
  {Levin}}\ and\ \bibinfo {author} {\bibfnamefont {M.}~\bibnamefont
  {Lublinsky}},\ }\href {\doibase 10.1016/S0370-2693(01)01217-5} {\bibfield
  {journal} {\bibinfo  {journal} {Phys. Lett. B}\ }\textbf {\bibinfo {volume}
  {521}},\ \bibinfo {pages} {233} (\bibinfo {year} {2001})},\ \Eprint
  {http://arxiv.org/abs/hep-ph/0108265} {arXiv:hep-ph/0108265} \BibitemShut
  {NoStop}%
\bibitem [{\citenamefont {Levin}\ and\ \citenamefont
  {Lublinsky}(2002{\natexlab{b}})}]{Levin:2002fj}%
  \BibitemOpen
  \bibfield  {author} {\bibinfo {author} {\bibfnamefont {E.}~\bibnamefont
  {Levin}}\ and\ \bibinfo {author} {\bibfnamefont {M.}~\bibnamefont
  {Lublinsky}},\ }\href {\doibase 10.1016/S0375-9474(02)01269-1} {\bibfield
  {journal} {\bibinfo  {journal} {Nucl. Phys. A}\ }\textbf {\bibinfo {volume}
  {712}},\ \bibinfo {pages} {95} (\bibinfo {year} {2002}{\natexlab{b}})},\
  \Eprint {http://arxiv.org/abs/hep-ph/0207374} {arXiv:hep-ph/0207374}
  \BibitemShut {NoStop}%
\bibitem [{\citenamefont {Hatta}\ \emph {et~al.}(2006)\citenamefont {Hatta},
  \citenamefont {Iancu}, \citenamefont {Marquet}, \citenamefont {Soyez},\ and\
  \citenamefont {Triantafyllopoulos}}]{Hatta:2006hs}%
  \BibitemOpen
  \bibfield  {author} {\bibinfo {author} {\bibfnamefont {Y.}~\bibnamefont
  {Hatta}}, \bibinfo {author} {\bibfnamefont {E.}~\bibnamefont {Iancu}},
  \bibinfo {author} {\bibfnamefont {C.}~\bibnamefont {Marquet}}, \bibinfo
  {author} {\bibfnamefont {G.}~\bibnamefont {Soyez}}, \ and\ \bibinfo {author}
  {\bibfnamefont {D.~N.}\ \bibnamefont {Triantafyllopoulos}},\ }\href {\doibase
  10.1016/j.nuclphysa.2006.04.003} {\bibfield  {journal} {\bibinfo  {journal}
  {Nucl. Phys. A}\ }\textbf {\bibinfo {volume} {773}},\ \bibinfo {pages} {95}
  (\bibinfo {year} {2006})},\ \Eprint {http://arxiv.org/abs/hep-ph/0601150}
  {arXiv:hep-ph/0601150} \BibitemShut {NoStop}%
\bibitem [{\citenamefont {Hatta}(2006)}]{Hatta:2006zz}%
  \BibitemOpen
  \bibfield  {author} {\bibinfo {author} {\bibfnamefont {Y.}~\bibnamefont
  {Hatta}},\ }\href {\doibase 10.22323/1.035.0037} {\bibfield  {journal}
  {\bibinfo  {journal} {PoS}\ }\textbf {\bibinfo {volume} {DIFF2006}},\
  \bibinfo {pages} {037} (\bibinfo {year} {2006})}\BibitemShut {NoStop}%
\bibitem [{\citenamefont {Contreras}\ \emph {et~al.}(2018)\citenamefont
  {Contreras}, \citenamefont {Levin}, \citenamefont {Meneses},\ and\
  \citenamefont {Potashnikova}}]{Contreras:2018adl}%
  \BibitemOpen
  \bibfield  {author} {\bibinfo {author} {\bibfnamefont {C.}~\bibnamefont
  {Contreras}}, \bibinfo {author} {\bibfnamefont {E.}~\bibnamefont {Levin}},
  \bibinfo {author} {\bibfnamefont {R.}~\bibnamefont {Meneses}}, \ and\
  \bibinfo {author} {\bibfnamefont {I.}~\bibnamefont {Potashnikova}},\ }\href
  {\doibase 10.1140/epjc/s10052-018-6179-0} {\bibfield  {journal} {\bibinfo
  {journal} {Eur. Phys. J. C}\ }\textbf {\bibinfo {volume} {78}},\ \bibinfo
  {pages} {699} (\bibinfo {year} {2018})},\ \Eprint
  {http://arxiv.org/abs/1806.10468} {arXiv:1806.10468 [hep-ph]} \BibitemShut
  {NoStop}%
\bibitem [{\citenamefont {Bendova}\ \emph {et~al.}(2021)\citenamefont
  {Bendova}, \citenamefont {Cepila}, \citenamefont {Contreras}, \citenamefont
  {Gon\c{c}alves},\ and\ \citenamefont {Matas}}]{Bendova:2020hkp}%
  \BibitemOpen
  \bibfield  {author} {\bibinfo {author} {\bibfnamefont {D.}~\bibnamefont
  {Bendova}}, \bibinfo {author} {\bibfnamefont {J.}~\bibnamefont {Cepila}},
  \bibinfo {author} {\bibfnamefont {J.~G.}\ \bibnamefont {Contreras}}, \bibinfo
  {author} {\bibfnamefont {t.~V.~P.}\ \bibnamefont {Gon\c{c}alves}}, \ and\
  \bibinfo {author} {\bibfnamefont {M.}~\bibnamefont {Matas}},\ }\href
  {\doibase 10.1140/epjc/s10052-021-09006-x} {\bibfield  {journal} {\bibinfo
  {journal} {Eur. Phys. J. C}\ }\textbf {\bibinfo {volume} {81}},\ \bibinfo
  {pages} {211} (\bibinfo {year} {2021})},\ \Eprint
  {http://arxiv.org/abs/2009.14002} {arXiv:2009.14002 [hep-ph]} \BibitemShut
  {NoStop}%
\bibitem [{\citenamefont {Le}(2021)}]{Le:2021afn}%
  \BibitemOpen
  \bibfield  {author} {\bibinfo {author} {\bibfnamefont {A.~D.}\ \bibnamefont
  {Le}},\ }\href {\doibase 10.1103/PhysRevD.104.014014} {\bibfield  {journal}
  {\bibinfo  {journal} {Phys. Rev. D}\ }\textbf {\bibinfo {volume} {104}},\
  \bibinfo {pages} {014014} (\bibinfo {year} {2021})},\ \Eprint
  {http://arxiv.org/abs/2103.07724} {arXiv:2103.07724 [hep-ph]} \BibitemShut
  {NoStop}%
\bibitem [{\citenamefont {Beuf}\ \emph {et~al.}(2022)\citenamefont {Beuf},
  \citenamefont {H\"anninen}, \citenamefont {Lappi}, \citenamefont {Mulian},\
  and\ \citenamefont {M\"antysaari}}]{Beuf:2022kyp}%
  \BibitemOpen
  \bibfield  {author} {\bibinfo {author} {\bibfnamefont {G.}~\bibnamefont
  {Beuf}}, \bibinfo {author} {\bibfnamefont {H.}~\bibnamefont {H\"anninen}},
  \bibinfo {author} {\bibfnamefont {T.}~\bibnamefont {Lappi}}, \bibinfo
  {author} {\bibfnamefont {Y.}~\bibnamefont {Mulian}}, \ and\ \bibinfo {author}
  {\bibfnamefont {H.}~\bibnamefont {M\"antysaari}},\ }\href {\doibase
  10.1103/PhysRevD.106.094014} {\bibfield  {journal} {\bibinfo  {journal}
  {Phys. Rev. D}\ }\textbf {\bibinfo {volume} {106}},\ \bibinfo {pages}
  {094014} (\bibinfo {year} {2022})},\ \Eprint
  {http://arxiv.org/abs/2206.13161} {arXiv:2206.13161 [hep-ph]} \BibitemShut
  {NoStop}%
\bibitem [{\citenamefont {Lappi}\ \emph {et~al.}(2023)\citenamefont {Lappi},
  \citenamefont {Le},\ and\ \citenamefont {M\"antysaari}}]{Lappi:2023frf}%
  \BibitemOpen
  \bibfield  {author} {\bibinfo {author} {\bibfnamefont {T.}~\bibnamefont
  {Lappi}}, \bibinfo {author} {\bibfnamefont {A.~D.}\ \bibnamefont {Le}}, \
  and\ \bibinfo {author} {\bibfnamefont {H.}~\bibnamefont {M\"antysaari}},\
  }\href@noop {} {\  (\bibinfo {year} {2023})},\ \Eprint
  {http://arxiv.org/abs/2307.16486} {arXiv:2307.16486 [hep-ph]} \BibitemShut
  {NoStop}%
\bibitem [{\citenamefont {Singh}\ and\ \citenamefont
  {Toll}(2023)}]{Singh:2023yvj}%
  \BibitemOpen
  \bibfield  {author} {\bibinfo {author} {\bibfnamefont {J.}~\bibnamefont
  {Singh}}\ and\ \bibinfo {author} {\bibfnamefont {T.}~\bibnamefont {Toll}},\
  }\href {\doibase 10.1016/j.cpc.2023.108872} {\bibfield  {journal} {\bibinfo
  {journal} {Comput. Phys. Commun.}\ }\textbf {\bibinfo {volume} {292}},\
  \bibinfo {pages} {108872} (\bibinfo {year} {2023})},\ \Eprint
  {http://arxiv.org/abs/2305.15880} {arXiv:2305.15880 [hep-ph]} \BibitemShut
  {NoStop}%
\bibitem [{\citenamefont {M\"antysaari}\ and\ \citenamefont
  {Schenke}(2016{\natexlab{a}})}]{Mantysaari:2016jaz}%
  \BibitemOpen
  \bibfield  {author} {\bibinfo {author} {\bibfnamefont {H.}~\bibnamefont
  {M\"antysaari}}\ and\ \bibinfo {author} {\bibfnamefont {B.}~\bibnamefont
  {Schenke}},\ }\href {\doibase 10.1103/PhysRevD.94.034042} {\bibfield
  {journal} {\bibinfo  {journal} {Phys. Rev. D}\ }\textbf {\bibinfo {volume}
  {94}},\ \bibinfo {pages} {034042} (\bibinfo {year} {2016}{\natexlab{a}})},\
  \Eprint {http://arxiv.org/abs/1607.01711} {arXiv:1607.01711 [hep-ph]}
  \BibitemShut {NoStop}%
\bibitem [{\citenamefont {M\"antysaari}\ and\ \citenamefont
  {Schenke}(2016{\natexlab{b}})}]{Mantysaari:2016ykx}%
  \BibitemOpen
  \bibfield  {author} {\bibinfo {author} {\bibfnamefont {H.}~\bibnamefont
  {M\"antysaari}}\ and\ \bibinfo {author} {\bibfnamefont {B.}~\bibnamefont
  {Schenke}},\ }\href {\doibase 10.1103/PhysRevLett.117.052301} {\bibfield
  {journal} {\bibinfo  {journal} {Phys. Rev. Lett.}\ }\textbf {\bibinfo
  {volume} {117}},\ \bibinfo {pages} {052301} (\bibinfo {year}
  {2016}{\natexlab{b}})},\ \Eprint {http://arxiv.org/abs/1603.04349}
  {arXiv:1603.04349 [hep-ph]} \BibitemShut {NoStop}%
\bibitem [{\citenamefont {M\"antysaari}\ \emph {et~al.}(2021)\citenamefont
  {M\"antysaari}, \citenamefont {Roy}, \citenamefont {Salazar},\ and\
  \citenamefont {Schenke}}]{Mantysaari:2020lhf}%
  \BibitemOpen
  \bibfield  {author} {\bibinfo {author} {\bibfnamefont {H.}~\bibnamefont
  {M\"antysaari}}, \bibinfo {author} {\bibfnamefont {K.}~\bibnamefont {Roy}},
  \bibinfo {author} {\bibfnamefont {F.}~\bibnamefont {Salazar}}, \ and\
  \bibinfo {author} {\bibfnamefont {B.}~\bibnamefont {Schenke}},\ }\href
  {\doibase 10.1103/PhysRevD.103.094026} {\bibfield  {journal} {\bibinfo
  {journal} {Phys. Rev. D}\ }\textbf {\bibinfo {volume} {103}},\ \bibinfo
  {pages} {094026} (\bibinfo {year} {2021})},\ \Eprint
  {http://arxiv.org/abs/2011.02464} {arXiv:2011.02464 [hep-ph]} \BibitemShut
  {NoStop}%
\bibitem [{\citenamefont {M\"antysaari}\ and\ \citenamefont
  {Penttala}(2022{\natexlab{a}})}]{Mantysaari:2022kdm}%
  \BibitemOpen
  \bibfield  {author} {\bibinfo {author} {\bibfnamefont {H.}~\bibnamefont
  {M\"antysaari}}\ and\ \bibinfo {author} {\bibfnamefont {J.}~\bibnamefont
  {Penttala}},\ }\href {\doibase 10.1007/JHEP08(2022)247} {\bibfield  {journal}
  {\bibinfo  {journal} {JHEP}\ }\textbf {\bibinfo {volume} {08}},\ \bibinfo
  {pages} {247} (\bibinfo {year} {2022}{\natexlab{a}})},\ \Eprint
  {http://arxiv.org/abs/2204.14031} {arXiv:2204.14031 [hep-ph]} \BibitemShut
  {NoStop}%
\bibitem [{\citenamefont {Kumar}\ and\ \citenamefont
  {Toll}(2022)}]{Kumar:2022aly}%
  \BibitemOpen
  \bibfield  {author} {\bibinfo {author} {\bibfnamefont {A.}~\bibnamefont
  {Kumar}}\ and\ \bibinfo {author} {\bibfnamefont {T.}~\bibnamefont {Toll}},\
  }\href {\doibase 10.1103/PhysRevD.105.114011} {\bibfield  {journal} {\bibinfo
   {journal} {Phys. Rev. D}\ }\textbf {\bibinfo {volume} {105}},\ \bibinfo
  {pages} {114011} (\bibinfo {year} {2022})},\ \Eprint
  {http://arxiv.org/abs/2202.06631} {arXiv:2202.06631 [hep-ph]} \BibitemShut
  {NoStop}%
\bibitem [{\citenamefont {Demirci}\ \emph {et~al.}(2022)\citenamefont
  {Demirci}, \citenamefont {Lappi},\ and\ \citenamefont
  {Schlichting}}]{Demirci:2022wuy}%
  \BibitemOpen
  \bibfield  {author} {\bibinfo {author} {\bibfnamefont {S.}~\bibnamefont
  {Demirci}}, \bibinfo {author} {\bibfnamefont {T.}~\bibnamefont {Lappi}}, \
  and\ \bibinfo {author} {\bibfnamefont {S.}~\bibnamefont {Schlichting}},\
  }\href {\doibase 10.1103/PhysRevD.106.074025} {\bibfield  {journal} {\bibinfo
   {journal} {Phys. Rev. D}\ }\textbf {\bibinfo {volume} {106}},\ \bibinfo
  {pages} {074025} (\bibinfo {year} {2022})},\ \Eprint
  {http://arxiv.org/abs/2206.05207} {arXiv:2206.05207 [hep-ph]} \BibitemShut
  {NoStop}%
\bibitem [{\citenamefont {M\"antysaari}\ and\ \citenamefont
  {Penttala}(2022{\natexlab{b}})}]{Mantysaari:2022bsp}%
  \BibitemOpen
  \bibfield  {author} {\bibinfo {author} {\bibfnamefont {H.}~\bibnamefont
  {M\"antysaari}}\ and\ \bibinfo {author} {\bibfnamefont {J.}~\bibnamefont
  {Penttala}},\ }\href {\doibase 10.1103/PhysRevD.105.114038} {\bibfield
  {journal} {\bibinfo  {journal} {Phys. Rev. D}\ }\textbf {\bibinfo {volume}
  {105}},\ \bibinfo {pages} {114038} (\bibinfo {year} {2022}{\natexlab{b}})},\
  \Eprint {http://arxiv.org/abs/2203.16911} {arXiv:2203.16911 [hep-ph]}
  \BibitemShut {NoStop}%
\bibitem [{\citenamefont {M\"antysaari}\ \emph {et~al.}(2023)\citenamefont
  {M\"antysaari}, \citenamefont {Schenke}, \citenamefont {Shen},\ and\
  \citenamefont {Zhao}}]{Mantysaari:2023qsq}%
  \BibitemOpen
  \bibfield  {author} {\bibinfo {author} {\bibfnamefont {H.}~\bibnamefont
  {M\"antysaari}}, \bibinfo {author} {\bibfnamefont {B.}~\bibnamefont
  {Schenke}}, \bibinfo {author} {\bibfnamefont {C.}~\bibnamefont {Shen}}, \
  and\ \bibinfo {author} {\bibfnamefont {W.}~\bibnamefont {Zhao}},\ }\href
  {\doibase 10.1103/PhysRevLett.131.062301} {\bibfield  {journal} {\bibinfo
  {journal} {Phys. Rev. Lett.}\ }\textbf {\bibinfo {volume} {131}},\ \bibinfo
  {pages} {062301} (\bibinfo {year} {2023})},\ \Eprint
  {http://arxiv.org/abs/2303.04866} {arXiv:2303.04866 [nucl-th]} \BibitemShut
  {NoStop}%
\bibitem [{\citenamefont {McLerran}\ and\ \citenamefont
  {Venugopalan}(1999)}]{McLerran:1998nk}%
  \BibitemOpen
  \bibfield  {author} {\bibinfo {author} {\bibfnamefont {L.~D.}\ \bibnamefont
  {McLerran}}\ and\ \bibinfo {author} {\bibfnamefont {R.}~\bibnamefont
  {Venugopalan}},\ }\href {\doibase 10.1103/PhysRevD.59.094002} {\bibfield
  {journal} {\bibinfo  {journal} {Phys. Rev. D}\ }\textbf {\bibinfo {volume}
  {59}},\ \bibinfo {pages} {094002} (\bibinfo {year} {1999})},\ \Eprint
  {http://arxiv.org/abs/hep-ph/9809427} {arXiv:hep-ph/9809427} \BibitemShut
  {NoStop}%
\bibitem [{\citenamefont {Venugopalan}(1999)}]{Venugopalan:1999wu}%
  \BibitemOpen
  \bibfield  {author} {\bibinfo {author} {\bibfnamefont {R.}~\bibnamefont
  {Venugopalan}},\ }\href@noop {} {\bibfield  {journal} {\bibinfo  {journal}
  {Acta Phys. Polon. B}\ }\textbf {\bibinfo {volume} {30}},\ \bibinfo {pages}
  {3731} (\bibinfo {year} {1999})},\ \Eprint
  {http://arxiv.org/abs/hep-ph/9911371} {arXiv:hep-ph/9911371} \BibitemShut
  {NoStop}%
\bibitem [{\citenamefont {Mueller}(1999)}]{Mueller:1999wm}%
  \BibitemOpen
  \bibfield  {author} {\bibinfo {author} {\bibfnamefont {A.~H.}\ \bibnamefont
  {Mueller}},\ }\href {\doibase 10.1016/S0550-3213(99)00394-6} {\bibfield
  {journal} {\bibinfo  {journal} {Nucl. Phys. B}\ }\textbf {\bibinfo {volume}
  {558}},\ \bibinfo {pages} {285} (\bibinfo {year} {1999})},\ \Eprint
  {http://arxiv.org/abs/hep-ph/9904404} {arXiv:hep-ph/9904404} \BibitemShut
  {NoStop}%
\bibitem [{\citenamefont {Xiao}\ \emph {et~al.}(2017)\citenamefont {Xiao},
  \citenamefont {Yuan},\ and\ \citenamefont {Zhou}}]{Xiao:2017yya}%
  \BibitemOpen
  \bibfield  {author} {\bibinfo {author} {\bibfnamefont {B.-W.}\ \bibnamefont
  {Xiao}}, \bibinfo {author} {\bibfnamefont {F.}~\bibnamefont {Yuan}}, \ and\
  \bibinfo {author} {\bibfnamefont {J.}~\bibnamefont {Zhou}},\ }\href {\doibase
  10.1016/j.nuclphysb.2017.05.012} {\bibfield  {journal} {\bibinfo  {journal}
  {Nucl. Phys. B}\ }\textbf {\bibinfo {volume} {921}},\ \bibinfo {pages} {104}
  (\bibinfo {year} {2017})},\ \Eprint {http://arxiv.org/abs/1703.06163}
  {arXiv:1703.06163 [hep-ph]} \BibitemShut {NoStop}%
\bibitem [{\citenamefont {Chirilli}\ \emph
  {et~al.}(2012{\natexlab{a}})\citenamefont {Chirilli}, \citenamefont {Xiao},\
  and\ \citenamefont {Yuan}}]{Chirilli:2011km}%
  \BibitemOpen
  \bibfield  {author} {\bibinfo {author} {\bibfnamefont {G.~A.}\ \bibnamefont
  {Chirilli}}, \bibinfo {author} {\bibfnamefont {B.-W.}\ \bibnamefont {Xiao}},
  \ and\ \bibinfo {author} {\bibfnamefont {F.}~\bibnamefont {Yuan}},\ }\href
  {\doibase 10.1103/PhysRevLett.108.122301} {\bibfield  {journal} {\bibinfo
  {journal} {Phys. Rev. Lett.}\ }\textbf {\bibinfo {volume} {108}},\ \bibinfo
  {pages} {122301} (\bibinfo {year} {2012}{\natexlab{a}})},\ \Eprint
  {http://arxiv.org/abs/1112.1061} {arXiv:1112.1061 [hep-ph]} \BibitemShut
  {NoStop}%
\bibitem [{\citenamefont {Chirilli}\ \emph
  {et~al.}(2012{\natexlab{b}})\citenamefont {Chirilli}, \citenamefont {Xiao},\
  and\ \citenamefont {Yuan}}]{Chirilli:2012jd}%
  \BibitemOpen
  \bibfield  {author} {\bibinfo {author} {\bibfnamefont {G.~A.}\ \bibnamefont
  {Chirilli}}, \bibinfo {author} {\bibfnamefont {B.-W.}\ \bibnamefont {Xiao}},
  \ and\ \bibinfo {author} {\bibfnamefont {F.}~\bibnamefont {Yuan}},\ }\href
  {\doibase 10.1103/PhysRevD.86.054005} {\bibfield  {journal} {\bibinfo
  {journal} {Phys. Rev. D}\ }\textbf {\bibinfo {volume} {86}},\ \bibinfo
  {pages} {054005} (\bibinfo {year} {2012}{\natexlab{b}})},\ \Eprint
  {http://arxiv.org/abs/1203.6139} {arXiv:1203.6139 [hep-ph]} \BibitemShut
  {NoStop}%
\bibitem [{\citenamefont {Parisi}\ and\ \citenamefont
  {Petronzio}(1979)}]{Parisi:1979se}%
  \BibitemOpen
  \bibfield  {author} {\bibinfo {author} {\bibfnamefont {G.}~\bibnamefont
  {Parisi}}\ and\ \bibinfo {author} {\bibfnamefont {R.}~\bibnamefont
  {Petronzio}},\ }\href {\doibase 10.1016/0550-3213(79)90040-3} {\bibfield
  {journal} {\bibinfo  {journal} {Nucl. Phys. B}\ }\textbf {\bibinfo {volume}
  {154}},\ \bibinfo {pages} {427} (\bibinfo {year} {1979})}\BibitemShut
  {NoStop}%
\bibitem [{\citenamefont {Collins}\ and\ \citenamefont
  {Soper}(1982)}]{Collins:1981va}%
  \BibitemOpen
  \bibfield  {author} {\bibinfo {author} {\bibfnamefont {J.~C.}\ \bibnamefont
  {Collins}}\ and\ \bibinfo {author} {\bibfnamefont {D.~E.}\ \bibnamefont
  {Soper}},\ }\href {\doibase 10.1016/0550-3213(82)90453-9} {\bibfield
  {journal} {\bibinfo  {journal} {Nucl. Phys. B}\ }\textbf {\bibinfo {volume}
  {197}},\ \bibinfo {pages} {446} (\bibinfo {year} {1982})}\BibitemShut
  {NoStop}%
\bibitem [{\citenamefont {Collins}\ \emph {et~al.}(1985)\citenamefont
  {Collins}, \citenamefont {Soper},\ and\ \citenamefont
  {Sterman}}]{Collins:1984kg}%
  \BibitemOpen
  \bibfield  {author} {\bibinfo {author} {\bibfnamefont {J.~C.}\ \bibnamefont
  {Collins}}, \bibinfo {author} {\bibfnamefont {D.~E.}\ \bibnamefont {Soper}},
  \ and\ \bibinfo {author} {\bibfnamefont {G.~F.}\ \bibnamefont {Sterman}},\
  }\href {\doibase 10.1016/0550-3213(85)90479-1} {\bibfield  {journal}
  {\bibinfo  {journal} {Nucl. Phys. B}\ }\textbf {\bibinfo {volume} {250}},\
  \bibinfo {pages} {199} (\bibinfo {year} {1985})}\BibitemShut {NoStop}%
\bibitem [{\citenamefont {Shi}\ \emph {et~al.}(2022)\citenamefont {Shi},
  \citenamefont {Wang}, \citenamefont {Wei},\ and\ \citenamefont
  {Xiao}}]{Shi:2021hwx}%
  \BibitemOpen
  \bibfield  {author} {\bibinfo {author} {\bibfnamefont {Y.}~\bibnamefont
  {Shi}}, \bibinfo {author} {\bibfnamefont {L.}~\bibnamefont {Wang}}, \bibinfo
  {author} {\bibfnamefont {S.-Y.}\ \bibnamefont {Wei}}, \ and\ \bibinfo
  {author} {\bibfnamefont {B.-W.}\ \bibnamefont {Xiao}},\ }\href {\doibase
  10.1103/PhysRevLett.128.202302} {\bibfield  {journal} {\bibinfo  {journal}
  {Phys. Rev. Lett.}\ }\textbf {\bibinfo {volume} {128}},\ \bibinfo {pages}
  {202302} (\bibinfo {year} {2022})},\ \Eprint
  {http://arxiv.org/abs/2112.06975} {arXiv:2112.06975 [hep-ph]} \BibitemShut
  {NoStop}%
\bibitem [{\citenamefont {Liu}\ \emph {et~al.}(2021{\natexlab{a}})\citenamefont
  {Liu}, \citenamefont {Melnitchouk}, \citenamefont {Qiu},\ and\ \citenamefont
  {Sato}}]{Liu:2020rvc}%
  \BibitemOpen
  \bibfield  {author} {\bibinfo {author} {\bibfnamefont {T.}~\bibnamefont
  {Liu}}, \bibinfo {author} {\bibfnamefont {W.}~\bibnamefont {Melnitchouk}},
  \bibinfo {author} {\bibfnamefont {J.-W.}\ \bibnamefont {Qiu}}, \ and\
  \bibinfo {author} {\bibfnamefont {N.}~\bibnamefont {Sato}},\ }\href {\doibase
  10.1103/PhysRevD.104.094033} {\bibfield  {journal} {\bibinfo  {journal}
  {Phys. Rev. D}\ }\textbf {\bibinfo {volume} {104}},\ \bibinfo {pages}
  {094033} (\bibinfo {year} {2021}{\natexlab{a}})},\ \Eprint
  {http://arxiv.org/abs/2008.02895} {arXiv:2008.02895 [hep-ph]} \BibitemShut
  {NoStop}%
\bibitem [{\citenamefont {Liu}\ \emph {et~al.}(2021{\natexlab{b}})\citenamefont
  {Liu}, \citenamefont {Melnitchouk}, \citenamefont {Qiu},\ and\ \citenamefont
  {Sato}}]{Liu:2021jfp}%
  \BibitemOpen
  \bibfield  {author} {\bibinfo {author} {\bibfnamefont {T.}~\bibnamefont
  {Liu}}, \bibinfo {author} {\bibfnamefont {W.}~\bibnamefont {Melnitchouk}},
  \bibinfo {author} {\bibfnamefont {J.-W.}\ \bibnamefont {Qiu}}, \ and\
  \bibinfo {author} {\bibfnamefont {N.}~\bibnamefont {Sato}},\ }\href {\doibase
  10.1007/JHEP11(2021)157} {\bibfield  {journal} {\bibinfo  {journal} {JHEP}\
  }\textbf {\bibinfo {volume} {11}},\ \bibinfo {pages} {157} (\bibinfo {year}
  {2021}{\natexlab{b}})},\ \Eprint {http://arxiv.org/abs/2108.13371}
  {arXiv:2108.13371 [hep-ph]} \BibitemShut {NoStop}%
\bibitem [{\citenamefont {Hou}\ \emph {et~al.}(2021)\citenamefont {Hou} \emph
  {et~al.}}]{Hou:2019efy}%
  \BibitemOpen
  \bibfield  {author} {\bibinfo {author} {\bibfnamefont {T.-J.}\ \bibnamefont
  {Hou}} \emph {et~al.},\ }\href {\doibase 10.1103/PhysRevD.103.014013}
  {\bibfield  {journal} {\bibinfo  {journal} {Phys. Rev. D}\ }\textbf {\bibinfo
  {volume} {103}},\ \bibinfo {pages} {014013} (\bibinfo {year} {2021})},\
  \Eprint {http://arxiv.org/abs/1912.10053} {arXiv:1912.10053 [hep-ph]}
  \BibitemShut {NoStop}%
\bibitem [{\citenamefont {Eskola}\ \emph {et~al.}(2022)\citenamefont {Eskola},
  \citenamefont {Paakkinen}, \citenamefont {Paukkunen},\ and\ \citenamefont
  {Salgado}}]{Eskola:2021nhw}%
  \BibitemOpen
  \bibfield  {author} {\bibinfo {author} {\bibfnamefont {K.~J.}\ \bibnamefont
  {Eskola}}, \bibinfo {author} {\bibfnamefont {P.}~\bibnamefont {Paakkinen}},
  \bibinfo {author} {\bibfnamefont {H.}~\bibnamefont {Paukkunen}}, \ and\
  \bibinfo {author} {\bibfnamefont {C.~A.}\ \bibnamefont {Salgado}},\ }\href
  {\doibase 10.1140/epjc/s10052-022-10359-0} {\bibfield  {journal} {\bibinfo
  {journal} {Eur. Phys. J. C}\ }\textbf {\bibinfo {volume} {82}},\ \bibinfo
  {pages} {413} (\bibinfo {year} {2022})},\ \Eprint
  {http://arxiv.org/abs/2112.12462} {arXiv:2112.12462 [hep-ph]} \BibitemShut
  {NoStop}%
\bibitem [{\citenamefont {Sun}\ \emph {et~al.}(2018)\citenamefont {Sun},
  \citenamefont {Isaacson}, \citenamefont {Yuan},\ and\ \citenamefont
  {Yuan}}]{Sun:2014dqm}%
  \BibitemOpen
  \bibfield  {author} {\bibinfo {author} {\bibfnamefont {P.}~\bibnamefont
  {Sun}}, \bibinfo {author} {\bibfnamefont {J.}~\bibnamefont {Isaacson}},
  \bibinfo {author} {\bibfnamefont {C.~P.}\ \bibnamefont {Yuan}}, \ and\
  \bibinfo {author} {\bibfnamefont {F.}~\bibnamefont {Yuan}},\ }\href {\doibase
  10.1142/S0217751X18410063} {\bibfield  {journal} {\bibinfo  {journal} {Int.
  J. Mod. Phys. A}\ }\textbf {\bibinfo {volume} {33}},\ \bibinfo {pages}
  {1841006} (\bibinfo {year} {2018})},\ \Eprint
  {http://arxiv.org/abs/1406.3073} {arXiv:1406.3073 [hep-ph]} \BibitemShut
  {NoStop}%
\bibitem [{\citenamefont {Prokudin}\ \emph {et~al.}(2015)\citenamefont
  {Prokudin}, \citenamefont {Sun},\ and\ \citenamefont
  {Yuan}}]{Prokudin:2015ysa}%
  \BibitemOpen
  \bibfield  {author} {\bibinfo {author} {\bibfnamefont {A.}~\bibnamefont
  {Prokudin}}, \bibinfo {author} {\bibfnamefont {P.}~\bibnamefont {Sun}}, \
  and\ \bibinfo {author} {\bibfnamefont {F.}~\bibnamefont {Yuan}},\ }\href
  {\doibase 10.1016/j.physletb.2015.09.064} {\bibfield  {journal} {\bibinfo
  {journal} {Phys. Lett. B}\ }\textbf {\bibinfo {volume} {750}},\ \bibinfo
  {pages} {533} (\bibinfo {year} {2015})},\ \Eprint
  {http://arxiv.org/abs/1505.05588} {arXiv:1505.05588 [hep-ph]} \BibitemShut
  {NoStop}%
\bibitem [{\citenamefont {Golec-Biernat}\ and\ \citenamefont
  {Wusthoff}(1998)}]{Golec-Biernat:1998zce}%
  \BibitemOpen
  \bibfield  {author} {\bibinfo {author} {\bibfnamefont {K.~J.}\ \bibnamefont
  {Golec-Biernat}}\ and\ \bibinfo {author} {\bibfnamefont {M.}~\bibnamefont
  {Wusthoff}},\ }\href {\doibase 10.1103/PhysRevD.59.014017} {\bibfield
  {journal} {\bibinfo  {journal} {Phys. Rev. D}\ }\textbf {\bibinfo {volume}
  {59}},\ \bibinfo {pages} {014017} (\bibinfo {year} {1998})},\ \Eprint
  {http://arxiv.org/abs/hep-ph/9807513} {arXiv:hep-ph/9807513} \BibitemShut
  {NoStop}%
\bibitem [{\citenamefont {Balitsky}(1996)}]{Balitsky:1995ub}%
  \BibitemOpen
  \bibfield  {author} {\bibinfo {author} {\bibfnamefont {I.}~\bibnamefont
  {Balitsky}},\ }\href {\doibase 10.1016/0550-3213(95)00638-9} {\bibfield
  {journal} {\bibinfo  {journal} {Nucl. Phys. B}\ }\textbf {\bibinfo {volume}
  {463}},\ \bibinfo {pages} {99} (\bibinfo {year} {1996})},\ \Eprint
  {http://arxiv.org/abs/hep-ph/9509348} {arXiv:hep-ph/9509348} \BibitemShut
  {NoStop}%
\bibitem [{\citenamefont {Kovchegov}\ and\ \citenamefont
  {Weigert}(2007{\natexlab{a}})}]{Kovchegov:2006wf}%
  \BibitemOpen
  \bibfield  {author} {\bibinfo {author} {\bibfnamefont {Y.~V.}\ \bibnamefont
  {Kovchegov}}\ and\ \bibinfo {author} {\bibfnamefont {H.}~\bibnamefont
  {Weigert}},\ }\href {\doibase 10.1016/j.nuclphysa.2007.03.008} {\bibfield
  {journal} {\bibinfo  {journal} {Nucl. Phys. A}\ }\textbf {\bibinfo {volume}
  {789}},\ \bibinfo {pages} {260} (\bibinfo {year} {2007}{\natexlab{a}})},\
  \Eprint {http://arxiv.org/abs/hep-ph/0612071} {arXiv:hep-ph/0612071}
  \BibitemShut {NoStop}%
\bibitem [{\citenamefont {Kovchegov}(1999)}]{Kovchegov:1999yj}%
  \BibitemOpen
  \bibfield  {author} {\bibinfo {author} {\bibfnamefont {Y.~V.}\ \bibnamefont
  {Kovchegov}},\ }\href {\doibase 10.1103/PhysRevD.60.034008} {\bibfield
  {journal} {\bibinfo  {journal} {Phys. Rev. D}\ }\textbf {\bibinfo {volume}
  {60}},\ \bibinfo {pages} {034008} (\bibinfo {year} {1999})},\ \Eprint
  {http://arxiv.org/abs/hep-ph/9901281} {arXiv:hep-ph/9901281} \BibitemShut
  {NoStop}%
\bibitem [{\citenamefont {Kovchegov}\ and\ \citenamefont
  {Weigert}(2007{\natexlab{b}})}]{Kovchegov:2006vj}%
  \BibitemOpen
  \bibfield  {author} {\bibinfo {author} {\bibfnamefont {Y.~V.}\ \bibnamefont
  {Kovchegov}}\ and\ \bibinfo {author} {\bibfnamefont {H.}~\bibnamefont
  {Weigert}},\ }\href {\doibase 10.1016/j.nuclphysa.2006.10.075} {\bibfield
  {journal} {\bibinfo  {journal} {Nucl. Phys. A}\ }\textbf {\bibinfo {volume}
  {784}},\ \bibinfo {pages} {188} (\bibinfo {year} {2007}{\natexlab{b}})},\
  \Eprint {http://arxiv.org/abs/hep-ph/0609090} {arXiv:hep-ph/0609090}
  \BibitemShut {NoStop}%
\bibitem [{\citenamefont {Albacete}\ \emph {et~al.}(2011)\citenamefont
  {Albacete}, \citenamefont {Armesto}, \citenamefont {Milhano}, \citenamefont
  {Quiroga-Arias},\ and\ \citenamefont {Salgado}}]{Albacete:2010sy}%
  \BibitemOpen
  \bibfield  {author} {\bibinfo {author} {\bibfnamefont {J.~L.}\ \bibnamefont
  {Albacete}}, \bibinfo {author} {\bibfnamefont {N.}~\bibnamefont {Armesto}},
  \bibinfo {author} {\bibfnamefont {J.~G.}\ \bibnamefont {Milhano}}, \bibinfo
  {author} {\bibfnamefont {P.}~\bibnamefont {Quiroga-Arias}}, \ and\ \bibinfo
  {author} {\bibfnamefont {C.~A.}\ \bibnamefont {Salgado}},\ }\href {\doibase
  10.1140/epjc/s10052-011-1705-3} {\bibfield  {journal} {\bibinfo  {journal}
  {Eur. Phys. J. C}\ }\textbf {\bibinfo {volume} {71}},\ \bibinfo {pages}
  {1705} (\bibinfo {year} {2011})},\ \Eprint {http://arxiv.org/abs/1012.4408}
  {arXiv:1012.4408 [hep-ph]} \BibitemShut {NoStop}%
\bibitem [{\citenamefont {Golec-Biernat}\ \emph {et~al.}(2002)\citenamefont
  {Golec-Biernat}, \citenamefont {Motyka},\ and\ \citenamefont
  {Stasto}}]{Golec-Biernat:2001dqn}%
  \BibitemOpen
  \bibfield  {author} {\bibinfo {author} {\bibfnamefont {K.~J.}\ \bibnamefont
  {Golec-Biernat}}, \bibinfo {author} {\bibfnamefont {L.}~\bibnamefont
  {Motyka}}, \ and\ \bibinfo {author} {\bibfnamefont {A.~M.}\ \bibnamefont
  {Stasto}},\ }\href {\doibase 10.1103/PhysRevD.65.074037} {\bibfield
  {journal} {\bibinfo  {journal} {Phys. Rev. D}\ }\textbf {\bibinfo {volume}
  {65}},\ \bibinfo {pages} {074037} (\bibinfo {year} {2002})},\ \Eprint
  {http://arxiv.org/abs/hep-ph/0110325} {arXiv:hep-ph/0110325} \BibitemShut
  {NoStop}%
\bibitem [{\citenamefont {Albacete}\ and\ \citenamefont
  {Kovchegov}(2007)}]{Albacete:2007yr}%
  \BibitemOpen
  \bibfield  {author} {\bibinfo {author} {\bibfnamefont {J.~L.}\ \bibnamefont
  {Albacete}}\ and\ \bibinfo {author} {\bibfnamefont {Y.~V.}\ \bibnamefont
  {Kovchegov}},\ }\href {\doibase 10.1103/PhysRevD.75.125021} {\bibfield
  {journal} {\bibinfo  {journal} {Phys. Rev. D}\ }\textbf {\bibinfo {volume}
  {75}},\ \bibinfo {pages} {125021} (\bibinfo {year} {2007})},\ \Eprint
  {http://arxiv.org/abs/0704.0612} {arXiv:0704.0612 [hep-ph]} \BibitemShut
  {NoStop}%
\bibitem [{\citenamefont {Balitsky}(2007)}]{Balitsky:2006wa}%
  \BibitemOpen
  \bibfield  {author} {\bibinfo {author} {\bibfnamefont {I.}~\bibnamefont
  {Balitsky}},\ }\href {\doibase 10.1103/PhysRevD.75.014001} {\bibfield
  {journal} {\bibinfo  {journal} {Phys. Rev. D}\ }\textbf {\bibinfo {volume}
  {75}},\ \bibinfo {pages} {014001} (\bibinfo {year} {2007})},\ \Eprint
  {http://arxiv.org/abs/hep-ph/0609105} {arXiv:hep-ph/0609105} \BibitemShut
  {NoStop}%
\bibitem [{\citenamefont {Gardi}\ \emph {et~al.}(2007)\citenamefont {Gardi},
  \citenamefont {Kuokkanen}, \citenamefont {Rummukainen},\ and\ \citenamefont
  {Weigert}}]{Gardi:2006rp}%
  \BibitemOpen
  \bibfield  {author} {\bibinfo {author} {\bibfnamefont {E.}~\bibnamefont
  {Gardi}}, \bibinfo {author} {\bibfnamefont {J.}~\bibnamefont {Kuokkanen}},
  \bibinfo {author} {\bibfnamefont {K.}~\bibnamefont {Rummukainen}}, \ and\
  \bibinfo {author} {\bibfnamefont {H.}~\bibnamefont {Weigert}},\ }\href
  {\doibase 10.1016/j.nuclphysa.2006.12.004} {\bibfield  {journal} {\bibinfo
  {journal} {Nucl. Phys. A}\ }\textbf {\bibinfo {volume} {784}},\ \bibinfo
  {pages} {282} (\bibinfo {year} {2007})},\ \Eprint
  {http://arxiv.org/abs/hep-ph/0609087} {arXiv:hep-ph/0609087} \BibitemShut
  {NoStop}%
\bibitem [{\citenamefont {Balitsky}\ and\ \citenamefont
  {Chirilli}(2008)}]{Balitsky:2007feb}%
  \BibitemOpen
  \bibfield  {author} {\bibinfo {author} {\bibfnamefont {I.}~\bibnamefont
  {Balitsky}}\ and\ \bibinfo {author} {\bibfnamefont {G.~A.}\ \bibnamefont
  {Chirilli}},\ }\href {\doibase 10.1103/PhysRevD.77.014019} {\bibfield
  {journal} {\bibinfo  {journal} {Phys. Rev. D}\ }\textbf {\bibinfo {volume}
  {77}},\ \bibinfo {pages} {014019} (\bibinfo {year} {2008})},\ \Eprint
  {http://arxiv.org/abs/0710.4330} {arXiv:0710.4330 [hep-ph]} \BibitemShut
  {NoStop}%
\bibitem [{\citenamefont {Berger}\ and\ \citenamefont
  {Stasto}(2011)}]{Berger:2010sh}%
  \BibitemOpen
  \bibfield  {author} {\bibinfo {author} {\bibfnamefont {J.}~\bibnamefont
  {Berger}}\ and\ \bibinfo {author} {\bibfnamefont {A.}~\bibnamefont
  {Stasto}},\ }\href {\doibase 10.1103/PhysRevD.83.034015} {\bibfield
  {journal} {\bibinfo  {journal} {Phys. Rev. D}\ }\textbf {\bibinfo {volume}
  {83}},\ \bibinfo {pages} {034015} (\bibinfo {year} {2011})},\ \Eprint
  {http://arxiv.org/abs/1010.0671} {arXiv:1010.0671 [hep-ph]} \BibitemShut
  {NoStop}%
\bibitem [{\citenamefont {Fujii}\ and\ \citenamefont
  {Watanabe}(2013)}]{Fujii:2013gxa}%
  \BibitemOpen
  \bibfield  {author} {\bibinfo {author} {\bibfnamefont {H.}~\bibnamefont
  {Fujii}}\ and\ \bibinfo {author} {\bibfnamefont {K.}~\bibnamefont
  {Watanabe}},\ }\href {\doibase 10.1016/j.nuclphysa.2013.06.011} {\bibfield
  {journal} {\bibinfo  {journal} {Nucl. Phys. A}\ }\textbf {\bibinfo {volume}
  {915}},\ \bibinfo {pages} {1} (\bibinfo {year} {2013})},\ \Eprint
  {http://arxiv.org/abs/1304.2221} {arXiv:1304.2221 [hep-ph]} \BibitemShut
  {NoStop}%
\bibitem [{\citenamefont {Lappi}\ and\ \citenamefont
  {M\"antysaari}(2013)}]{Lappi:2013zma}%
  \BibitemOpen
  \bibfield  {author} {\bibinfo {author} {\bibfnamefont {T.}~\bibnamefont
  {Lappi}}\ and\ \bibinfo {author} {\bibfnamefont {H.}~\bibnamefont
  {M\"antysaari}},\ }\href {\doibase 10.1103/PhysRevD.88.114020} {\bibfield
  {journal} {\bibinfo  {journal} {Phys. Rev. D}\ }\textbf {\bibinfo {volume}
  {88}},\ \bibinfo {pages} {114020} (\bibinfo {year} {2013})},\ \Eprint
  {http://arxiv.org/abs/1309.6963} {arXiv:1309.6963 [hep-ph]} \BibitemShut
  {NoStop}%
\bibitem [{\citenamefont {Torales-Acosta}(2023)}]{Torales-Acosta:202302}%
  \BibitemOpen
  \bibfield  {author} {\bibinfo {author} {\bibfnamefont {F.}~\bibnamefont
  {Torales-Acosta}},\ }\href@noop {} {}\bibinfo {howpublished} {private
  communication} (\bibinfo {year} {2023})\BibitemShut {NoStop}%
\bibitem [{\citenamefont {Anselmino}\ \emph {et~al.}(2011)\citenamefont
  {Anselmino}, \citenamefont {Barone},\ and\ \citenamefont
  {Kotzinian}}]{Anselmino:2011ss}%
  \BibitemOpen
  \bibfield  {author} {\bibinfo {author} {\bibfnamefont {M.}~\bibnamefont
  {Anselmino}}, \bibinfo {author} {\bibfnamefont {V.}~\bibnamefont {Barone}}, \
  and\ \bibinfo {author} {\bibfnamefont {A.}~\bibnamefont {Kotzinian}},\ }\href
  {\doibase 10.1016/j.physletb.2011.03.067} {\bibfield  {journal} {\bibinfo
  {journal} {Phys. Lett. B}\ }\textbf {\bibinfo {volume} {699}},\ \bibinfo
  {pages} {108} (\bibinfo {year} {2011})},\ \Eprint
  {http://arxiv.org/abs/1102.4214} {arXiv:1102.4214 [hep-ph]} \BibitemShut
  {NoStop}%
\bibitem [{\citenamefont {Barone}\ and\ \citenamefont
  {Predazzi}(2002)}]{Barone:2002cv}%
  \BibitemOpen
  \bibfield  {author} {\bibinfo {author} {\bibfnamefont {V.}~\bibnamefont
  {Barone}}\ and\ \bibinfo {author} {\bibfnamefont {E.}~\bibnamefont
  {Predazzi}},\ }\href@noop {} {\emph {\bibinfo {title} {{High-Energy Particle
  Diffraction}}}},\ \bibinfo {series} {Texts and Monographs in Physics}, Vol.\
  \bibinfo {volume} {v.565}\ (\bibinfo  {publisher} {Springer-Verlag},\
  \bibinfo {address} {Berlin Heidelberg},\ \bibinfo {year} {2002})\BibitemShut
  {NoStop}%
\bibitem [{\citenamefont {Diehl}\ and\ \citenamefont
  {Sapeta}(2005)}]{Diehl:2005pc}%
  \BibitemOpen
  \bibfield  {author} {\bibinfo {author} {\bibfnamefont {M.}~\bibnamefont
  {Diehl}}\ and\ \bibinfo {author} {\bibfnamefont {S.}~\bibnamefont {Sapeta}},\
  }\href {\doibase 10.1140/epjc/s2005-02242-9} {\bibfield  {journal} {\bibinfo
  {journal} {Eur. Phys. J. C}\ }\textbf {\bibinfo {volume} {41}},\ \bibinfo
  {pages} {515} (\bibinfo {year} {2005})},\ \Eprint
  {http://arxiv.org/abs/hep-ph/0503023} {arXiv:hep-ph/0503023} \BibitemShut
  {NoStop}%
\bibitem [{\citenamefont {Caldwell}\ and\ \citenamefont
  {Kowalski}(2010)}]{Caldwell:2010zza}%
  \BibitemOpen
  \bibfield  {author} {\bibinfo {author} {\bibfnamefont {A.}~\bibnamefont
  {Caldwell}}\ and\ \bibinfo {author} {\bibfnamefont {H.}~\bibnamefont
  {Kowalski}},\ }\href {\doibase 10.1103/PhysRevC.81.025203} {\bibfield
  {journal} {\bibinfo  {journal} {Phys. Rev. C}\ }\textbf {\bibinfo {volume}
  {81}},\ \bibinfo {pages} {025203} (\bibinfo {year} {2010})}\BibitemShut
  {NoStop}%
\end{thebibliography}%

\end{document}